\documentclass[twocolumn,pra,showpacs,final,superscriptaddress,10pt,aps,floatfix]{revtex4}
\usepackage[usenames]{color}
\usepackage{amsmath}
\usepackage{amssymb}
\usepackage{mathrsfs}
\usepackage{bbold}
\usepackage{euscript}
\usepackage{graphicx}
\usepackage{graphics}
\usepackage{bm}

\newcommand{\delsq}{\bigtriangledown^2}

\definecolor{RED}{rgb}{1,0,0}
\begin{document}
\title{Multiconfiguration Time-Dependent Hartree-Fock Treatment of Electronic and Nuclear Dynamics in Diatomic Molecules}

\author{D.  J. Haxton}
\affiliation{Chemical Sciences and Ultrafast X-ray Science Laboratory, Lawrence Berkeley National Laboratory, Berkeley CA 94720}
\author{K. V. Lawler}
\affiliation{Chemical Sciences and Ultrafast X-ray Science Laboratory, Lawrence Berkeley National Laboratory, Berkeley CA 94720}
\author{C. W.  McCurdy}
\affiliation{Chemical Sciences and Ultrafast X-ray Science Laboratory, Lawrence Berkeley National Laboratory, Berkeley CA 94720}
\affiliation{Departments of Applied Science and Chemistry, Davis CA 95616}


\begin{abstract}
The multiconfiguration time-dependent Hartree-Fock (MCTDHF) method is formulated for treating the coupled electronic and nuclear dynamics of diatomic molecules without the Born-Oppenheimer approximation. The method treats the full dimensionality of the electronic motion, uses no model interactions, and is in principle capable of an exact nonrelativistic description of diatomics in electromagnetic fields.  An expansion of the wave function in terms of configurations of orbitals whose dependence on internuclear distance is only that provided by the underlying prolate spheroidal coordinate system is demonstrated to provide the key simplifications of the working equations that allow their practical solution.  Photoionization cross sections are also computed from the MCTDHF wave function in calculations using short pulses. 
\end{abstract}

\maketitle

\section{Introduction}

New sources of short radiation pulses, in particular high-harmonic generation~\cite{krausz2009,Sekikawa} and free-electron lasers~\cite{Ayvazyan}, promise to enable a new generation of pump/probe experiments on molecules in which the central frequencies of both pulses are in the ultraviolet or X-ray.  
The short time scales that have recently become practical for  these measurements extend to subfemtosecond pulses with delays between them on the order of femtoseconds or tens of femtoseconds.  This generation of experiments can also involve probe pulses which ionize or dissociate the target molecule and thus add the dimension of the time-resolved measurement of the electron and molecular fragment energy distributions.   To accurately interpret and describe the results of these experiments,  \textit{ab initio} methods must be developed to treat highly electronically excited and strongly nonadiabatic molecular dynamics. 
Coincidence experiments~\cite{doernercoltrims}, for instance, demand the capability to describe multiple ionization and dissociation, and nonlinear effects~\cite{stolow_nmd, mukamelxray} may entail the excitation of multiple electrons.  In general, most of these phenomena currently remain beyond the reach of accurate \textit{ab initio} theoretical descriptions.  

However, significant headway has been made.
 Approaches employing
classical trajectories on coupled Born-Oppenheimer surfaces~\cite{martinezrev, mckoyno2} are well suited for dynamics on low lying excited bound
electronic states, and have been applied to molecules as large as DNA bases~\cite{martinezUT}.  With an approximate treatment of the coupling to the ionization 
continuum, such trajectory methods may describe time-resolved photoelectron signals in pump-probe experiments~\cite{mckoyno2,martinezUT}; such a treatment
also permitted the study of the quantum nuclear dynamics of Auger decay~\cite{cederbaum_h2o_auger}.

While these methods have already shown great progress in describing a range of non-Born-Oppenheimer and excited-state effects, 
their utility is greatest for situations in which ionization may be treated approximately and in which the interacting electronic states may be explicitly
identified.  
Several other approaches avoid the use of Born-Oppenheimer states altogether
and show promise for treating highly electronically excited or nonadiabatic electronic and nuclear dynamics~\cite{nakai_nomo_2007, nagashima_mcmo_2009}.  In a similar context, ionization has been included in a variational treatment that explicitly includes electron-nuclear correlation~\cite{kreibich_leeuwen_gross2004}  and has also been treated with coupled electronic and semiclassical nuclear wave packets~\cite{takatsuka}.

The time-dependent multi-configuration Hartree Fock (MCTDHF) approach would seem to be a natural and widely applicable starting point to the electronic part of this problem, because it is capable in principle of exactly describing the dynamics of many-electron motion. There is a considerable literature on this subject already including analysis of the formulation of MCTDHF and its application to small or
model systems~\cite{Cederbaum2007,Cederbaum2009,Scrinzi_MCTDHF_2004,Scrinzi_MCTDHF_2005,Kato_Kono2008,KatoYamanouchi2009,Nest2006,Nest2007,Levine2008}, and also attacks on the combination of electronic and nuclear motion~\cite{Nest2009}. 
On the basis of this literature, the fundamental idea can be said to be well established that a 
time-dependent linear combination of determinants of time dependent orbitals
should be flexible enough to describe the electronic response a molecule to short intense pulses in any part of the spectrum.     

Similar ideas in a different context have underpinned the development of the multiconfiguration time-dependent Hartree (MCTDH) method~\cite{kucar_tdrh,mey90:73, bec97:113,Meyer_review,Meyer_feature,MCTDH_package},  which has had considerable success in treating problems of nuclear dynamics, including vibronic coupling and reactive scattering.
 However, by comparison, the MCTDHF method for electrons has still not delivered its full potential, particularly in the presence of ionizing fields or including nuclear motion.  The reason appears to lie in several serious technical barriers to its implementation and general application:

\begin{list}{\labelitemi}{\leftmargin=1em}

\item{Electronic and nuclear motion are strongly correlated; the cusps in the electronic wave function at the positions of the nuclei must be accurately represented for all geometries, and the basis set error in the electronic part of the calculation must not depend strongly on nuclear coordinates.}

\item{The evaluation of the two-electron integrals over the time-dependent orbitals must be numerically efficient, otherwise it will dominate the computational time required.}

\item{The ionization continuum must be properly treated within the MCTDHF description, if it is relevant to the problem at hand.}

\item{The nonlinear, unitary, stiff differential equation involving orbitals and configuration coefficients must efficiently numerically integrated.}

\end{list}

Here we address all these difficulties for the case of diatomic molecules.   The key to overcoming the first three is the use of the Finite Element Method and Discrete Variable Representation (FEM-DVR) in prolate spheroidal coordinates. The electronic basis is a set of piecewise interpolating polynomials with cusps at the nuclei, parametrically dependent upon the bond distance.
Such an atom-centered, parametrized basis has been used before~\cite{kreibich_leeuwen_gross2004}, but the choice of prolate spheroidal  DVR has several important advantages.  

\begin{list}{\labelitemi}{\leftmargin=1em}

\item{Orbitals defined in these coordinates are orthogonal for all values of the internuclear distance $R$, and therefore a single set of orbitals may be used for all nuclear
geometries, radically reducing numerical effort.  The prolate spheroidal DVR basis allows for a sparse representation of the primitive two-electron integrals and their rapid contraction
into orbital matrix elements, and it leads to rapid convergence of both bound and continuum wave functions, as we have found in previous fixed-nucei calculations  on single and double photoionization of H$_2$ \cite{prolate1_2009,prolate2_2009,prolate3_2010}.  }

\item{The basis enables rigorous inclusion of the ionization continuum via Exterior Complex Scaling (ECS)~\cite{moirev}.  The implementation of the ECS formalism with the FEM-DVR approach is well established as a formally sound and computationally efficient treatment of both photoionization and electron-impact ionization~\cite{TopicalReview04}, because it imposes correct outgoing scattering boundary conditions for both single and double ionization
~\cite{Horner2008a, palacios08,palacios09, H2_DPI_vanroose,prolate1_2009,prolate2_2009,prolate3_2010, Yip_2010}.}

\item{Finally, inclusion of nuclear motion in the case of diatomics, without the necessity of the Born-Oppenheimer approximation, is also greatly simplified by the prolate spheroidal FEM-DVR. 
We employ a good approximation to the exact nonrelativistic molecular Hamiltonian including the interaction with the radiation field, omitting only the Coriolis coupling and mass polarization terms.  }
\end{list}

Thus, the prolate spheroidal DVR basis is ideally suited for the study time dependent excited state  dynamics of diatomic molecules.  Its matrix elements appear in a nonlinear differential equation of potentially large dimension, however, and the integration of this equation is a formidable challenge in its own right.  Several different methods of integrating the MCTDHF equations have been described, and we have found an efficient and stable generalization of the method of Ref.~\cite{bec97:113}. 

The choice of  a single set of electronic orbitals with parametric dependence on $R$ that is used in this approach might appear to be, from the traditional perspective of the Born Oppenheimer approximation, an unnatural starting point,  and raises questions regarding the convergence of the wave function for coupled electronic and nuclear motion with respect to the numbers of configurations included, which we must address.  However, this choice also offers important advantages, simplifying the working equations in a critical way and allowing the slow, nuclear degree of freedom to be distributed across supercomputer processors. 

The outline of this paper is as follows.  In  Sec.~\ref{sec:MCTDHF}, we  first review the working formalism for MCTDHF without nuclear motion.  In Sec.~\ref{sec:prolate} we describe the formulation of the diatomic problem in prolate spheroidal coordinates, and the form of the Hamiltonian appropriate to those coordinates when the underlying electronic basis has a parametric dependence on internuclear distance.    Secs.~\ref{sec:ham_nuc_motion}  and \ref{sec:nuc_motion} describe the inclusion of nuclear motion using a DVR basis in the nuclear coordinate $R$ in combination with the MCTDHF treatment of electronic motion. We then discuss the application of the ECS transformation to treat ionization within the MCTDHF framework in Sec.~\ref{sec_ECS}.   The remaining sections discuss computational details and some preliminary results for both bound and continuum electronic motion.  We use atomic units throughout.

\section{MCTDHF formalism for fixed nuclei} 
\label{sec:MCTDHF}

The MCTDHF working equations have been formulated previously~\cite{Scrinzi_MCTDHF_2004,Nest2007,Cederbaum2007, Levine2008, Kato_Kono2008, Cederbaum2009}, and we will give here only a brief description of the working equations in order to establish the starting point for the inclusion of nuclear motion using this approach, and to indicate how exterior complex scaling of the electronic coordinates is implemented in this context to treat ionization.   The MCTDHF approach begins with an expansion of the electronic wave function in antisymmetrized products (determinants) of time-dependent spin orbitals
\begin{equation}
\vert \Psi(t) \rangle = \sum_a  A_a(t) |{\vec{n}}_a(t) \rangle
\label{eq:trialfcn}
\end{equation}
in which each antisymmetrized product of $N$ spin orbitals is specified by the vector ${\vec{n}}_a$  and is defined by 
\begin{equation}
\vert {\vec{n}}_a(t) \rangle =  \mathscr{A}\left( \vert \phi_{n_{a1}}(t) \rangle \times ... \vert \phi_{n_{aN}}(t) \rangle \right) \, .
\end{equation}
We use spin restricted orbitals which are product functions of the space and spin coordinates $\vec{q}_i = \{ \vec{r}_i, \Xi_i \}$ of a single electron,
\begin{equation}
\langle \vec{q} \vert \phi_n(t) \rangle = \phi_{\alpha_n}(\vec{r}) \, \Omega_{ms_n}(\Xi)
\end{equation}
where $\Omega$ is a spinor, $\Omega=\alpha$ or $\Omega=\beta$.  Alternatively, in second quantization we can write
\begin{equation}
\vert \phi_n (t) \rangle = a^\dag_{\alpha_n ms_n}(t) \vert 0 \rangle
\end{equation}
in terms the creation operator, $a^\dag$, for spin orbitals.
The space part of each  spin orbital is expanded in a set of time-independent DVR  basis functions, $f_j$, which we will specify in Sec.~\ref{sec:prolate} below,
\begin{equation}
 \phi_{\alpha}(\vec{r},t) = \sum_j c^\alpha_j(t) f_j(\vec{r}) \, ,
 \label{eq:spfexpansion}
\end{equation}
so each spatial orbital is associated with a time-dependent coefficient vector $\vec{c}_\alpha(t)$; the vector of all $\vec{c}_\alpha$ is denoted $\vec{c}$.

The MCTDHF equations are based on the application of the Dirac-Frenkel variational principle for the time-dependent Schr\"odinger equation to the trial function in Eq. (\ref{eq:trialfcn}),
\begin{equation}\
\begin{split}
 \Big\langle \delta \Psi(t) \Big\vert \hat{H} - i \frac{\partial}{\partial t} \Big\vert \Psi(t) \Big\rangle 
 - \sum_{\alpha \le \beta} \lambda_{\alpha\beta} \Big( \left\langle \delta \phi_\alpha \vert \phi_\beta \right\rangle - \delta_{\alpha\beta} \Big)= 0 \, ,
\end{split}
\label{eq:diracfrenkel}
\end{equation}
where we include the constraint that the orbitals remain orthonormal, $\left\langle \phi_\alpha \vert \phi_\beta \right\rangle - \delta_{\alpha\beta} = 0$, along with the corresponding Lagrange multipliers $\lambda_{\alpha\beta}$.  The variations in Eq. (\ref{eq:diracfrenkel}) are variations in the coefficients $A_{\vec{n}}$ and $\vec{c}_\alpha$.

In this work we employ a full configuration interaction (CI) representation of the wave function in Eq. (\ref{eq:trialfcn}), although further application of the method could entail restricted CI wave functions such as the
``complete active space'' approach that are standard in modern quantum chemistry.  An important point is that in the full CI case addressed here the solution of Eq.(\ref{eq:diracfrenkel}) yields $\lambda_{\alpha\beta}=0$.    The wave function is invariant with respect to rotations among the orbitals, which may be compensated for by rotations among the A-coefficients.
The solution of Eq.(\ref{eq:diracfrenkel}) is therefor not uniquely defined and unique time propagation requires an additional 
constraint besides orthogonality of the orbitals.  The simplest constraint is to set  the  time-derivative matrix 
$g$ in the orbital basis to zero,
\begin{equation}
g_{\alpha\beta} = \left\langle \phi_\alpha \left\vert i \frac{\partial}{\partial t} \right \vert \phi_\beta \right\rangle = 0 \quad .
\label{eq:constraint}
\end{equation}
For the orbitals one obtains the equation of motion
\begin{equation}
\begin{split}
 i \frac{\partial}{\partial t}& \vec{c}_\alpha = 
(\mathbf{1}-{\bf P}  )  \sum_\beta   \left[    {\bf h^{(1)}} \delta_{\alpha\beta}
+ \sum_{\gamma} \rho^{-1}_{\alpha  \gamma } {\bf\widetilde{W}^{\gamma\beta}} \right]   
   \vec{c}_{\beta} \, ,
\end{split}
\label{eq:orbEOM}
\end{equation}
where the projector $\mathbf{P}$ is the matrix representation of the projection operator, $\hat{P} = \sum_\alpha |\phi_\alpha (t)\rangle \langle \phi_\alpha(t) |$, onto the space spanned by the orbitals at time $t$, 
\begin{equation}
\mathbf{P}_{j,j'} =  \sum_\alpha c^\alpha_j(t) \,  c^{\alpha *}_{j'}(t) \, ,
\end{equation}
so that $\mathbf{1}-\mathbf{P}$ projects this equation on to the space orthogonal to that spanned by the orbitals. 
Our convention is that boldface symbols are matrices in either the orbital ($\vec{c}$) or configuration ($\vec{A}$) basis, 
and ${\bf O} \vec{v}$ stands for matrix multiplication. 
In Eq. (\ref{eq:orbEOM}),  $\rho_{\alpha  \gamma }$ is the reduced one-electron density matrix for the wave function in Eq. (\ref{eq:trialfcn}), 
\begin{equation}
\rho_{\alpha\beta} = \sum_{a\ b\ ms} A_a^* A_b \langle \vec{n}_a \vert a^\dag_{\alpha \, ms} \, a_{\beta \, ms} \vert \vec{n}_b \rangle \quad ,
\end{equation}
and  ${\bf h^{(1)}}$ is the matrix of one-body operators in the Hamiltonian with respect to the underlying time-independent basis in Eq. (\ref{eq:spfexpansion}).  All quantities in  Eq. (\ref{eq:orbEOM}) are time-dependent except for the identity and ${\bf h^{(1)}}$ matrices (in the absence of an external time-dependent field).  The reduced two-electron operator $\widetilde{{\bf W}}$ is defined~\cite{Cederbaum2007,Cederbaum2009}  in terms of the reduced two-particle density matrix, $\Gamma_{\gamma s \alpha l}$ and the  electron repulsion, $ \mathbf{W}_{sl}$, expressed as its matrix representation in $\mathbf{r}_1$ in the underlying time-independent basis, 
\begin{eqnarray}
 {\bf\widetilde{W}^{\gamma\beta}}&=&\frac{1}{R}\sum_{sl} \Gamma_{\gamma s  \alpha l} \mathbf{W}_{sl} (t)  \, , \\
  \label{eq:two_elec_op}
 \mathbf{W}_{s l}(t) &=& \int \phi_s^*(\vec{r}_2,t) \frac{R}{|\vec{r}_1-\vec{r}_2|} \phi_l(\vec{r}_2,t) d\vec{r}_2 \, .
\end{eqnarray}
 In Eq.(\ref{eq:two_elec_op}) the electron repulsion operator appears as $R/ {|\vec{r}_1-\vec{r}_2|}$ because its matrix elements then have no $R$ dependence in the underlying prolate spheroidal DVR, and therefore also in the orbital basis.  That fact significantly simplifies the implementation of the MCTDHF equations in these coordinates.

The time derivative of the orbital coefficients being specified, one then obtains the equations of motion for the A-coefficients
\begin{equation}
i \frac{\partial}{\partial t} \vec{A} =  {\bf H}  \vec{A}     \qquad \quad \mathbf{H}_{a,a'} = \langle \vec{n}_a \vert H \vert \vec{n}_{a'} \rangle \, , 
\label{eq:Aequation}
\end{equation}
where ${\mathbf H}$ is the matrix of the Hamiltonian in the configuration basis, consistent with Eq.(\ref{eq:constraint}).

For Hermitian Hamiltonians, the MCTDHF equations conserve the norm and the expectation value of the energy~\cite{mey90:73}. We next turn to the specification of the underlying basis in Eq.(\ref{eq:spfexpansion}) in prolate spheroidal coordinates and the implementation of ECS in that basis for the treatment of ionization during the propagation of Eqs. (\ref{eq:Aequation}) and (\ref{eq:orbEOM}).

\section{Fixed-nuclei Hamiltonian and wave function}
\label{sec:prolate}

\subsection{Hamiltonian}

Prolate spheroidal coordinates were used in early quantum chemical calculations on diatomic molecules, because analytic basis sets, such as Slater-type orbitals, in those coordinates could exactly satisfy the cusp conditions on the two nuclei, and because their scaling properties with internuclear distance offered additional computational advantages~\cite{Roothaan1964,McCullough1975}.   We use them here for some of the same reasons, but also because the implementation of the FEM-DVR approach in these coordinates dramatically simplifies the calculation of the two-electron integrals.

If the nuclei of our diatomic molecule are at $\vec{R}_A$ and $\vec{R}_B$, we can define prolate spheroidal coordinates $(\xi,\eta,\varphi)$ for each electron in the usual way by rotating a two-dimensional elliptical coordinate system $(\xi,\eta)$ about the focal axis of the ellipse,
\begin{equation}
\begin{split}
\xi&= \frac{|\vec{r}-\vec{R}_A|+|\vec{r}-\vec{R}_B|}{R}\quad (1\le\xi\le\infty)\\
\eta&=  \frac{|\vec{r}-\vec{R}_A|-|\vec{r}-\vec{R}_B|}{R}\quad (-1\le\eta\le 1)\,, 
\end{split}
\label{eq:prolate}
\end{equation}
and the remaining coordinate, $\varphi$,  $(0\le  \varphi  \le 2\pi)$ is the azimuthal angle. The one-body operators in our Hamiltonian in these coordinates are specified by the Laplacian, 
\begin{equation}
\begin{split}
\nabla ^2  = &\frac{4}{{R^2 (\xi ^2  - \eta ^2 )}}\left[ 
 \frac{\partial }{{\partial \xi_i }}(\xi ^2  - 1)\frac{\partial }{{\partial \xi }} + \frac{\partial }{{\partial \eta }}(1 - \eta ^2 )\frac{\partial }{{\partial \eta }} \right. \\ 
  &\left. + \left( {\frac{1}{{(\xi^2  - 1)}} + \frac{1}{{(1 - \eta^2 )}}} \right)\frac{{\partial ^2 }}{{\partial \varphi ^2 }}\right]\,,
\end{split}
  \label{Laplacian}
\end{equation}
and the electron-nuclear attraction 
\begin{equation}
-\frac{Z_A}{|\vec{r}-\vec{R}_A|}-\frac{Z_B}{|\vec{r}- \vec{R}_B|}=-Z_A Z_B\frac{4\xi}{R(\xi^2-\eta^2)}  \, ,
\end{equation}
while  the one-electron volume element is
\begin{equation}
dV = (R/2)^3 (\xi^2-\eta^2) \, d \xi \,  d \eta \, d \varphi \, .
\label{eq:vol-elem}
\end{equation}
For an N-electron problem then, a factor of $R^3$ appears in the volume element in Eq.(\ref{eq:vol-elem})  for integration over the coordinates of each electron. We eliminate that factor in a fixed-nuclei calculation by solving for $R^{3N/2}$ times the electronic wave function.  In Sec.\ref{sec:nuc_motion} when we consider nuclear motion we will solve for  $R^{3N/2 +1}$ times the total wave function to further simplify the form of the nuclear kinetic energy operator.

\subsection{Wavefunction}

We make use of an FEM-DVR in both $\xi$ and $\eta$ for each electron, and our primitive basis functions are  products of those DVR functions with a factor describing the $\phi$ motion with a particular angular momentum projection, $\mathcal{M}$,  along the molecular axis.  As we have noted in previous studies~\cite{prolate2_2009,prolate3_2010}  on H$_2$,  the specification of the FEM-DVR depends on whether $\mathcal{M}$ is even or odd, in order to properly represent the analytic dependence of the wave function as $\xi$ or $\eta$ approach the singularity at $\pm1$.   Since we are constructing a DVR in each finite element, we use Gauss-Radau quadrature on in the element in $\xi$ beginning at $\xi=1$, and Gauss-Lobatto quadrature for the others.  We use Gauss-Legendre quadrature to define the DVR in $\eta$.  So defining the basic DVR interpolating functions in terms of the quadrature points and weights by (with $x=\eta$ or $\xi$)
\begin{equation}
\displaystyle \chi_{i}(x)= \frac{1}{\sqrt{w_i}}\prod_{j\neq i}^N\frac{x-x_j}{x_i-x_j} 
\times
\begin{cases}
\displaystyle  1 & \text{even} \ \mathcal{M} \\
\displaystyle  \sqrt{\frac{x^2-1}{x_i^2-1}} & \text{odd} \ \mathcal{M} \\
\end{cases} \ \ .
\label{eq:prim_DVR}
\end{equation}
We define the one-electron primitive FEM-DVR functions according to  
\begin{equation} 
f_{ia}^\mathcal{M} =
 \sqrt{\frac{8}{\xi_a^2 - \eta_i^2} } \chi_i(\eta) \chi_a(\xi) \frac{e^{i\mathcal{M}\varphi}}{\sqrt{2\pi}} 
\label{eq:DVRbasis}
\end{equation}
%

Note that if we use the underlying quadrature to calculate the overlaps, because for fixed nuclei we solve for $R^{3N/2}$ times the wave function, these functions are orthonormal with respect to the volume element $\frac{1}{8}(\xi^2-\eta^2) \, d \xi \,  d \eta \, d \varphi$.

We have discussed the simplifications of the integrals of $ 1/r_{12} $ in a similar basis before~\cite{prolate3_2010},  in which we used spherical harmonics in the variables $\eta$ and $\varphi$ instead of the DVR in $\eta$  we are using here.   In this basis the simplifications are even more powerful.  By making use of the Neumann expansion of  $1/r_{12} $ in prolate spheroidal coordinates, and solving Poisson's equation in $\xi$ in the FEM-DVR basis  followed by using the Gauss quadrature in $\eta$ we arrive ultimately at expressions for the two electron integrals of the general form
\begin{equation}
\begin{split}
& \left\langle f_{ia}^{\mathcal{M}_1}(\vec{r}_1) f_{jb}^{\mathcal{M}_2}(\vec{r}_2) \left| \frac{1}{r_{12}} \right| f_{i'a'}^{\mathcal{M}_3}(\vec{r}_1) f_{j'b'}^{\mathcal{M}_4}(\vec{r}_2) \right\rangle  =  \\
&\qquad \delta_{i,i'}\delta_{j,j'} \delta_{a,a'} \delta_{b,b'}      F_{i,j,a,b}^{m}
\end{split}
\label{eq:two_e_int}
\end{equation}
Where in addition we have the requirement that ${m = \mathcal{M}_3-\mathcal{M}_1 = \mathcal{M}_2-\mathcal{M}_4}$.    We give the explicit formula in Appendix~\ref{sec:two-elec-ints}.  We can see therefore that the two-electron integrals in the FEM-DVR basis are diagonal in the indices corresponding to  both $\xi$ and $\eta$.   This is an immense simplification over the use of analytic basis functions, like Gaussians, and dramatically reduces the effort in transforming the two electron integrals from the FEM-DVR basis to the time-dependent orbital basis, and thereby simplifies and speeds up the both the construction of the two-electron portion of reduced Hamiltonian in Eq.(\ref{eq:two_elec_op}) and its operation in Eq.(\ref{eq:orbEOM}).

\section{Hamiltonian and wave function for nuclear motion}
\label{sec:ham_nuc_motion}

\subsection{Hamiltonian}

When we include nuclear motion, we must account for the fact that this FEM-DVR basis is $R$-dependent through the dependence of the electronic coordinates on internuclear distance.  The DVR quadrature points that define the basis in Eq.(\ref{eq:DVRbasis}) move with $R$.   As has been discussed before -- for example by Esry and Sadgepour~\cite{Esry1999} -- in a basis of 
functions of prolate spheroidal coordinates, the derivatives with respect to $R$ in the Hamiltonian must be calculated holding the cartesian coordinates, 
$x, y$ and $z$  of the electrons fixed instead of 
holding $\xi, \eta$ and $\phi$ fixed.  The relation between those derivates is
\begin{equation}
\left(\frac{\partial}{\partial R}\right)_{xyz} = \left(\frac{\partial}{\partial R}\right)_{\xi\eta\phi} - \sum_{i=1}^N \frac{1}{R} \hat{Y_i}
\end{equation}
where the sum is  over electrons,  and the operator $\hat{Y}$ is
\begin{equation}
\begin{split}
\hat{Y} =  \frac{1}{\xi^2 - \eta^2} &\left((\xi+\alpha\eta)(\xi^2-1) \frac{\partial}{\partial \xi} \right. \\
 & \left. + (\eta+\alpha\xi)(1-\eta^2) \frac{\partial}{\partial \eta} \right)
\end{split}
\end{equation}
with  $\alpha = (M_A-M_B)/(M_A+M_B)$ being the mass asymmetry parameter.  We must use these relations when constructing the second derivative term, $\left(\partial^2/\partial R^2 \right)_{x,y,z}$, in the nuclear kinetic energy.  When including nuclear motion, we calculate $R^{3N/2+1}$ times the wave function, and so the radial part of the nuclear kinetic energy operator then becomes
\begin{equation}
\begin{split} 
K_R=&-\frac{1}{2\mu_{R}}
\Bigg[ 
\left( \frac{\partial^2}{\partial R^2} \right)_{\xi\eta\phi} 
+ \frac{1}{R^2} \left(\sum_i \hat{Y_i}+\frac{3N}{2}\right)^2 \\
& - \left(\sum_i \hat{Y_i}+\frac{3N}{2}\right) \left( \frac{2}{R} \left( \frac{\partial}{\partial R} \right)_{\xi\eta\phi} - \frac{1}{R^2} \right) 
\Bigg]
\end{split}
\label{eq:nucke}
\end{equation}

For more than one electron, we do not employ an exact treatment, which might be accomplished in polyspherical coordinates~\cite{polynote, gatti} for example, but instead employ a straightforward adaptation of the one-electron Hamiltonian~\cite{Esry1999} that omits several minor terms unimportant for a host of nonadiabatic dynamics relevant to attosecond physics.  In particular, our electron position vectors are represented in prolate spheroidal coordinates relative to the center of mass of the nuclei.  We therefore omit
terms relating of mass polarization of the electrons, i.e., for N electrons, the deviation of the center of mass of the nuclei from the center of mass of the nuclei and any N-1 subset of the N electrons.  In the chosen coordinate system, such terms would be represented as two-electron derivative operators.  In contrast, an exact polyspherical treatment such as  that using heliocentric Radau coordinates~\cite{smithradau}, in prolate spheroidal form, would entail
nonseparable corrections to the electron-nuclear potential which, without further approximation, would be intractable in the present framework. 

We additionally omit Coriolis coupling and write the Hamiltonian for rotational quantum number $J$ as
\begin{equation}
H =  K_R+
\frac{1}{2\mu_R R^2} \left[ J(J+1) -2J_z^2 + \hat{l}^2 \right]
-\frac{1}{2{\mu_e}} \sum_i \delsq_i + V
\label{eq:nucham}
\end{equation}
The reduced masses are defined as $\mu_R = m_A m_B / (m_A + m_B)$ where A and B are the two nuclei, and (in atomic units) $\mu_e$ for an N-electron system is $\mu_e = (m_A + m_B + N-1)/ (m_A + m_B + N)$. 
$J_z$ is the projection of the electronic angular momentum on the internuclear axis and equals the sum over the $l_z$
eigenvalues of the individual electrons,   $\sum_{i=1}^N \mathcal{M}_i$.  The operator $\hat{l}^2$ is the square of the electronic angular momentum operator.  The potential, $V$, includes the electron-electron repulsion, electron-nuclear attraction, internuclear repulsion, 
and 
time-dependent dipole interaction term, if an external field is being applied.  At present, we also omit 
the two-electron terms in $\hat{l}^2$ (proportional to $\hat l_i \hat l_j$ and in $(\sum_i \hat Y_i)^2$ (proportional to $\hat Y_i \hat Y_j)$. 
These terms are in general similar in magnitude to the mass polarization terms that we have already omitted, and also not relevant to the processes of immediate interest.
Usual nonadiabatic effects such as curve crossing transitions are driven by the cross term in Eq.(\ref{eq:nucke}) involving products of electronic and vibrational momenta, as opposed to the terms that we have omitted containing terms quadratic in the electronic momenta.  They and the Coriolis coupling may be included in future applications, and their omission here accelerates the numerical implementation.

\subsection{Wave function}

To include nuclear motion and electronic motion simultaneously, and also avoid the Born-Oppenheimer Approximation, we begin with a trial function  in which we use the same configurations specified in Eq.(\ref{eq:trialfcn}), expressing the time-dependent orbitals in the prolate spheroidal FEM-DVR, and taking advantage of the implicit dependence on the internuclear distance $R$ of those orbitals,
\begin{equation}
\langle R \vert \Psi(t) \rangle = \sum_{a,\kappa}  A_{a,\kappa} (t) |{\vec{n}}_a(t;R) \rangle \chi_\kappa(R) \, .
\label{eq:nuc_trialfcn}
\end{equation}
In Eq.(\ref{eq:nuc_trialfcn}) the function $\chi_\kappa(R)$ is an ordinary FEM-DVR basis function based on Gauss-Lobatto quadrature within finite elements in $R$, and is labeled by the grid point $R_\kappa$ at which it is nonzero.  The coefficients, $A_{a,\kappa} (t)$, of the configurations depend explicitly on the index $\kappa$ as well as the configuration index $a$.  So this representation of the wave function uses MCTDHF for the electrons while using a full primitive basis DVR representation of nuclear motion.  
 The configurations have a parametric dependence on $R$, which we emphasize with the notation $  |{\vec{n}}_a(t;R) \rangle$, and that dependence comes entirely through the $R$ dependence of the prolate coordinates, and thus of the FEM-DVR grid for the electrons,
\begin{equation}
\langle \vec{q} \vert \phi_n(t;R) \rangle =  \phi_{\alpha_n}(\vec{r}(R),t) \Omega_{ms_n}(\Xi)
\end{equation}
%
where $\vec{r}(R) = (\xi(R),\eta(R),\varphi)$ are the prolate spheroidal coordinates.

Using this trial function in the variational principle in Eq.(\ref{eq:diracfrenkel}) means that a single set of orbitals is used to describe the electronic motion for all $R$, and thus that the coefficients $c^\alpha_{i}(t)$
\begin{equation}
 \phi_{\alpha}(\vec{r}    ,t) = \sum_{i} c^\alpha_{i}(t) f_{i}(\vec{r}(R)) \, ,
 \label{eq:spfexpansion_R}
\end{equation}
do not depend on $R$. 
 This fact simplifies the resulting MCTDHF equations in a fundamental way, and is key to the practically of this approach.  
The ansatz in Eq.(\ref{eq:nuc_trialfcn}) is largely motivated by this fact, but  it is clear nonetheless that it is capable in principle  of representing the exact wave function if a sufficient number of orbitals is included.  Our numerical tests below will test the efficiency of  this expansion.  The coefficients of the configurations depend explicitly on the $R$ index, and thus are capable of weighting the configurations constructed from those orbitals differently at different internuclear distances, and thus different orbitals can contribute differently at different values of $R$. 
The nuclear cusps of the wave function are represented accurately at all values of $R$ in this approach, and the orbitals are automatically orthogonal for all internuclear distances.  


\section{MCTDHF working equations including nuclear motion}
\label{sec:nuc_motion}

Using the following notation for combining parts of the Hamiltonian in Eq(\ref{eq:nucham}):
\begin{equation}
\begin{split}
\hat{T}^{el} & = \sum_i \left[ -\frac{R^2}{2\mu_e} \nabla_i^2 - \frac{1}{2\mu_R} \left(\hat{Y_i} + \frac{3}{2}\right)^2 + \frac{1}{2\mu_R} \hat{l_i}^2 \right] \, , \\
\hat{T}^R & = -\frac{1}{2} \frac{\partial^2}{\partial R^2} \qquad \hat{D}^{el}  =  \sum_i \left[  \hat{Y_i}+\frac{3}{2}\right] \, , \\
\hat{D}^R &  = \left( \frac{2}{R} \left( \frac{\partial}{\partial R} \right)_{\xi\eta\phi} - \frac{1}{R^2} \right) \quad , \\
\end{split}
\label{eq:shorthand}
\end{equation}
the Hamiltonian including a radiation field employed here for $R^{3N/2+1}$ times the wave function, omitting Coriolis and two-electron terms in $\hat{l}^2$ and $(\sum_i \hat Y_i)^2$, may be compactly expressed as 
\begin{equation}
\hat{H_0} = \frac{1}{R^2} \hat{T}^{el} + \frac{1}{R} V + \frac{1}{\mu_R} \left( \hat{T}^R + \hat{D}^R \hat{D}^{el} \right)
\end{equation}
\begin{equation}
\begin{split}
H(t) & = \hat{H_0} + R \, \mathcal{E}(t) \,  \hat{\mu} \quad \mathrm{(length)} \\
  & = \hat{H_0} + \frac{1}{R} \frac{\mathcal{A}(t)}{c} \hat{\mu} \quad \mathrm{(velocity)}
\end{split}
\label{eq:compact}
\end{equation}
where $\mathcal{E}(t)$ and $\mathcal{A}(t)$ are the electric field and vector potential, and $\hat{\mu}$ is a coordinate or derivative operator in the electronic prolate spheroidal coordinates.  The operators  $\hat{D}^R$ and $\hat{D}^{el}$ are antihermitian, first order differential operators, and the potential is a separable product of ${1}/{R}$ times a potential  for the nuclear repulsion plus two electron repulsion that is a function of only the prolate spheroidal coordinates:
\begin{equation}
V = \left(Z_1Z_2 + v_{12}(\xi_1,\eta_1,\varphi_1,\xi_2,\eta_2,\varphi_2)\right)
\end{equation}
with $v_{12} = {R}/{r_{12}}$.

The working equations for MCTDHF with nuclear motion are similar to the Born-Oppenheimer version.  The inverse of the reduced single-particle (electronic) reduced density matrix still appears, and that matrix is a sum over nuclear grid points, 
\begin{equation}
\rho_{\alpha \beta}  = \sum_\kappa \rho_{\kappa \alpha \kappa \beta}  \,, 
\end{equation}
where
\begin{equation}
\rho_{\kappa \alpha \tau \beta}  = \sum_{a\ b\ ms} A_{a,\kappa}^* A_{b, \tau} \langle \vec{n}_a \vert a^\dag_{\alpha \, ms} \, a_{\beta \, ms} \vert \vec{n}_b \rangle \, , 
\end{equation}
and  the reduced two-electron matrix $\mathbf{W^{\gamma\beta}}$ is defined similarly.  We also  have the reduced two-electron density matrix defined as
\begin{equation}
\begin{split}
& \Gamma_{\kappa \ \alpha  \ \alpha'   \ \kappa \ \beta  \ \beta'  }  = \\
& \ \ \sum_{a\ b\ ms \ ms'} A_{a,\kappa}^* A_{b, \kappa} \langle \vec{n}_a \vert  \,  a^\dag_{\alpha \, ms}  \, a^\dag_{\alpha' \, ms' }  \, a_{\beta \, ms}  \, a_{\beta' \, ms'} \vert \vec{n}_b \rangle \ .
\end{split}
\end{equation}
By defining 
reduced matrices for ${1}/{R}$, ${1}/{R^2}$, and  $R$ (which appears the length gauge dipole operator), and a reduced derivative operator,
\begin{eqnarray}
&\displaystyle \mathscr{R}^{(Q)}_{\alpha\beta} =  \sum_{\kappa} \rho_{\kappa \ \alpha  \ \kappa \ \beta } R_\kappa^Q \qquad Q=\text{1,-1,-2} \quad , \nonumber \\
&\displaystyle \mathscr{R}^{(-1)}_{\alpha\beta\alpha'\beta'} =  \sum_{\kappa} \Gamma_{\kappa \ \alpha  \ \alpha'   \ \kappa \ \beta  \ \beta'  } \frac{1}{R_\kappa}\quad  ,  \nonumber \\
&\displaystyle \mathscr{D^R}_{\alpha\beta} =  \sum_{\kappa \ \tau } \rho_{\kappa \ \alpha  \ \tau \ \beta } {\bf D^R}_{\kappa \tau} \quad , 
\label{scriptrs}
\end{eqnarray}
along with matrix elements of the differential operators in Eq.(\ref{eq:shorthand}), we arrive at the MCTDHF equations for the orbital coefficients,
\begin{equation}
\begin{split}\displaystyle
 \quad & i\frac{\partial}{\partial t} \vec{c_\alpha} = \sum_{\beta} 
 ({\bf 1}-{\bf P})  \rho^{-1}_{\alpha\beta} \sum_\gamma
\Bigg[ \mathscr{R}^{(-2)}_{\beta\gamma} {\bf T^{el}}  \\
&  + \mathscr{D^R}_{\beta\gamma} \  \bf{D^{el}} 
 +\sum_{\beta'\gamma'} \mathscr{R}^{(-1)}_{\beta\gamma\beta'\gamma'} \ {\bf \bar{W}}_{\beta'\gamma'}\Bigg] \, .
 \vec{c_\gamma}  \\
  \end{split}
  \label{working0}
\end{equation}
The equation for the coefficients of the configurations still has the form  $i \frac{\partial}{\partial t} \vec{A} =  {\bf H}  \vec{A} $.
We provide expressions for the various matrix elements of the one-electron operators appearing in this section 
in  Appendix \ref{sec:1ematrices}.  

\section{Exterior complex scaling and the treatment of ionzation}
\label{sec_ECS}

In the application of the MCTDHF approach to molecules subject to short UV pulses, there is in general some component of the wave function that is ionized.  As the wave function is propagated the outgoing electron flux will inevitably reach the end of the FEM-DVR grid in $\xi$ and reflect back from it.  In a number of other time-dependent applications of grid methods it has been shown that the ECS transformation is capable of perfectly extinguishing those reflections, both in the absence of an external field~\cite{Stroud_McCurdy1991}  and in the presence of an time-varying field~\cite{Tao_ECS,Scrinzi2010}.  

 In a time-dependent calculation the ECS transformation  enforces outgoing wave boundary conditions for the ionized part of the wave function, even at very long times.
 In the application of the ECS method in prolate spheroidal coordinates  the electronic coordinates are scaled only beyond a radius $\xi_0$ by a complex phase factor according to $\xi \rightarrow \xi_0+(\xi-\xi_0)e^{i\Theta}$, where $0< \Theta< \pi/2$ is an angle on which the results do not depend formally.   The value of $\xi_0$ is chosen large enough that the physical quantities of interest can be calculated from the wave function for all electronic coordinates satisfying $\xi<\xi_0$.   This is a formally exact procedure, and in a converged calculation does not alter the wave function in the inner region from its exact value.  In fact, we may calculate accurate bound state energies even with $\xi_0=1$, as shown in
Table~\ref{tab:hf}.    However, to extract ionization information, we choose a larger $\xi_0$ and perform analysis inward of that value, on the real $\xi$ axis.  

The analytic continuation of the Hamiltonian under the ECS transformation leads to a complex symmetric matrix representation when the basis functions are real at real values of the coordinates, and the DVR implementation of ECS, detailed previously~\cite{dvr00,TopicalReview04} also leads to complex symmetric matrices.  So for example, the matrix representation of the one-body operators in the Hamiltonian, $ {\mathbf h^{(1)}} $ in the present FEM-DVR is complex symmetric, and the DVR basis functions, as defined in Eq.(\ref{eq:prim_DVR}) are orthonormal when their overlap is quadratured along the ECS contour~\cite{dvr00}.   

Once the MCTDHF orbitals are expanded in terms of the  orthonormal FEM-DVR basis in Eq.(\ref{eq:DVRbasis}), they are represented by the coefficient vectors $\vec{c}_\alpha$  in Eq.(\ref{eq:spfexpansion_R}).    We may define the inner product of a pair of those orbitals in terms of the expansion coefficients in the usual, Hermitian way,
\begin{equation}
\langle \phi_a \vert \phi_b \rangle = \vec{c}_a^{ \,\,  \dagger}  \cdot \vec{c}_b
\label{eq:hermprod}
\end{equation}
making use of the orthonormality of the DVR functions on the ECS contour. 
We  can then arrange the coefficients defining the MCTDHF orbitals as the matrix $C_{i,\alpha}$, and use this inner product when transforming the operators from the FEM-DVR basis to the orbital basis.  So for example we would transform the ECS-scaled one-body Hamiltonian according to
\begin{equation}
  \tilde{\mathbf{h}}^{(1)} = \mathbf{C}^\dagger  {\mathbf h^{(1)}_\textrm{ECS} } \mathbf{C}
\end{equation}
where $\dagger$ denotes the Hermitian conjugate of the matrix.  Because it is unitary (in the limit that there are the same number of orbitals as FEM-DVR basis functions), this transformation does not change the spectrum of the ECS-scaled matrix representation of $ {\mathbf h^{(1)}} $.  Doing every transformation to the orbital basis that is involved in constructing the matrices in the working equations, Eqs. (\ref{eq:Aequation}) and (\ref{eq:orbEOM}), in this way preserves the analytic properties of the ECS solutions.  

The implementation of ECS in this manner in the MCTDHF equations has another, very important advantage.  As the solutions are propagated forward in time, the constraint that the orbitals remain orthogonal, and the constraint on their variations in time imposed by Eq.(\ref{eq:constraint}), are then imposed with the usual sense of the Hermitian inner product in Eq.(\ref{eq:hermprod}).   This procedure is essential to the numerical robustness of the MCTDHF method, because if an inner product without  complex conjugation  were used instead, as it is frequently in the complex scaling literature,  an orbital could in principle have zero overlap with itself.  While there may be no formal reason to choose one implementation over the other, there is therefore a compelling numerical reason to chose the Hermitian inner product.    
This choice does result in matrices $\mathbf{H}$ and $\mathbf{W}$ that appear in the working equations, Eqs. (\ref{eq:Aequation}) and (\ref{eq:orbEOM})  for example, with no symmetry, but nonetheless the overall properties of the solutions, in particular their outgoing wave character, under ECS are correctly reproduced.

\section{Numerical integration}

The integration of the coupled nonlinear differential equations of Eqs.(\ref{eq:Aequation}) and (\ref{eq:orbEOM}) has been the source of several theoretical and numerical studies over the past years.  Splitting of the orbital equation by separating the one-body, stiff kinetic energy terms from the two-body, local, nonstiff potential terms has received considerable attention~\cite{kochlubich,kochapprox,lubichvar}, but we do not pursue this avenue here.

In the context of nuclear motion only within the MCTDH method, it has been shown~\cite{bec97:113} that it is useful to decouple the orbital and A-vector equations for short times, which is enabled by the fact that the product of the inverse density matrix and the reduced operator $\rho^{-1}{\bf\widetilde{W}} $ in Eq.(\ref{eq:orbEOM}) is in general slowly changing with time.  Within the MCTDH implementation, $\rho^{-1}{\bf\widetilde{W}} $ and the A-vector Hamiltonian $\mathbf{H}$ are taken as constant over a short time step, which approximation is denoted as ``constant mean field'' (CMF).  The orbital and A-vector equations are integrated separately over the constant mean field time step, which is typically much bigger than, for instance, the time step used in the integration of the nonlinear equation for the orbitals.  The error is determined by backwards propagation and the CMF time step is adjusted to keep it within a specified tolerance.  

We employ a similar method, but without intelligent error control and with constant stepsize, and with a predictor/corrector scheme that appears numerically robust.   The predictor step is identical to the CMF step in the MCTDH implementation, and the corrector step incorporates a linear approximation to the matrices $\textbf{H}(t)$ and $\rho^{-1}\mathbf{W}(t)$.  In the MCTDH terminology, this would be ``linear mean field'' (LMF).  Thus, in the CMF predictor step starting at $t_0$, with the matrices $\rho_0$, $\mathbf{W_0}$, and $\mathbf{H_0}$ for that time in hand, the wave function for the next time $t_1 = t_0 + \delta t$ is obtained as
\begin{equation}
\begin{split}
\vec{A}_1 & = e^{-i\mathbf{H_0}\delta t} \vec{A}_0 \\
i \frac{\partial}{\partial t} \vec{c} & = (\mathbf{1}-{\bf P(t) }  )   \left[   {\bf h^{(1)}}  + 
\rho_0^{-1} {\bf\widetilde{W}_0} \right]      \vec{c} ,  \quad t=t_0...t_1
\end{split}
\label{eq:cmf}
\end{equation}
This first, predictor step in the propagation yields a first guess for $\vec{A}(t_1)$ and $\vec{c}(t_1)$, which yields first
guesses for the matrices  $\rho_1$, $\mathbf{W_1}$, and $\mathbf{H_1}$ at $t_1$.  These first guesses are used to propagate the wave function
in a LMF corrector step, in which the first order Magnus approximation is used for the A-vector and a linear approximation for the product 
$\rho^{-1} \mathbf{W}$ is used.  
\begin{equation}
\begin{split}
\vec{A}_1 & = e^{-i(\mathbf{H_0} + \mathbf{H_1})\delta t/2} \vec{A}_0 \\
i \frac{\partial}{\partial t} \vec{c} & = (\mathbf{1}-{\bf P(t) }  )   \Big[   {\bf h^{(1)}}  + 
\big(
	\frac{t_1-t}{\delta t} \rho_0^{-1}{\bf\widetilde{W}_0} \\
& + \frac{t-t_0}{\delta t} \rho_1^{-1}{\bf\widetilde{W}_1} \big)
 \Big]      \vec{c} ,  \quad t=t_0...t_1
\end{split}
\label{eq:lmf}
\end{equation}
The splitting of the orbital and A-vector equations over the mean field step is beneficial for, among other things, ensuring unitarity and parallelizing the algorithm.
Although we have implemented versions of this method of higher order than linear (LMF), none exhibited its nearly unconditional stability.
We use the \verb#expokit# package~\cite{expokit} to calculate both the matrix exponential for the A-vector equation and the solution of the orbital equation.   The exponential propagation of the orbital equation was the fastest method tried in this study, although we note that the explicit, basic Verlet method also gave good results.  

Our wave functions are Slater determinants and are not spin adapted; it is most efficient to calculate the high-spin case, so for a triplet we
include projections of total spin $M_s$ = 1, but therefore higher multiplets are present in the configuration basis.  However, we intermittently project the wave function on the proper spin subspace to ensure that it is not contaminated by numerical error. A full description of the integration method will be presented in a forthcoming publication.

\section{Ground electronic states from imaginary time propagation\label{sec:bound}}

\begin{table}[t]
\begin{tabular}{|c|cccc|l|}
\hline
              & $R_0$  &$N_\eta$ & $n_\xi$ & $\xi$ elements & Energy  \\
\hline
H$_2$   & 1.4    &  9               & 14         &    3.0, 10.0, 10.0                         &  -1.13362957146  \\
 & \multicolumn{3}{c}{$\textrm{ same with}\, \theta=15^\circ$} & \multicolumn{1}{r}{1.1$\times$10$^{-9}$ $i$ } & -1.133629573   \\
 & \multicolumn{3}{c}{$\textrm{ same with}\, \theta=30^\circ$} & \multicolumn{1}{r}{1.2$\times$10$^{-9}$ $i$ } & -1.133629572 \\
               &           &                  &               &    HF limit  &  -1.1336295715~\cite{jensen1999}  \\
\hline
Li$_2$  &  5.051 & 25            & 20        &   0.75, 3$\times$ 4.0               & -14.8715620178     \\ 
               &           &                  &               & eliptic basis HF& -14.8715619~\cite{ThompsonWilliams} \\
\hline
LiH        &    3.015     &   21           & 19        &   1.0, 3$\times$ 5.0  &  -7.987352237     \\
               &           &                  &               &   numerical HF   &  -7.987352237~\cite{Kobus} \\
\hline
CO         &   2.132    & 21             & 19         &    1.5, 7.5, 7.5            &  -112.79090718\\ 
 & \multicolumn{3}{c}{$\textrm{ same with}\, \theta=15^\circ$} &\multicolumn{1}{r}{1.1$\times$10$^{-8}$ $i$ } & -112.79090714   \\
  & \multicolumn{3}{c}{$\textrm{ same with}\, \theta=30^\circ$} & \multicolumn{1}{r}{8$\times$10$^{-8}$ $i$ } & -112.79090714  \\
& & & & numerical HF& -112.790907~\cite{Kobus} \\
\hline
N$_2$   &  2.068 & 21 & 19   &  1.5, 7.5, 7.5 & -108.99382563 \\ 
&& & &numerical HF& -108.9938257~\cite{Kobus} \\
\hline
\hline
N$_2$& \multicolumn{4}{c|}{same basis, {(14/10) CAS-SCF}} & -109.14184793(5) \\
& \multicolumn{4}{c|}{ (14/10) Columbus ccpvtz} &  -109.132509251 \\
& \multicolumn{4}{c|}{ (14/10) Columbus ccpvqz} &  -109.140039408\\
\hline
  \end{tabular}
\caption{Converged  Hartree-Fock energies from MCTDHF relaxation calculations and the FEM-DVR  basis sets required to converge them, compared with literature values. For H$_2$ and CO, the calculation is repeated with complex scaling with $\xi_0=1$, for
two scaling angles. In these ECS results the last real digit and the imaginary components are not converged with respect to primitive basis.    Also in the last entry, an MCTDHF relaxation calculation equivalent to 
a 10 orbital full CI MCSCF result for $N_2$,  compared with results computed with cartesian Gaussian functions and a triple or quadruple zeta basis set.
\label{tab:hf}}
  \end{table}

Of course, one requires initial state eigenfunctions to be used as a starting
point for a time dependent calculation.  While some aspects of the present method have been well established in
the literature, others -- in particular, the use of prolate spheroidal orbitals shared among all points in $R$ -- have not,
and for this reason here we
provide various ground, metastable, and excited vibrational state properties calculated with the present method.
These are obtained by ``improved relaxation''~\cite{mey06:179, dor08:224109},
in which the orbitals are propagated forward in imaginary time and the CI Hamiltonian (except when there is only one configuration) is diagonalized at every time step. 
We have also implemented a state-averaged version analagous to a state-averaged
multiconfiguration self-consistent field (MCSCF) calculation in which the orbitals are optimized to minimize the
average energy of the first $N$ eigenfunctions of the A-vector Hamiltonian.
This procedure requires averaging the density matrices and reduced operators for the first $N$ eigenfunctions of the $A$-vector  Hamiltonian and propagating their shared orbitals in imaginary time.

\subsection{Fixed nuclei ground electronic states}

First, as a measure of the performance of the primitive basis in representing electronic wave functions, we list 
calculated fixed-nuclei Hartree-Fock energies for a variety of molecules in Table~\ref{tab:hf}.  An MCTDHF calculation using full CI in the space of 10 orbitals  space for N$_2$ is also reported and  compared with the corresponding calculation using the
Columbus quantum chemistry suite~\cite{col3} and the correlation-consistent triple and quadruple-zeta bases of Dunning~\cite{dunning}.  We also include results for
a few molecules including  complex scaling with $\xi_0 = 0$, in which the coordinates of all electrons are continued
into the complex plane.  These latter complex scaling results were obtained by using the c-norm, not the hermitian norm, as explained in Section~\ref{sec_ECS}, and demonstrate that the electronic Hamiltonian has been accurately 
analytically continued to complex $\xi$.  To achieve a given accuracy, these in general require slightly more DVR basis functions because  of the oscillatory nature of the solutions under complex scaling.

\subsection{Nuclear motion: HD$^+$ and natural orbitals for electrons and nuclei}

\begin{table}
\begin{tabular}{cl | cll }
 \multicolumn{2}{c} {Energy}  &\multicolumn{3}{c}{Natural occ.} \\
no.  orbitals & E (hartree) &orb. & B.O. & N.B.O. \\
\hline
1 & -0.5946128688        		   			    & 1 & .99634 & 0.99629 \\
2 & -0.5978526051       				             & 2 & 3.64 $\times 10\bm{^{-3}}$ & 3.68 $\times 10\bm{^{-3}}$ \\
3 & -0.5978974489       				             & 3 & 2.44 $\times 10\bm{^{-5}}$ & 2.48 $\times 10\bm{^{-5}}$\\
4 & -0.5978979622   				             & 4 & 2.36 $\times 10\bm{^{-7}}$ & 2.42 $\times 10\bm{^{-7}}$\\
5 & -0.5978979683    		   		             & 5 & 2.50 $\times 10\bm{^{-9}}$ & 2.6 $\times 10\bm{^{-9}}$\\
6 & -0.5978979683         					    & 6 & 2.94 $\times 10\bm{^{-11}}$ & 3.1 $\times 10\bm{^{-11}}$\\
exact & -0.5978979686    					   & 7 & 4.4 $\times 10\bm{^{-13}}$ &  5 $\times 10\bm{^{-13}}$\\
Ref.~\cite{Balint-Kurti1990} & -0.5978979686     & 8 & 7 $\times 10\bm{^{-15}}$ &  1 $\times 10\bm{^{-14}}$\\
\end{tabular}
\caption{Left: Energies of MCTDHF wave functions for ground state HD$^+$ as a function of orbitals, along with the exact result using our prolate spheroidal basis
and the exact $J=0$ nonadiabatic result of Balint-Kurti et al~\cite{Balint-Kurti1990}.  Right: natural prolate spheroidal occupation numbers for Born-Oppenheimer and non-Born-Oppenheimer  HD$^+$ calculations.  }
 \label{tab:hd+}
\end{table}

In our treatment the orbitals are used to span the entire range of internuclear distances $R$.  Because the prolate spheroidal coordinates do 
not mimic the behavior of molecular orbitals -- which asymptotically limit to  atomic orbitals with constant size, whereas the prolate spheroidal coordinates
continue to expand with increasing $R$ -- a greater number of orbitals is required to represent a wave function with nuclear motion than one without.
To precisely quantify this behavior, in Table~\ref{tab:hd+} we give ground state HD$^+$ energies calculated both with a numerically exact, converged 
calculation we performed using a large primitive FEM-DVR basis, and with the MCTDHF method for an increasing number of orbitals.  One can see that sub-microhartree accuracy is achieved with
three orbitals, and essentially the exact result is achieved with 5 orbitals. 
For HD$^+$,
our Hamiltonian is exact for $J=0$ and our exact result agrees with that of Balint-Kurti~\cite{Balint-Kurti1990} to all significant figures given.

\begin{figure}
\begin{center}
\begin{tabular}{cc}
\resizebox{0.4\columnwidth}{!}{\includegraphics{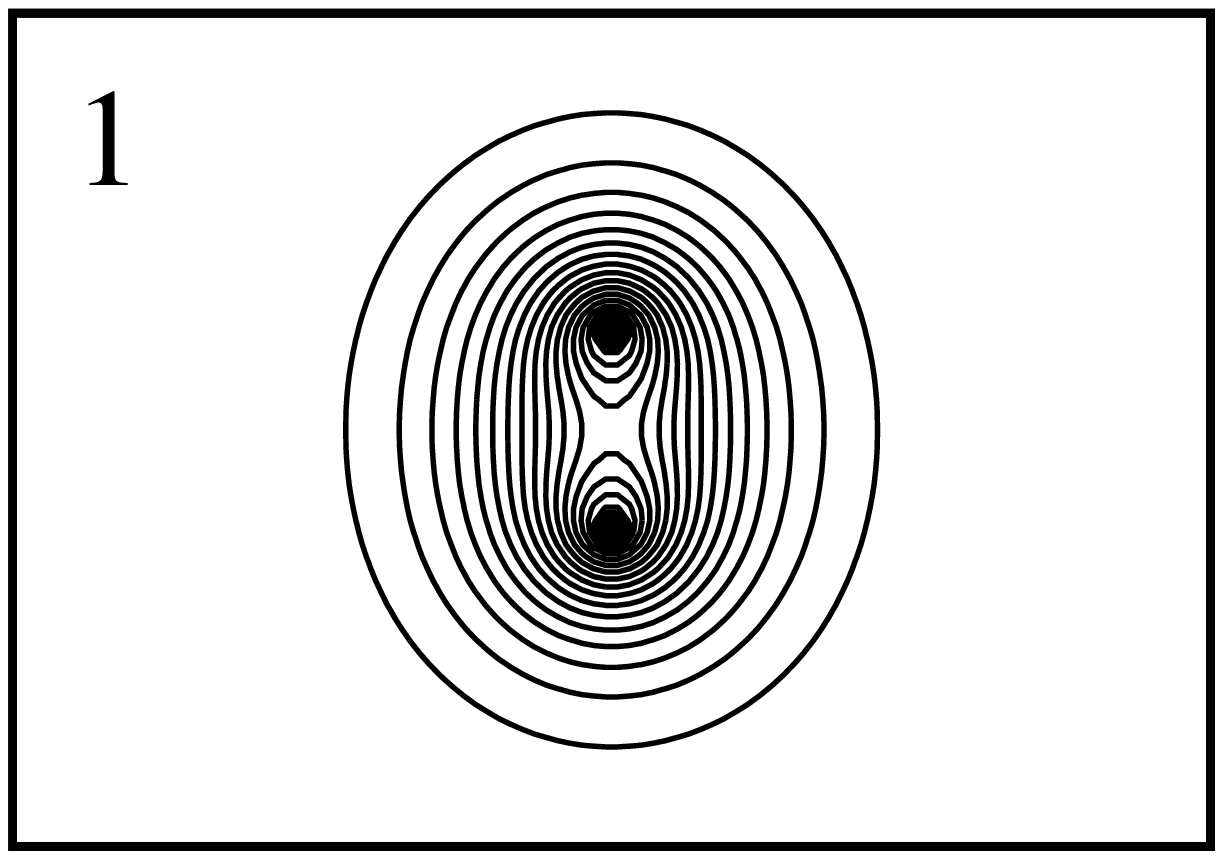}} & \resizebox{0.4\columnwidth}{!}{\includegraphics{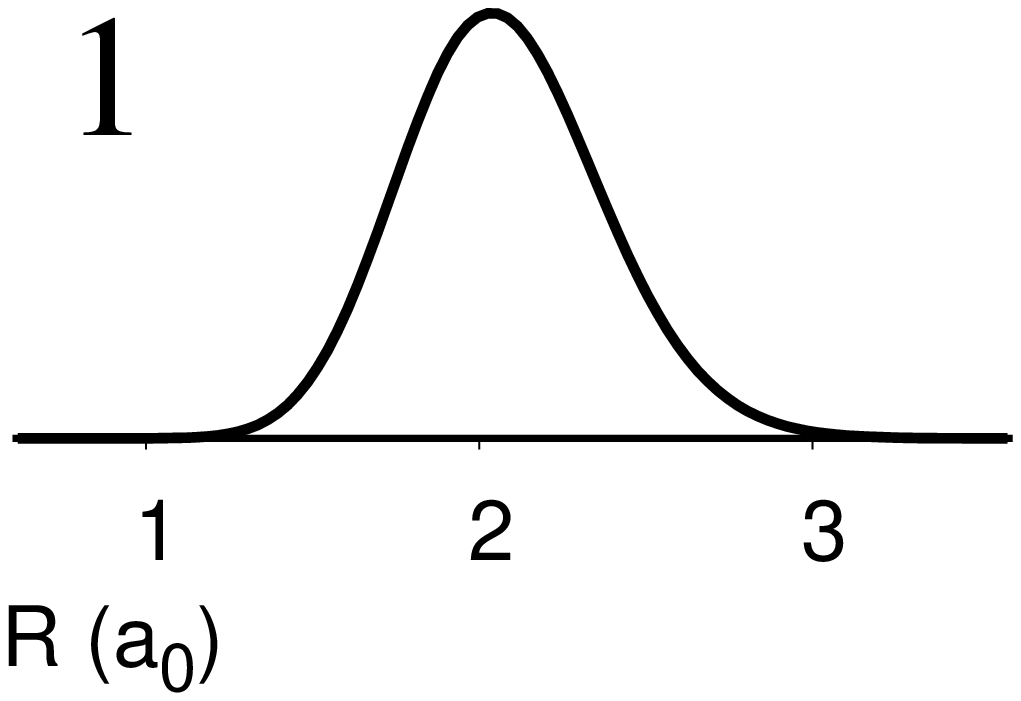}} \\
\resizebox{0.4\columnwidth}{!}{\includegraphics{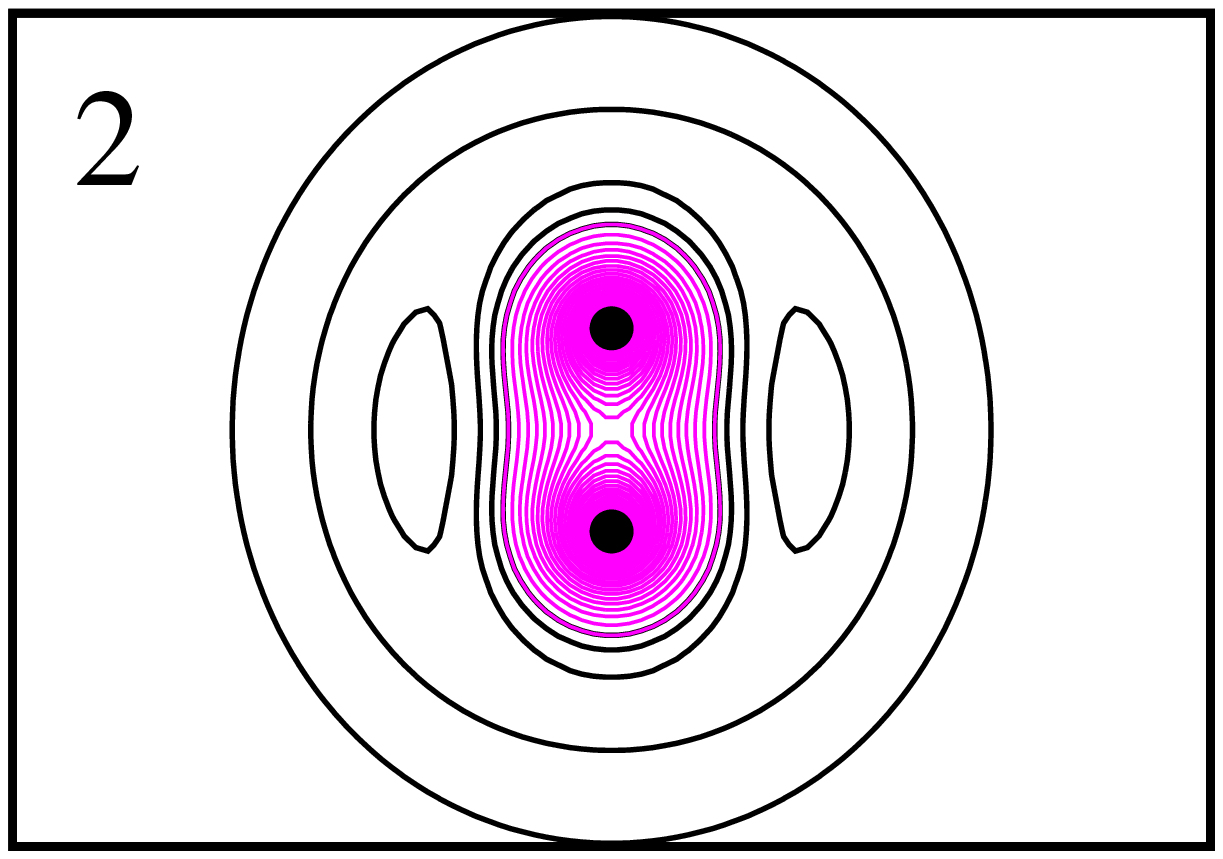}} & \resizebox{0.4\columnwidth}{!}{\includegraphics{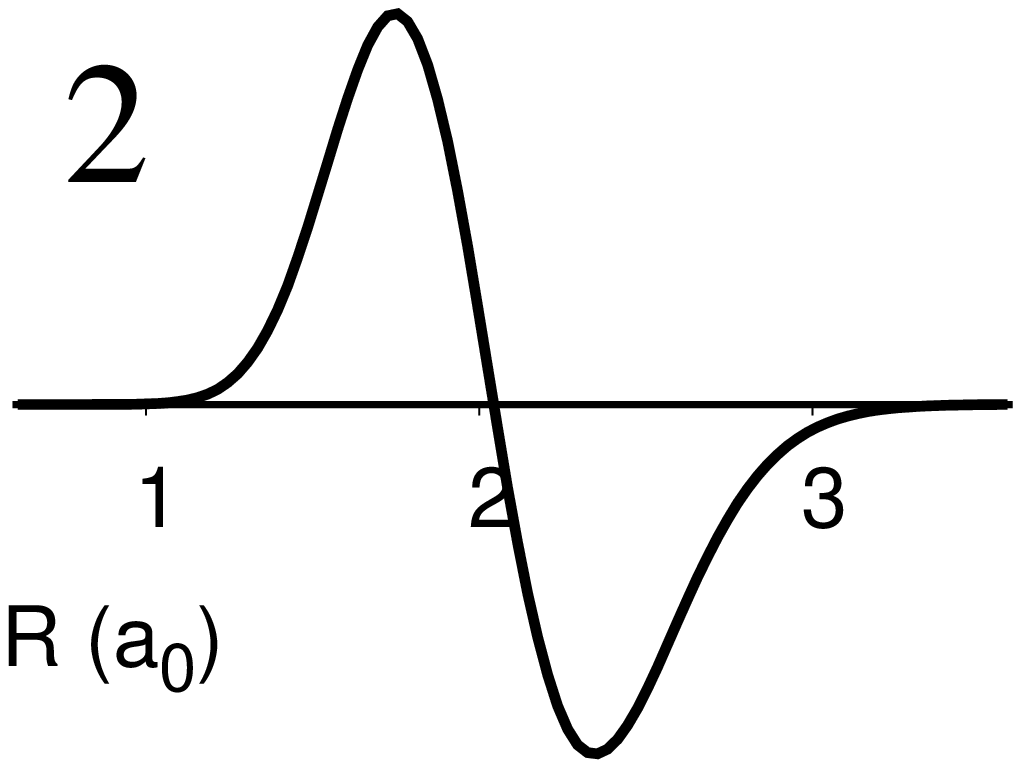}} \\
\resizebox{0.4\columnwidth}{!}{\includegraphics{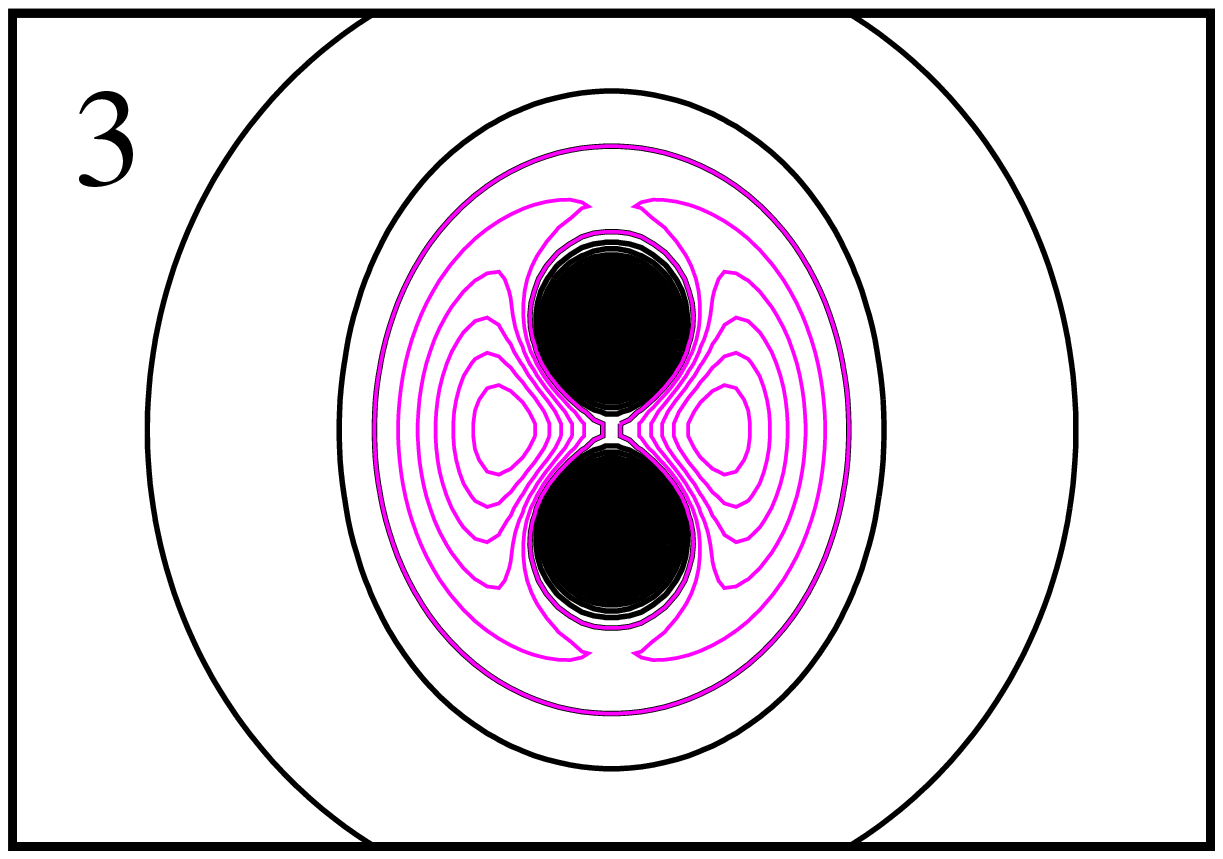}} & \resizebox{0.4\columnwidth}{!}{\includegraphics{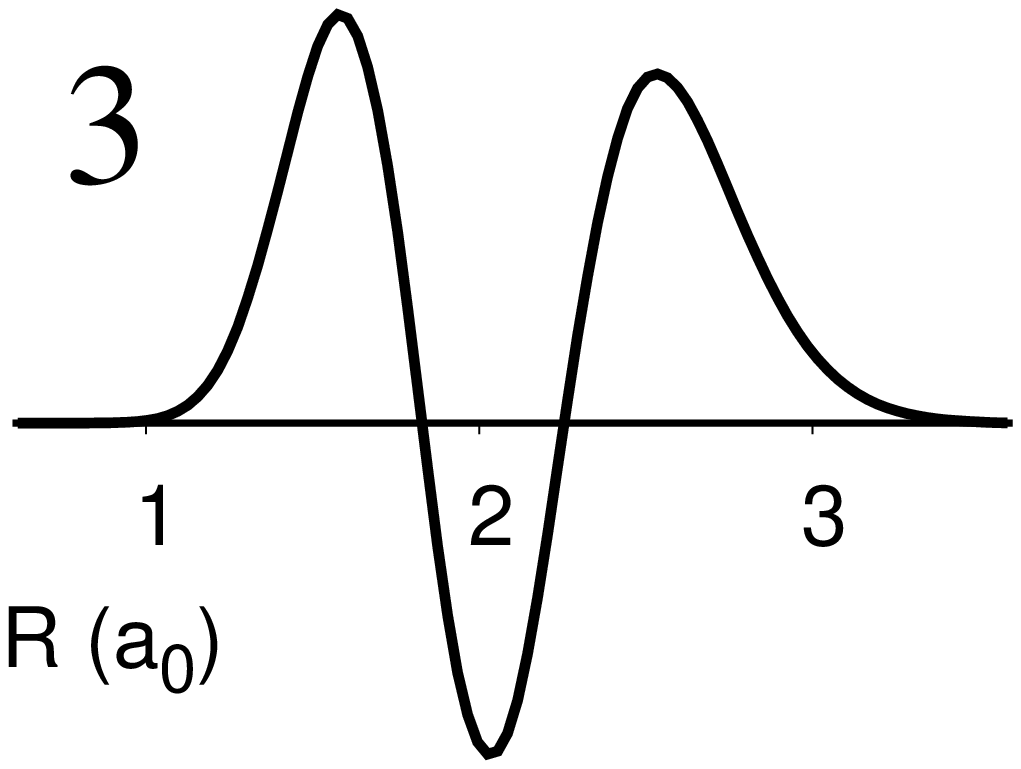}} \\
\resizebox{0.4\columnwidth}{!}{\includegraphics{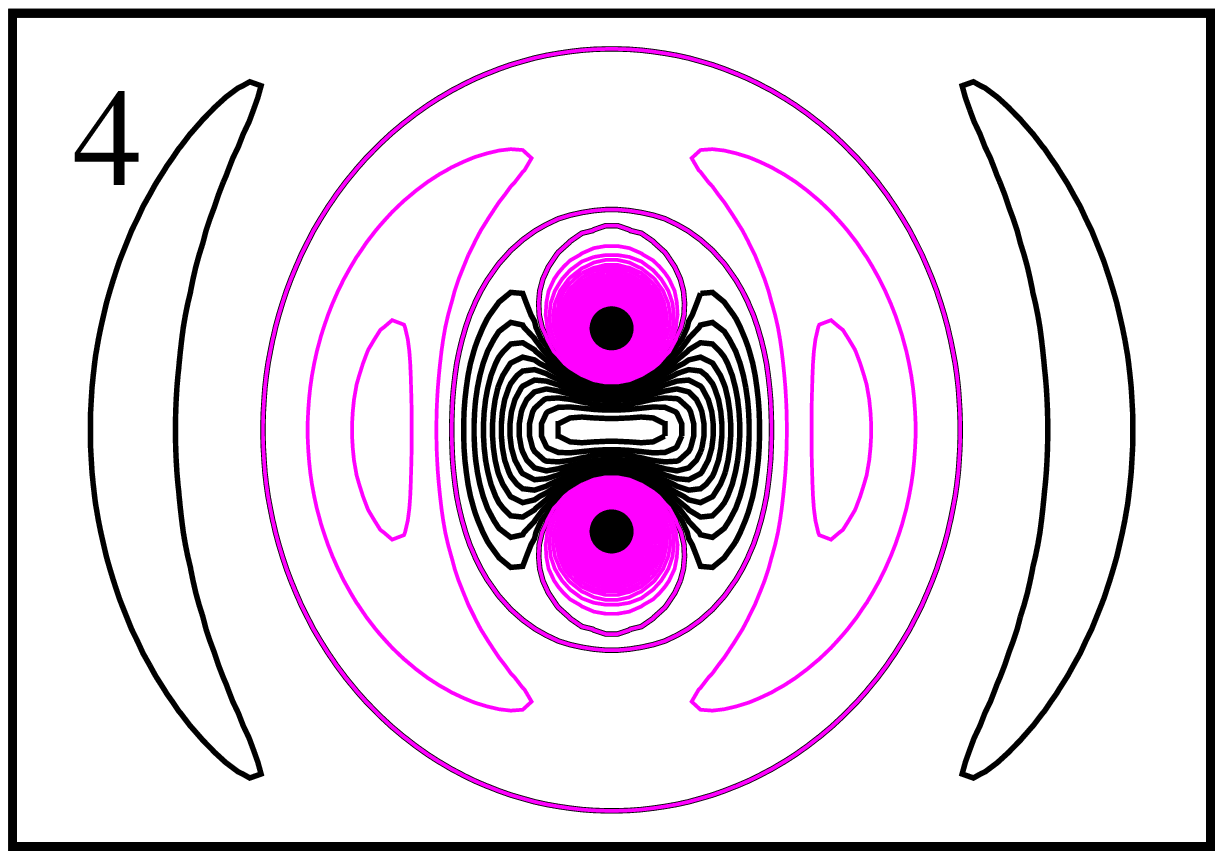}} & \resizebox{0.4\columnwidth}{!}{\includegraphics{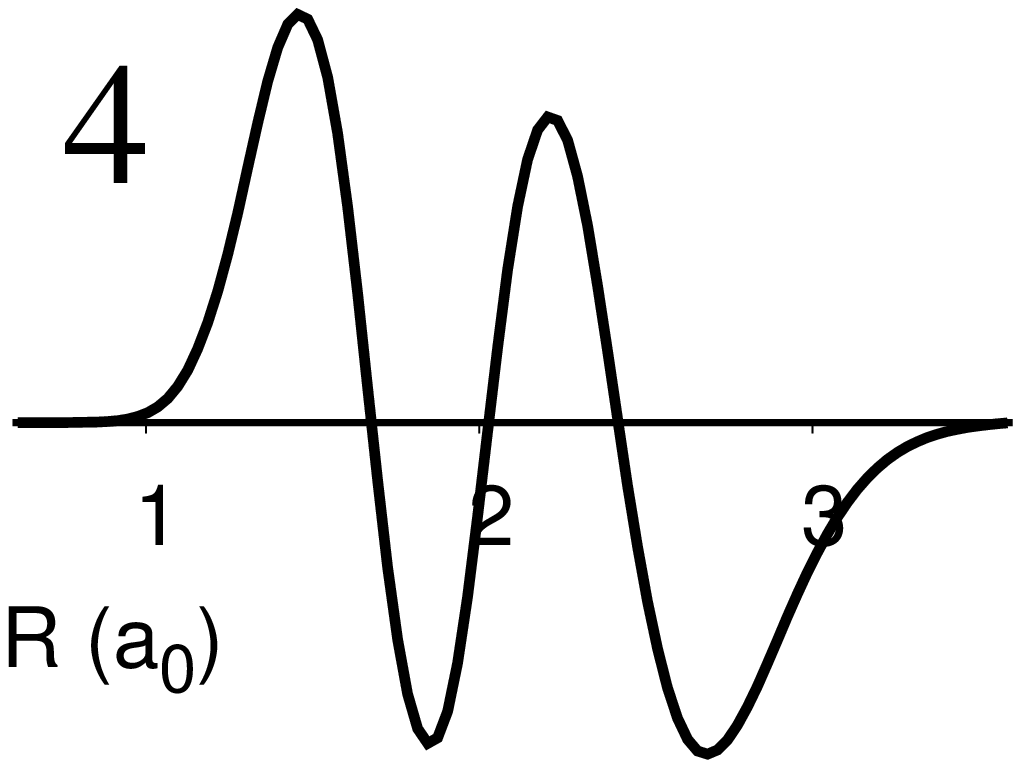}} \\
\end{tabular}
\end{center}
\caption{(Color online) Schmidt decompositon of the HD$^+$ ground state wave function:  Natural prolate spheroidal (elliptic) orbitals and conjugate natural orbitals in $R$ for J=0 ground state HD$^+$, diagonalizing the reduced density
matricers in Eqs.~(\ref{eq:proreduced}) and (\ref{eq:rreduced}), corresponding to the occupation numbers listed in Table~\ref{tab:hd+}.  The origins of the prolate spheroidal
coordinate system are denoted by black dots.
 \label{fig:hdorb}}
\end{figure}

\begin{table*}
\begin{tabular}{ c  c | c c c c c c c c c c c }
& $\nu$ & $<E>$ &  $\langle T\rangle$ & $\langle V\rangle$ & $\langle r_1\rangle$ & $\langle r_1^2\rangle$ & $\langle r_2\rangle$ & $\langle r_2^2\rangle $ & $\langle R \rangle$ & $\langle R^2 \rangle$ & $D$ \\
\hline
5$\sigma$ 1$\pi$ H$_2$ & 0 & -1.16008 & 1.16006 & -2.32014 & 1.5834 & 3.1894 & 1.5834 & 3.1893 & 1.4553 & 2.1451 & 0.0000 \\
8$\sigma$ 1$\pi$ H$_2$ & 0 & -1.16088 & 1.16085 & -2.32174 & 1.5784 & 3.1620 & 1.5784 & 3.1620 & 1.4527 & 2.1383 & 0.0000 \\
Ref. \cite{bubin2003}$^a$, \cite{kinghorn2000}$^b$ & 0 & -1.16403$^{a,b}$ & 1.16403$^b$ & 2.32805$^b$ & & & & & 1.4487$^a$ & 2.1270$^a$ &\\
5$\sigma$ 1$\pi$ B.O. & 0 &  &  & -2.32242 & 1.5773 & 3.1557 & 1.5773 & 3.1557 & 1.4520 & 2.1367 & 0.0000 \\
\hline
5$\sigma$ 1$\pi$ H$_2$ & 1 & -1.14047 & 1.14040 & -2.28087 & 1.6270 & 3.3691 & 1.6270 & 3.3691 & 1.5482 & 2.4787 & 0.0000 \\
8$\sigma$ 1$\pi$ H$_2$ & 1 & -1.14156 & 1.14167 & -2.28324 & 1.6295 & 3.38286& 1.6295 & 3.3826 & 1.5482 & 2.4817 & 0.0000 \\
Ref. \cite{bubin2003} & 1 & -1.14506 & & & & & & & 1.5453 & 2.4740 & \\
\hline
5$\sigma$ 1$\pi$ HD & 0 &  -1.16164  & 1.16158 & -2.32322    & 1.57520 &  3.14928 & 1.57547 & 3.15029 & 1.44634 &  2.11511 & -.0005391 \\
Ref.~\cite{stankeHD2009} & 0 & -1.16547 & & &                         1.57119 & 3.13009 & 1.57148 & 3.13120 & 1.44223 & 2.10432 & \\
5$\sigma$ 1$\pi$ B.O. & 0 &    & & -2.32506 &  1.5738  & 3.1409 & 1.5738 & 3.1409 & 1.4456 & 2.1142 & 0.0 \\
\hline
6-$\sigma$ 1$\pi$ LiH & 0 &   -8.03762  & 8.03765 & -16.0753 &   2.5808  & 7.8354  & 1.9864 &  6.6936 &  3.0834  & 9.5398   & 2.3458 \\
Ref.\cite{bubin2005} & 0 & -8.06644 & &                                            & 2.5651 & 7.74517 & 1.9719 & 6.5857 & 3.0610 & 9.4197 & \\
6-$\sigma$ 1$\pi$ BO & 0 &   &  & -16.08629 & 2.6219  & 8.1238 & 2.0066 & 6.8593 &  3.1285 & 9.8404 & 2.3306 \\
\hline
\end{tabular}
\caption{Properties of vibronic states. The H$_2$ calculation is from a state averaged calculation on the $\nu=0$ and $\nu=1$ states.  Otherwise the energy of the ground vibrational state has been minimized.
With six $\sigma$ and one $\pi$ orbital for LiH, fixed nuclei at 3.015, the dipole moment calculated was 2.2856 atomic units as may be compared with 
the prior result of 2.306~\cite{cooperlih}.\label{tab:vib}}
\end{table*}

Also shown in Table~\ref{tab:hd+} are the occupation numbers corresponding to the eigenfunctions of reduced density matrices for the ground $J=0$ state of HD$^+$.  These are the natural orbitals for electronic and nuclear motion in this coupled system.   These natural orbitals and their eigenvalues provide a compact representation of the wave function known in the quantum information literature as a Schmidt decomposition~\cite{Ekert_Knight_1995, peres_1993,nielsen_chuang}.

According to the theorems on which the Schmidt decomposition is based we may divide the HD$^+$ molecule with $J=0$  into two subsystems, namely that represented by (1) the coordinates of the electron and (2) by the nuclear separation, $R$.  Then the two reduced density matrices, that for electronic motion,
\begin{equation}
\begin{split}
\rho_{\textrm{el}}(\xi',\eta',\phi',\xi,\eta,\phi )=  
\int  dR R^2 \psi(\xi,\eta,\phi,R) \,  \psi^*(\xi',\eta',\phi', R)  
\end{split}
\label{eq:proreduced}
\end{equation}
and for nuclear motion
\begin{equation}
\begin{split}
\rho_{\textrm{nuc}}& (R,R') = \\
&\int  d\xi   d\eta  d \phi  \frac{R^3(\xi^2-\eta^2)}{8} \psi(\xi,\eta,\phi,R)  \psi^*(\xi,\eta,\phi, R')  
\end{split}
\label{eq:rreduced}
\end{equation}
have exactly the same same nonzero eigenvalues, $\rho_i$.  The complete wave function may be expressed in terms of the eigenfunctions, $\varphi^\textrm{el}_i(\xi,\eta,\phi)$ and $ \varphi^\textrm{nuc}_i(R)$ of these matrices
as
\begin{equation}
\psi(\xi,\eta,\phi,R) = \sum_i \rho_i^{1/2} \varphi^\textrm{el}_i(\xi,\eta,\phi) \, \varphi^\textrm{nuc}_i(R)
\label{eq:Schmidt}
\end{equation}

The $\rho_i$ are the natural occupations, and are a measure of the
degree to which the parametric dependence of the prolate spheroidal coordinates upon the bond length follows the change in the 
electronic wave function within the Franck-Condon region.  
In contrast, beyond $\rho_1$ the occupation numbers in cartesian coordinates $x,y,z$ that do not follow the nuclei
would be much higher.  
In Table~\ref{tab:hd+} we show two sets of occupation numbers, those for the Born Oppenheimer approximation to the ground vibrational state
and those for the the numerically exact wave function whose energy agrees with Ref.~\cite{Balint-Kurti1990}, and the occupations are comparable.  In Figure~\ref{fig:hdorb}
we plot the pairs of corresponding natural orbitals in $\xi,\eta$ (independent of $\phi$ since $m=0$) and in $R$, obtained from 
the exact wave function natural orbitals.  In this example, only slight differences exist between these and those from the Born-Oppenheimer or 
improved adiabatic~\cite{Esry1999} wave functions in the same coordinate system.

For a more complicated system, the concept of these coordinate-system-dependent natural occupations can be generalized.  For a multielectron
wave function, the generalization of the natural orbitals in $R$ is straightforward.  In this case, for the electronic degrees of freedom, we would have natural multielectron wave functions, not just orbitals, corresponding to the same set density matrix elgenvalues.   For a polyatomic system, we expect that the number of terms needed to converge the Schmidt decomposition analogous to Eq.(\ref{eq:Schmidt}), as indicated by the the $R$-natural orbital or natural wave function occupation numbers
will be a measure of the suitability of a hypothetical geometry-dependent electronic coordinate system.  One could compare two different choices of coordinate systems for the electronic degree of freedom (which like prolate spheroidal coordinates need not be orthogonal to $R$) by computing only the eigenvalues of the reduced density matrix, $\rho_\textrm{nuc}$, for a suitable wave function.

\subsection{Vibrational states}

\begin{figure}
\begin{center}
\begin{tabular}{c}
\resizebox{0.85\columnwidth}{!}{\includegraphics{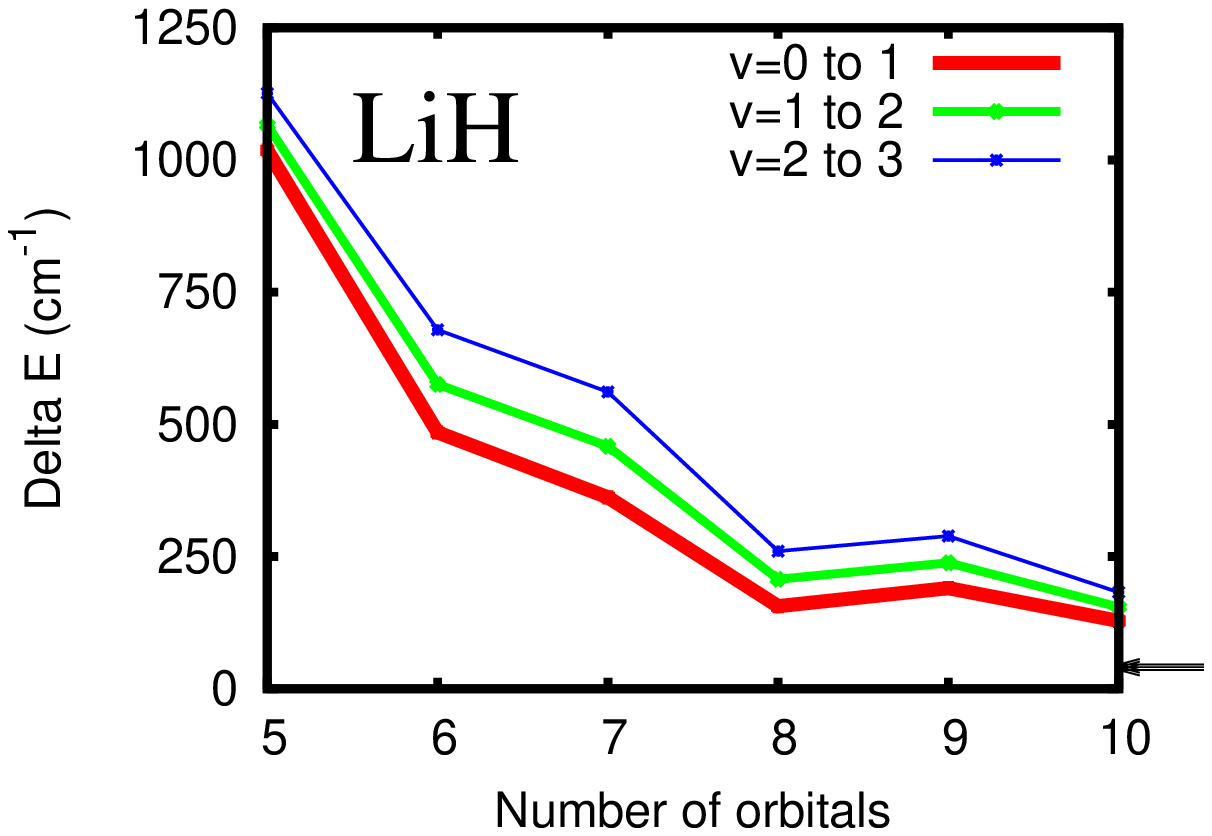}} \\
\resizebox{0.85\columnwidth}{!}{\includegraphics{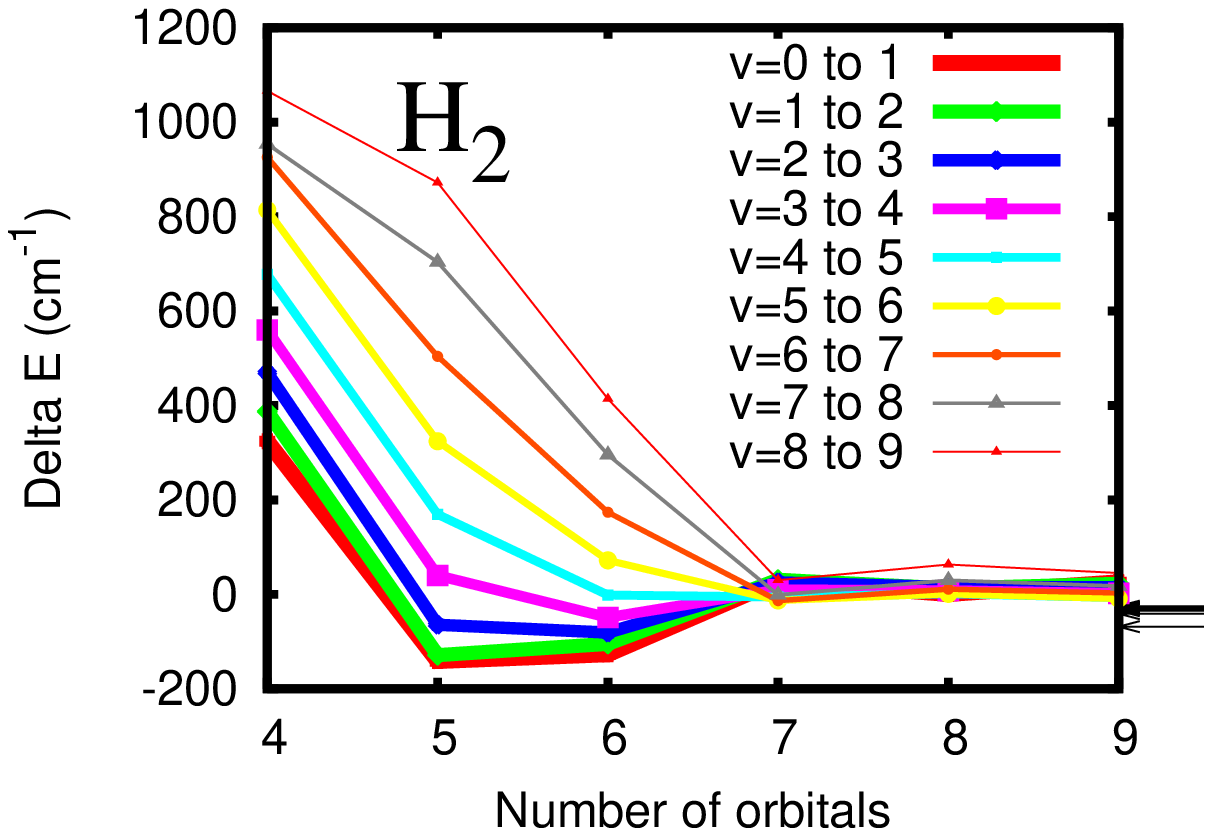}}
\end{tabular}
\end{center}
\caption{(Color online) Convergence of vibrational transition frequencies for H$_2$ and LiH from state averaged MCTDHF calculations using one 
$\pi$ orbital while optimizing 4 vibrational states  for LiH and 10 vibrational states  for H$_2$.  The difference between the calculated transition frequency and the experimental one is plotted with respect to the total number of orbitals, varying the number of $\sigma$ orbitals.  For LiH, the errors using the Hartree-Fock Born-Oppenheimer curve are also plotted, with atomic masses as the arrows on the right side.  For H$_2$, the errors for the same transitions from a Born-Oppenheimer calculation with five $\sigma$, one $\pi$ orbitals are also plotted as arrows to the right. 
\label{fig:vib}}
\end{figure}

\begin{figure}
\begin{center}
\begin{tabular}{cc}
{\large Born-Op.} & {\large $\nu=0$} \\ 
\resizebox{0.35\columnwidth}{!}{\includegraphics[viewport=0 0 350 310, clip]{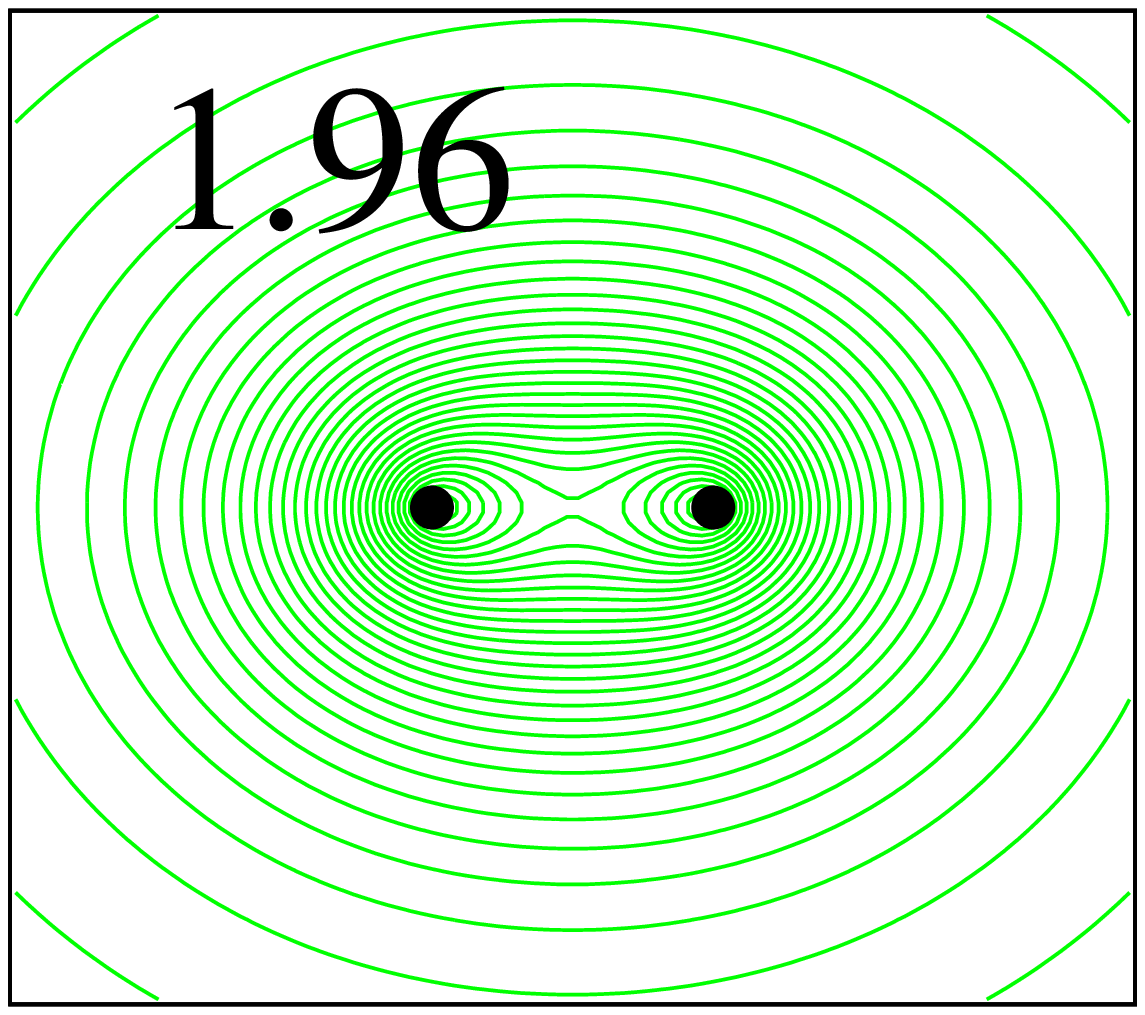}} &
\resizebox{0.35\columnwidth}{!}{\includegraphics[viewport=0 0 350 310, clip]{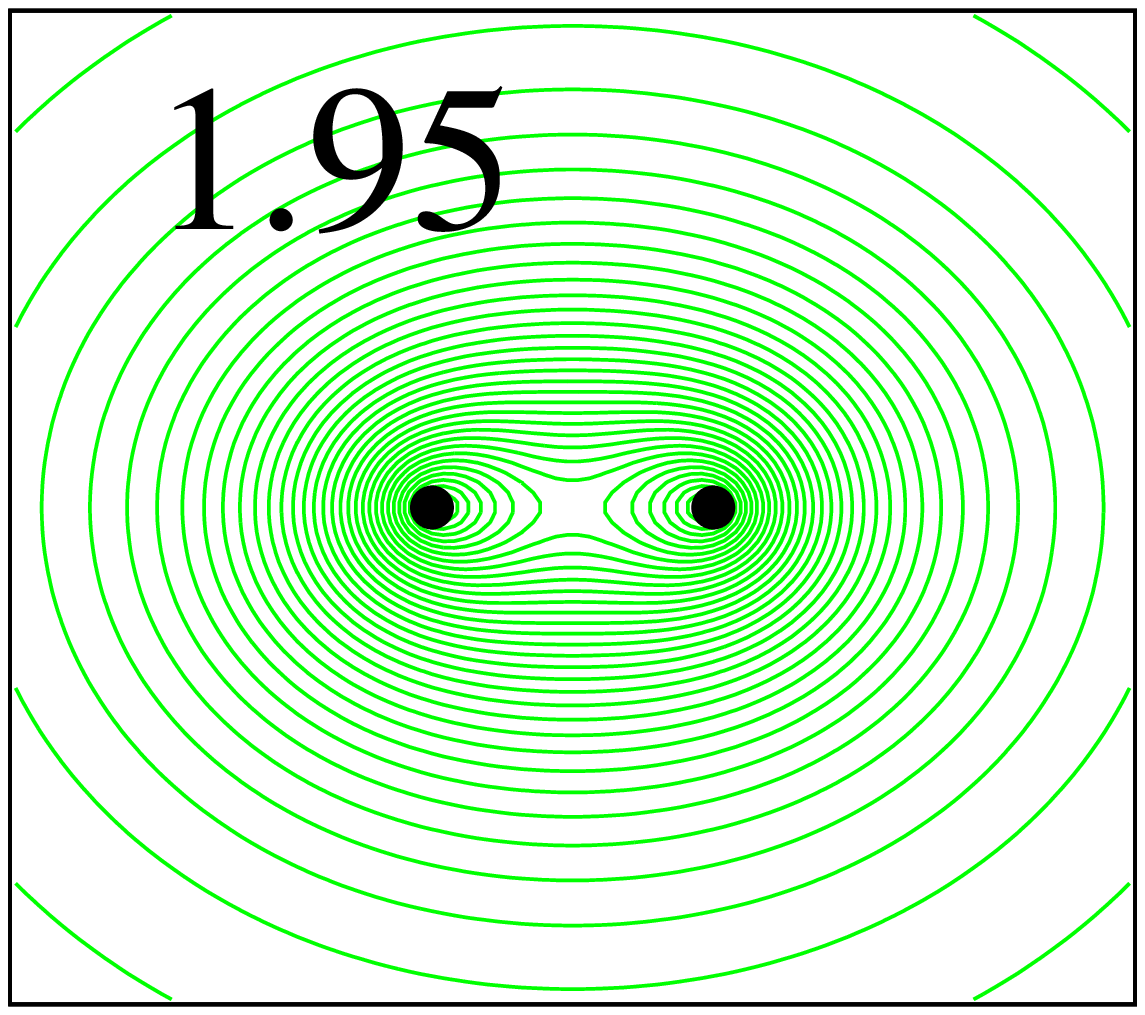}} \\
\resizebox{0.35\columnwidth}{!}{\includegraphics[viewport=0 0 350 310, clip]{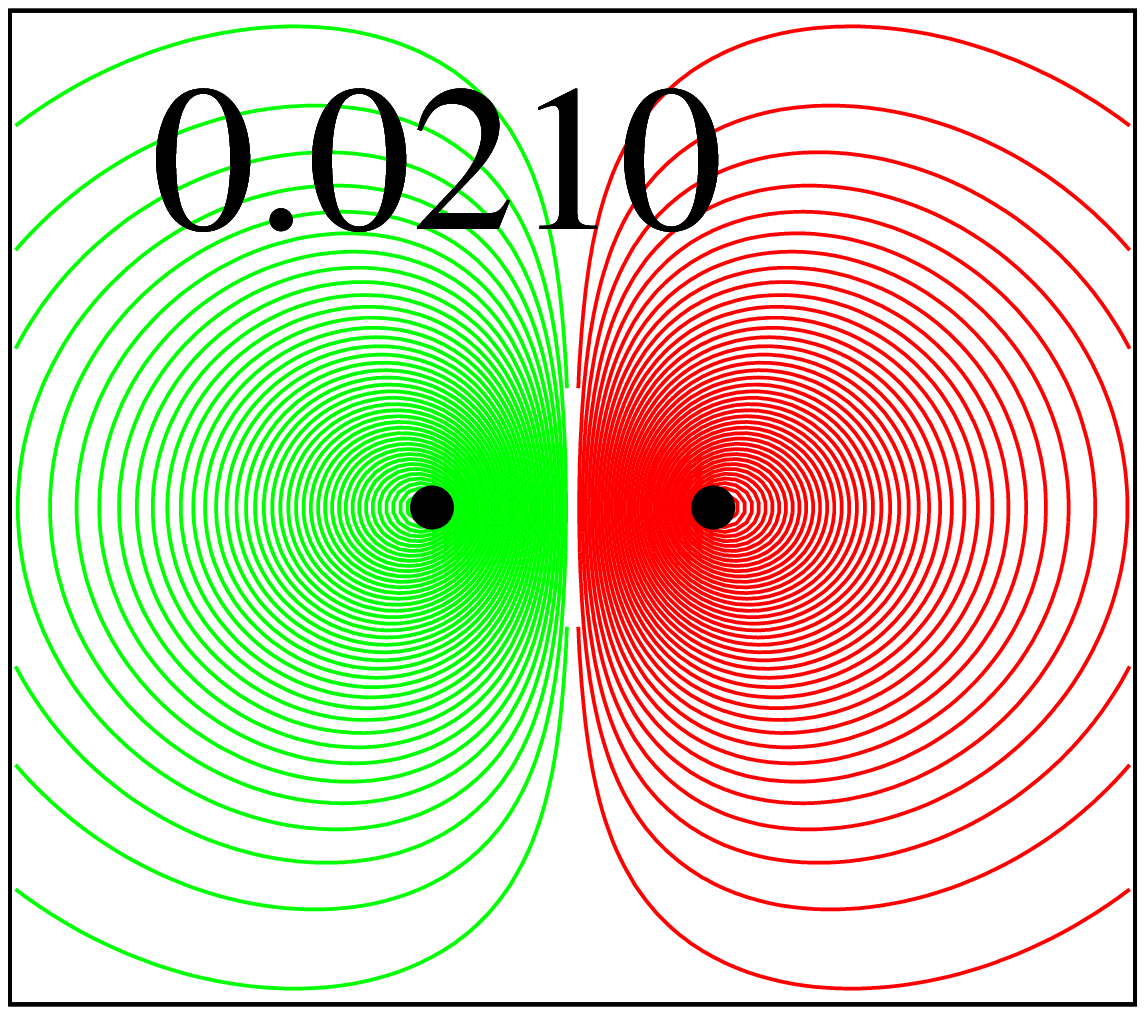}} &
\resizebox{0.35\columnwidth}{!}{\includegraphics[viewport=0 0 350 310, clip]{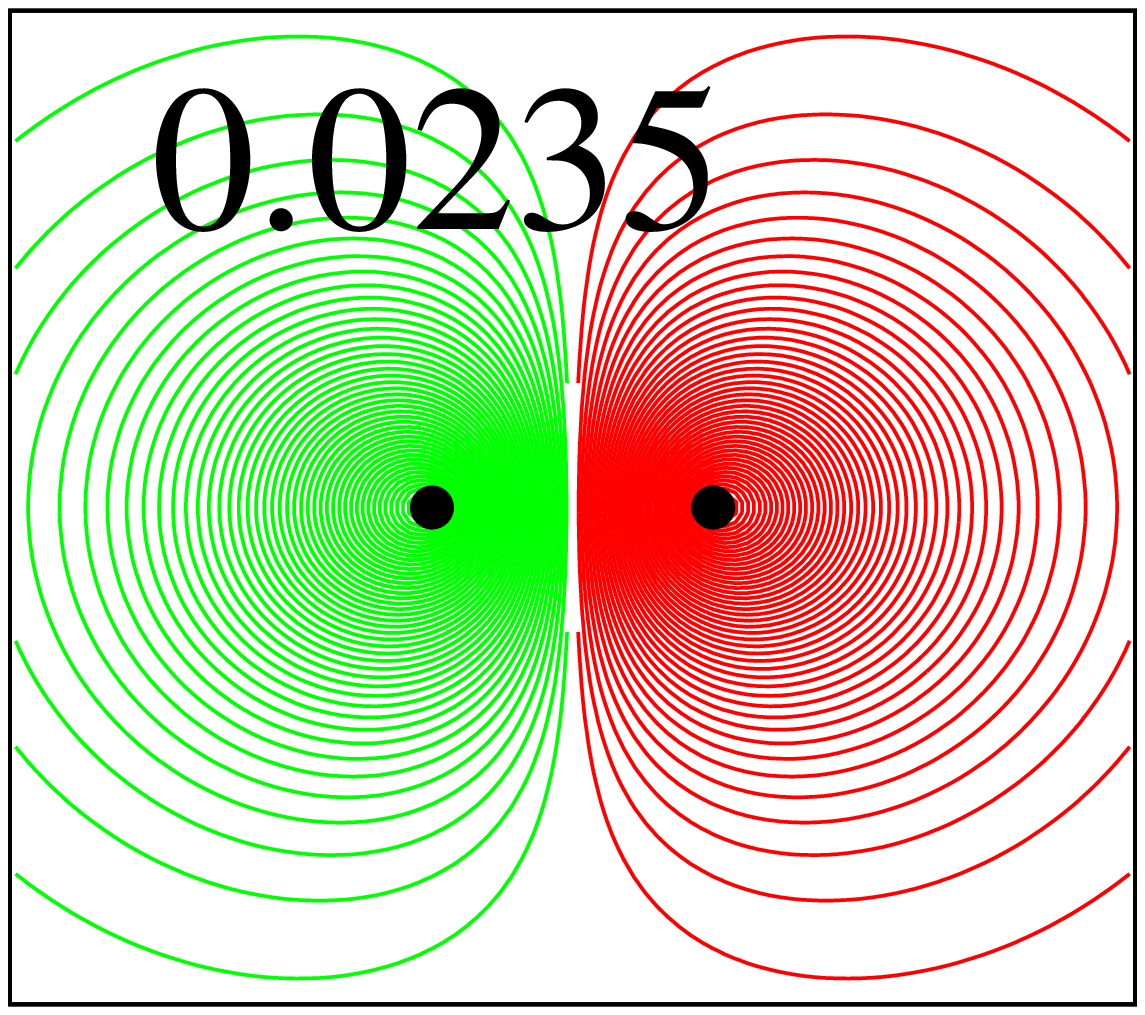}} \\
\resizebox{0.35\columnwidth}{!}{\includegraphics[viewport=0 0 350 310, clip]{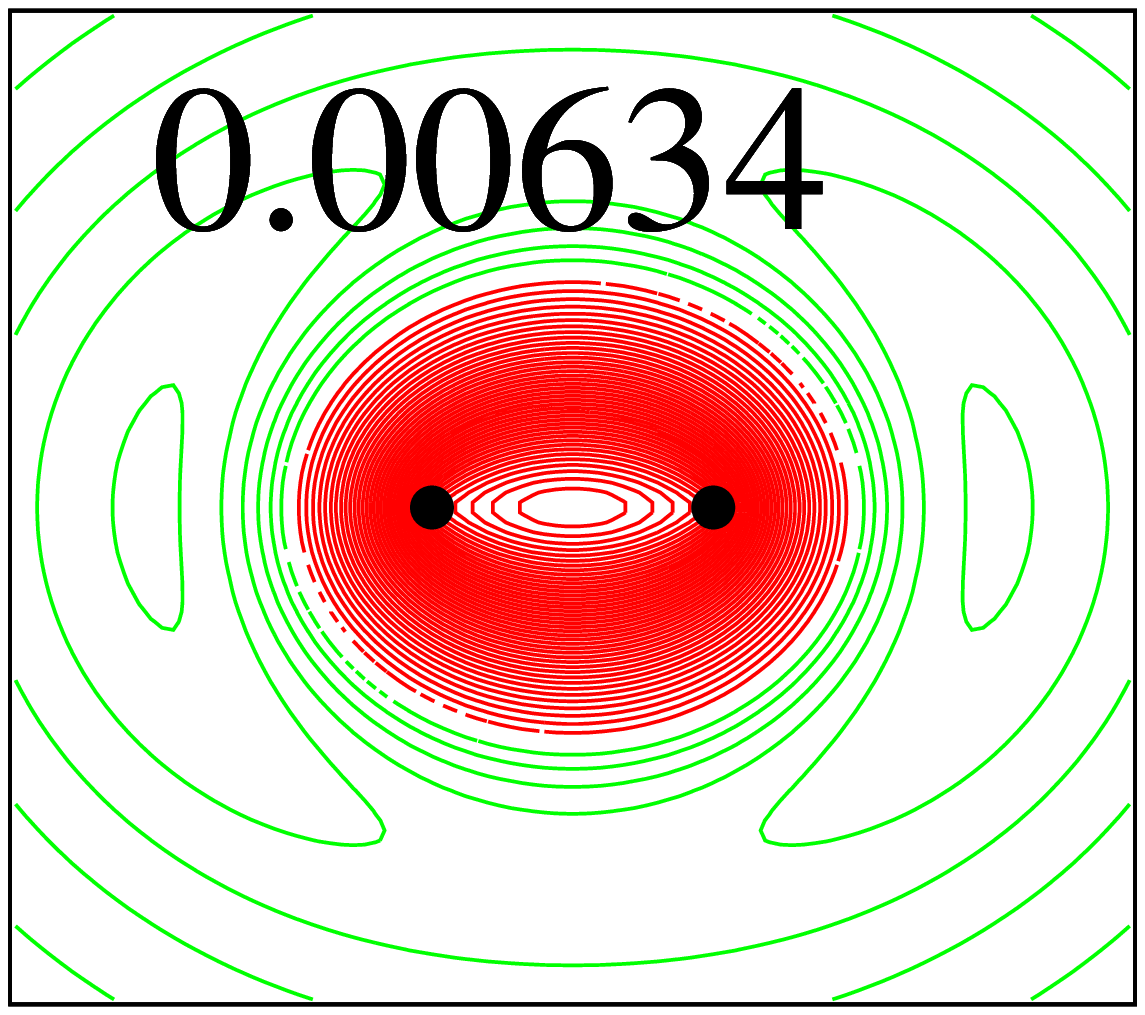}} &
\resizebox{0.35\columnwidth}{!}{\includegraphics[viewport=0 0 350 310, clip]{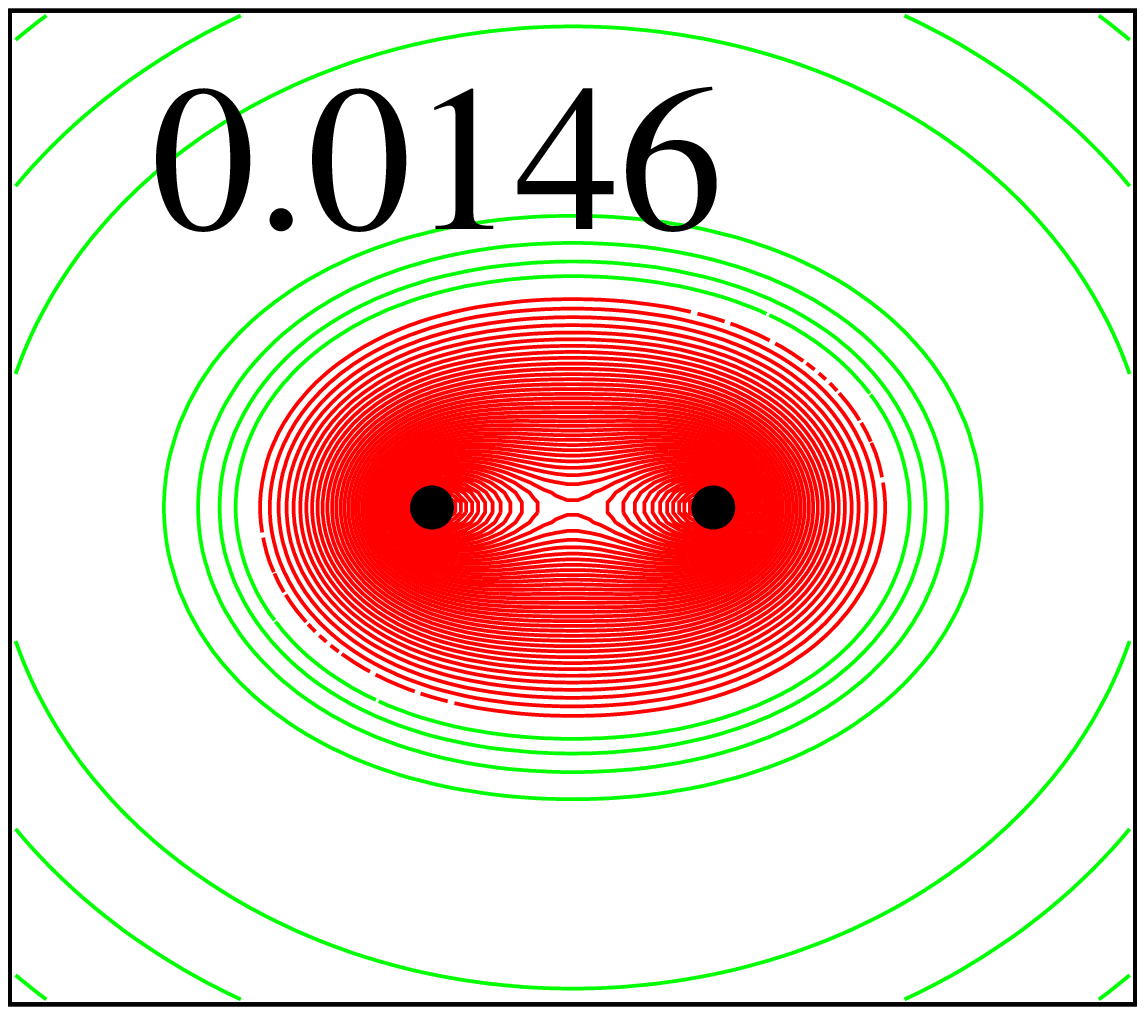}} \\
\resizebox{0.35\columnwidth}{!}{\includegraphics[viewport=0 0 350 310, clip]{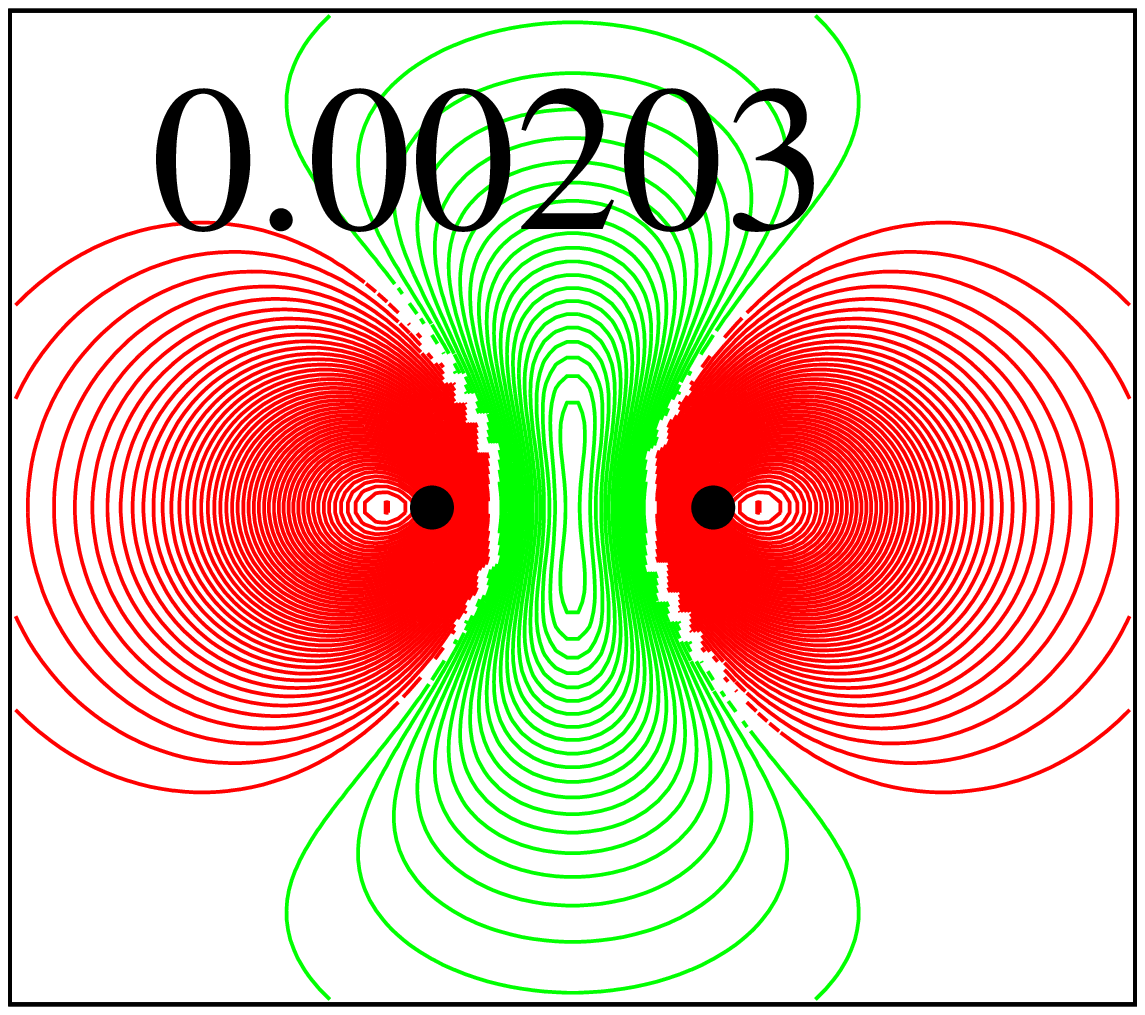}} &
\resizebox{0.35\columnwidth}{!}{\includegraphics[viewport=0 0 350 310, clip]{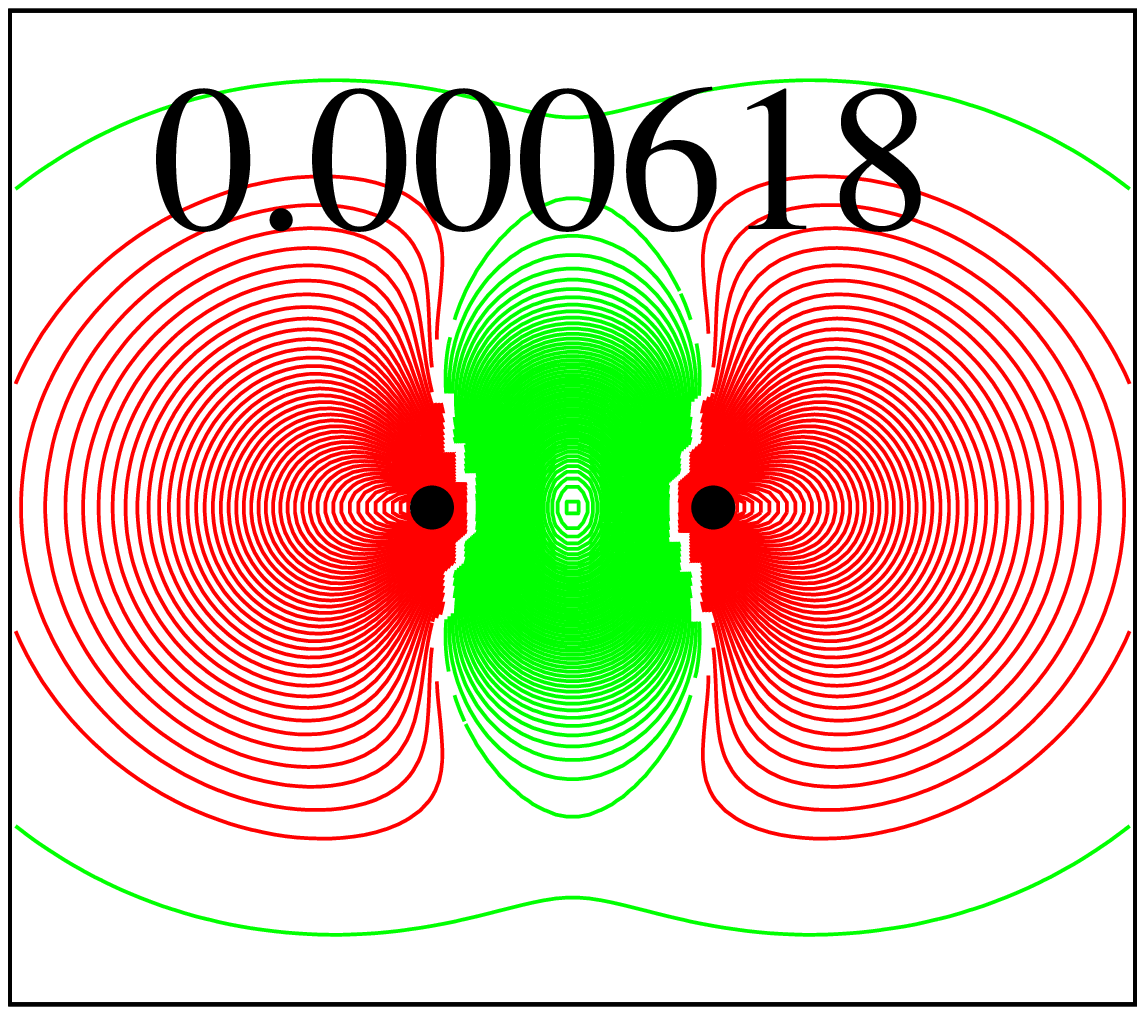}} \\
\resizebox{0.35\columnwidth}{!}{\includegraphics[viewport=0 0 350 310, clip]{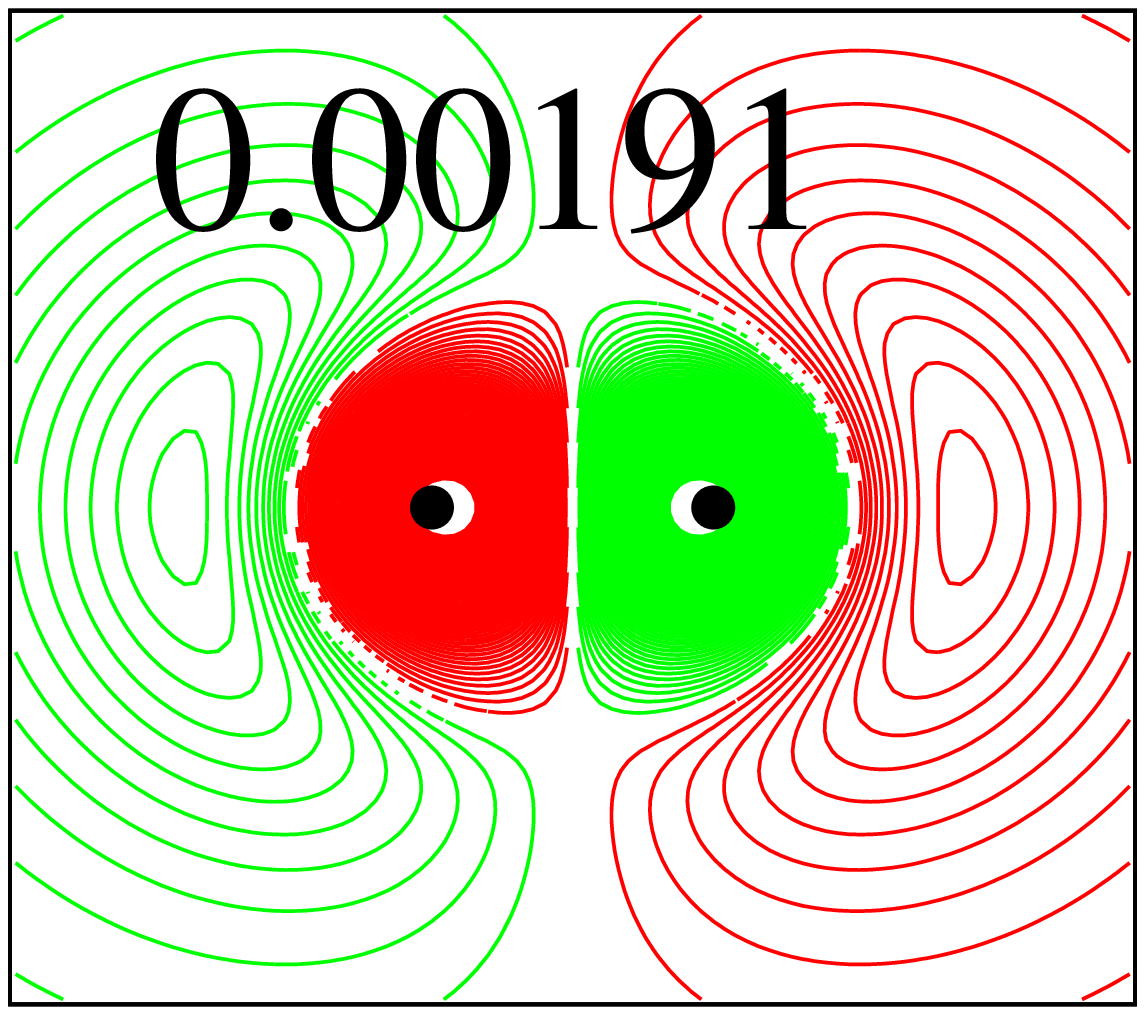}} &
\resizebox{0.35\columnwidth}{!}{\includegraphics[viewport=0 0 350 310, clip]{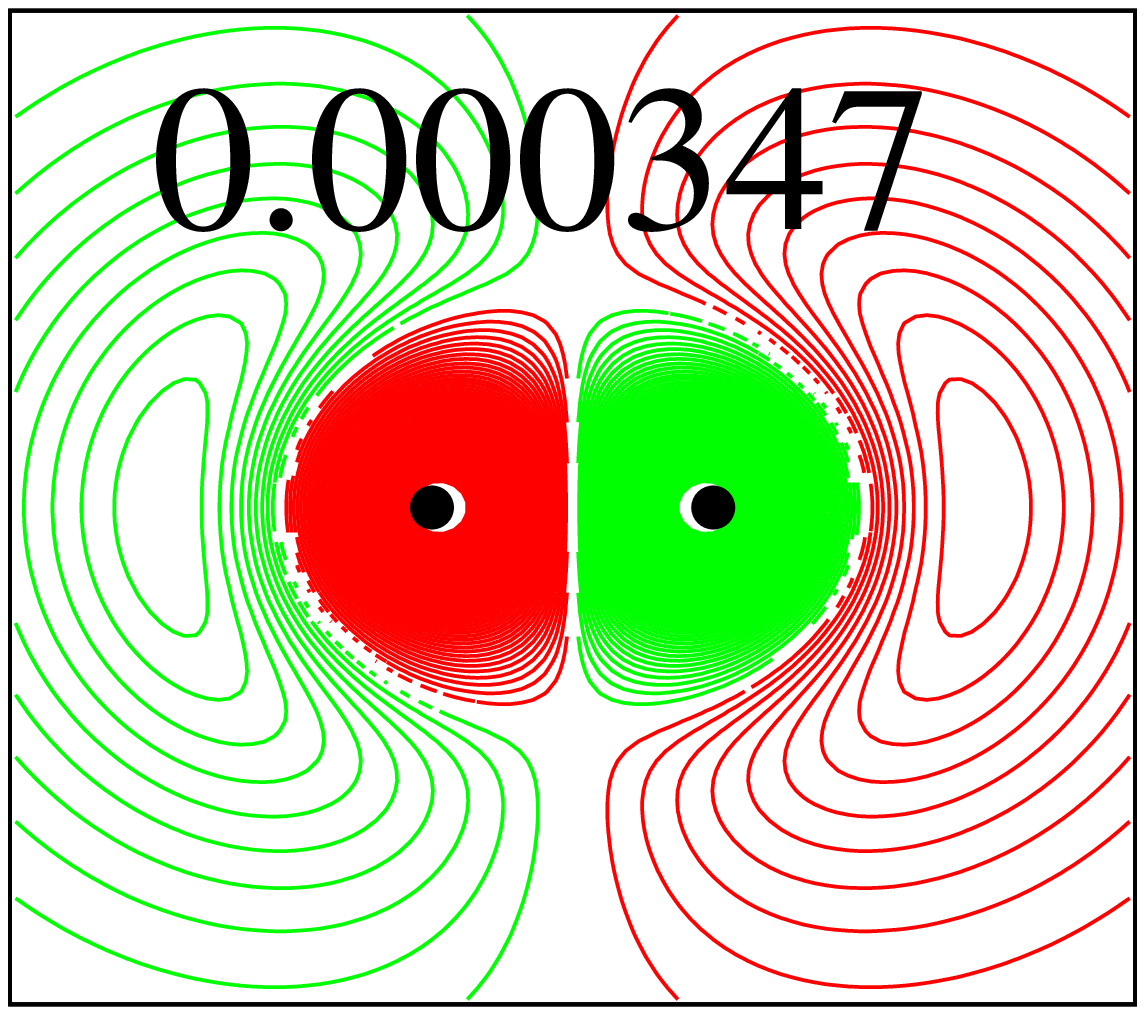}} \\
\resizebox{0.35\columnwidth}{!}{\includegraphics[viewport=0 0 350 310, clip]{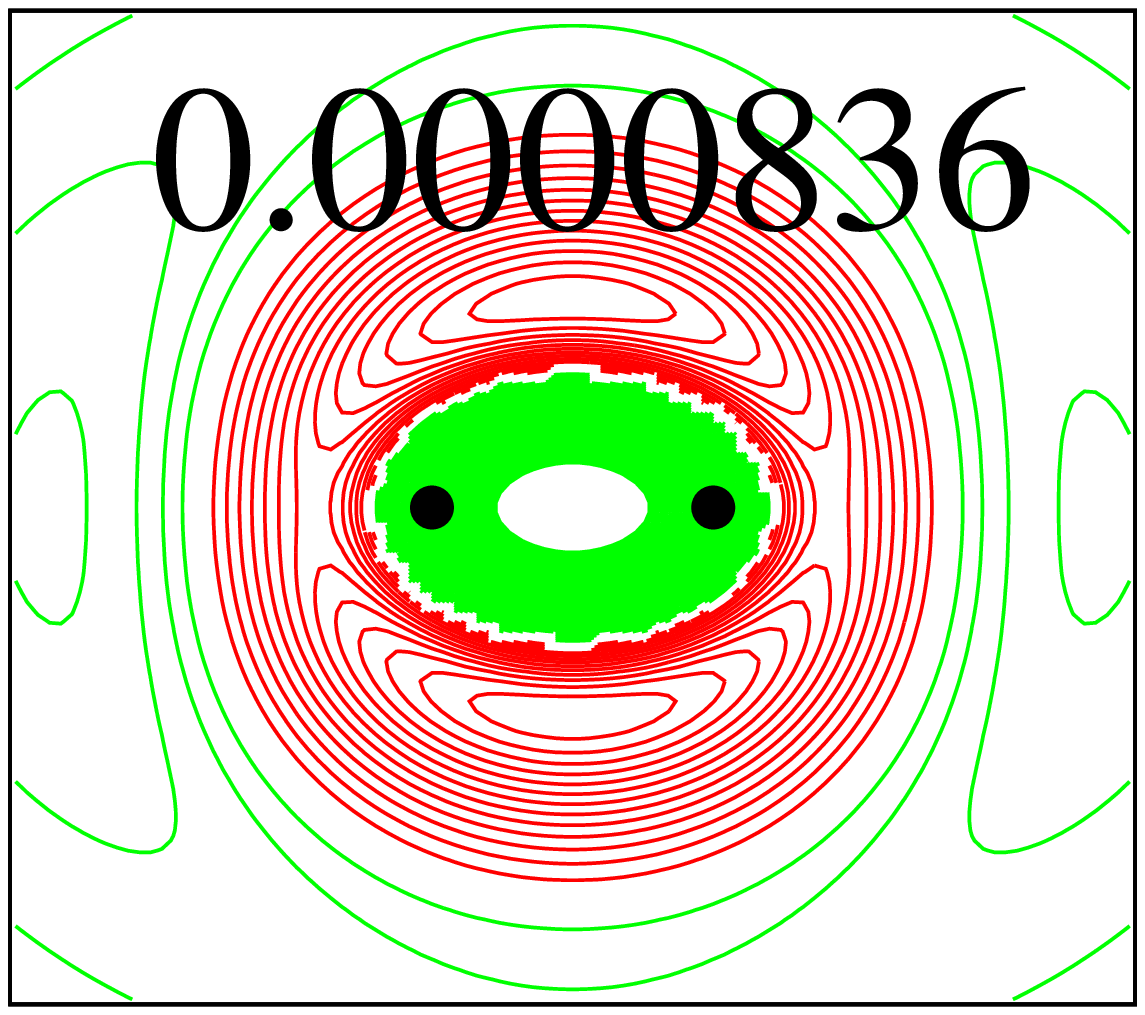}} &
\resizebox{0.35\columnwidth}{!}{\includegraphics[viewport=0 0 350 310, clip]{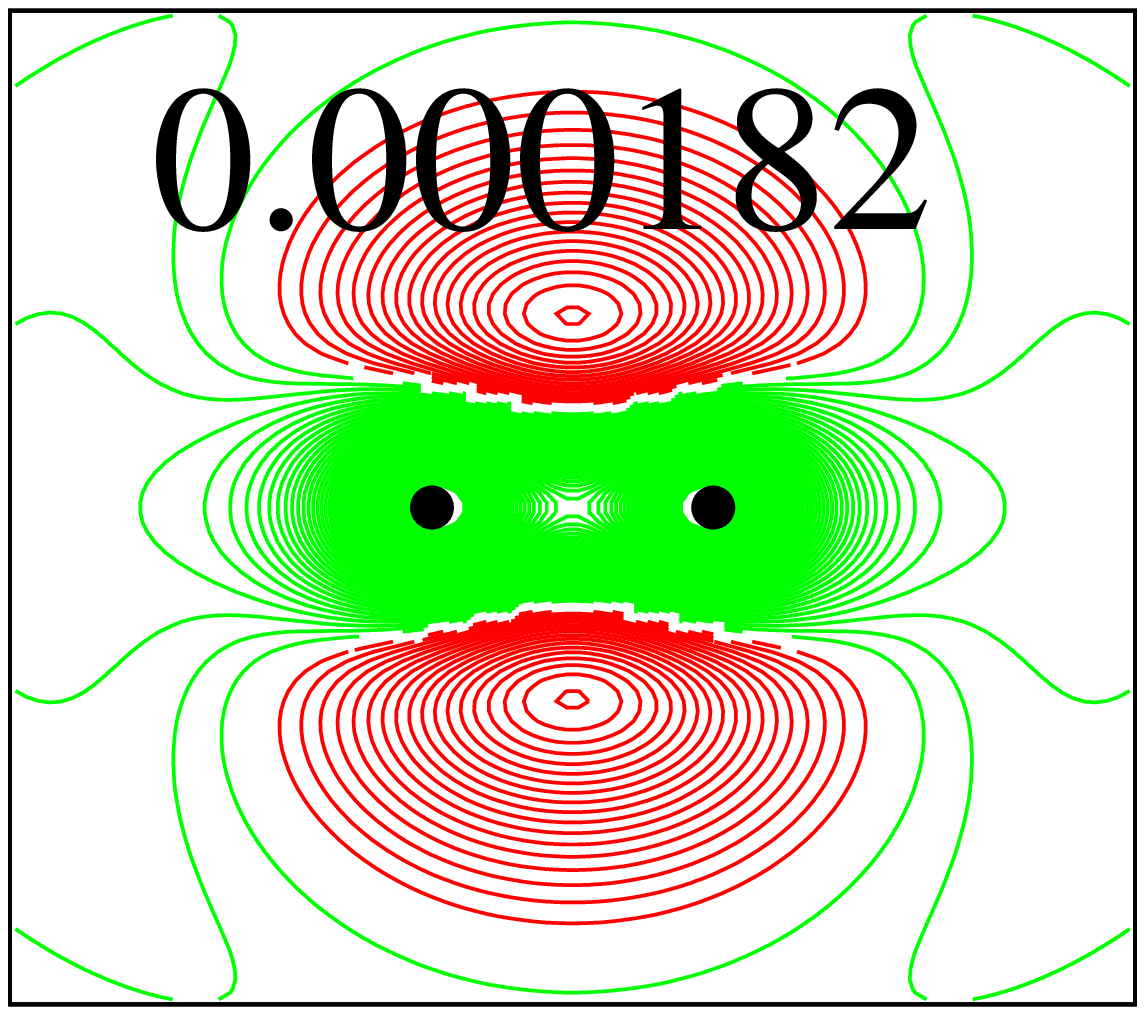}} \\
\resizebox{0.35\columnwidth}{!}{\includegraphics[viewport=0 0 350 310, clip]{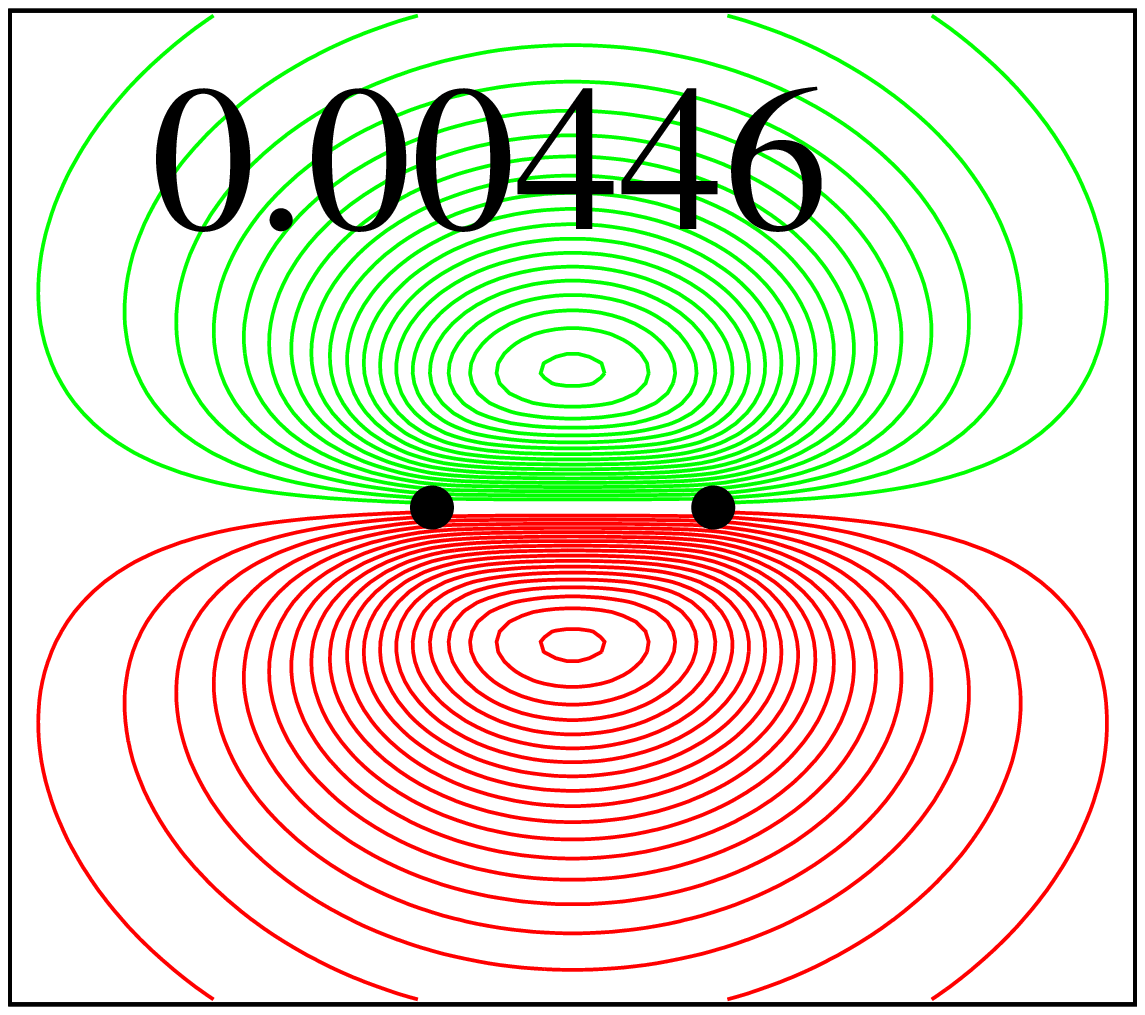}} &
\resizebox{0.35\columnwidth}{!}{\includegraphics[viewport=0 0 350 310, clip]{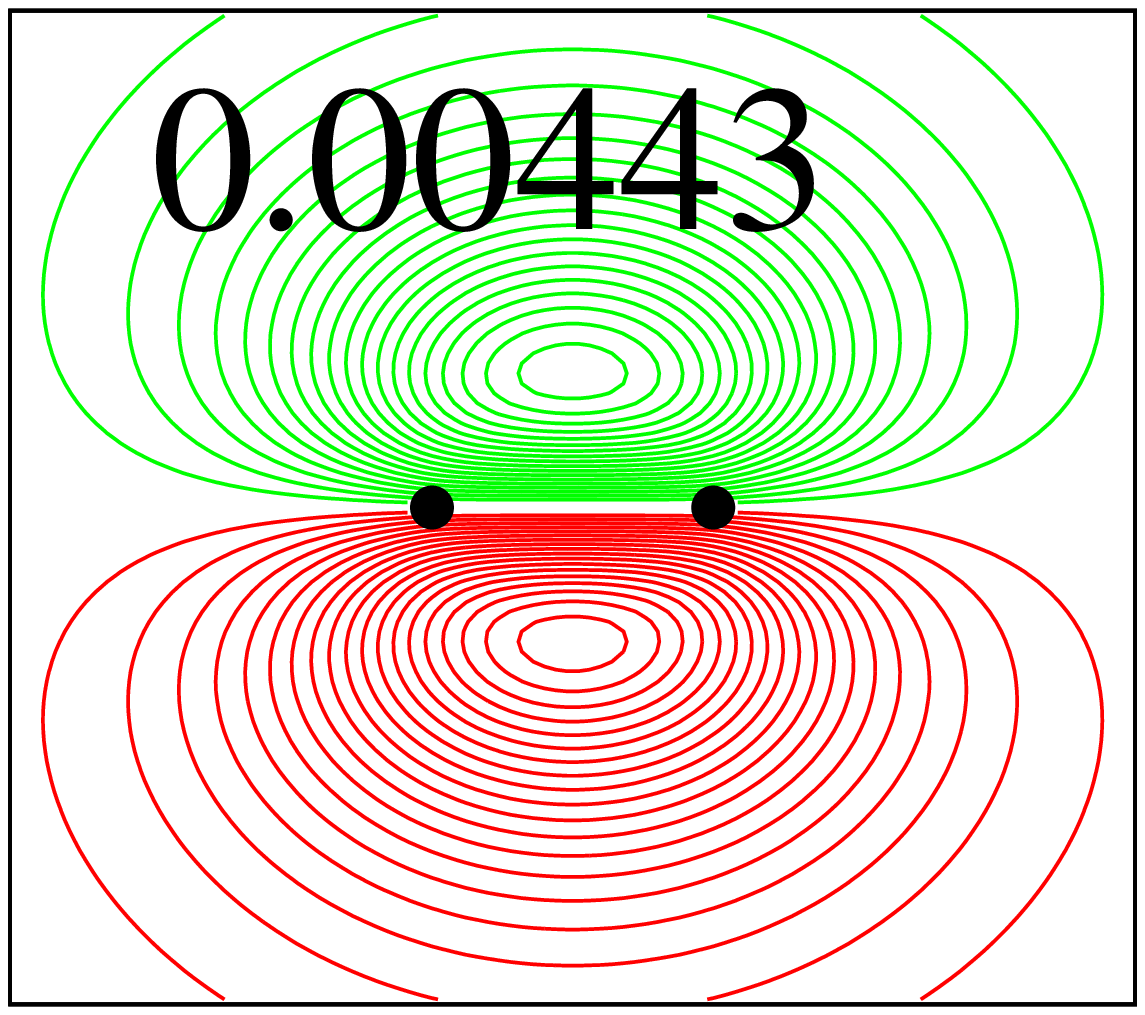}} \\
\end{tabular}
\end{center}
\caption{(Color online) Electronic natural orbitals of ground state H$_2$, with occupation numbers, from calculations with one $\pi$ and six $\sigma$ orbitals.  
Left column: Born-Oppenheimer natural orbitals at $R=1.4 \, a_0$; right column: natural orbitals from nonadiabatic calculation of the $\nu=0$ state.\label{fig:h2natorb}}
\end{figure}

In Table~\ref{tab:vib} we give properties calculated using MCTDHF  for the J=0 ground vibrational states of LiH,
for H$_2$, and HD, using a modest number of orbitals, with comparison to exact nonadiabatic results from the literature.  
The reported values for H$_2$ are
obtained from a calculation in which the first two vibrational states are simultaneously optimized using the same set of orbitals, whereas the other
results are from optimizing the ground $\nu=0$ state only. 
These calculations all use a single
$\pi$ orbital and varying numbers of $\sigma$ orbitals.
Differences in energies from the exact result on the order
of several millihartree for HD and H$_2$ or tens of millihartree for LiH are apparent.  The various expectation values differ by approximately one
percent or less from their exact values, even though in our multiconfiguration wave function a relatively small number of electronic orbitals have been used
to span all gridpoints in R which  number from 36 for H$_2$ to 48 for LiH.  For the calculations Tables \ref{tab:hd+} and \ref{tab:vib} and Fig.~\ref{fig:vib} we use nuclear masses $m_H=1836.152701$, $m_D = 3670.483014$, $m_{Li} = 12789.395862$.

 In Table~\ref{tab:vib} we also report properties calculated for the Born-Oppenheimer wave function, i.e., 
the solution $\chi_0(R)\psi_1(\vec{r};R)$ obtained by diagonalizing the Born-Oppenheimer vibrational Hamiltonian using the ground Born-Oppenheimer
electronic state $\psi_1(\vec{r};R)$ with orbitals optimized for each $R_i$ separately, and using the atomic masses.  For H$_2$, the error in these
results is comparable to that of our nonadiabatic MCTDHF wave functions.  
For LiH, we achieve better agreement with the previously computed accurate values using the full nonadiabatic treatment than we do  with the Born-Oppenheimer calculation using fixed masses.  

There is a striking observation to be made about these results concerning the convergence of our approach in which a single set of electronic orbitals is used for all $R$.  By accounting for the nonadiabatic coupling terms in the Hamiltonian and using one small set of orbitals for all R, we achieve a better representation of the wave function than we do using the Born-Oppenheimer wave function with orbitals optimized separately at each R.

In Fig.~\ref{fig:vib} we show the performance of the method in representing the vibrational spectrum of H$_2$ and LiH.  For LiH, the first four  vibrational states are simultaneously optimized using the same orbitals in these calculations, and for H$_2$ the first 10 are optimized.   The errors in the vibrational transition frequencies are plotted with respect to the total number of  orbitals.   The corresponding errors in the transition frequencies for the vibrational states of the Born-Oppenheimer 
curves (again, computed with atomic masses) are plotted as arrows on the right.  These errors are comparable so the arrows overlap.

Because in these calculations we are simultaneously optimizing a set of vibrational states spanning 
a larger range of internuclear distances than $\nu=0$ and $\nu=1$, the errors in Fig.~\ref{fig:vib} are greater than those in Table~\ref{tab:vib}.   However, despite this fact the  errors in the vibrational transitions may be made to be quite small even in a state averaged calculation.  We obtain 4173.4cm$^{-1}$ versus the correct value of 
4161.1 for the $\nu=0$ to 1 transition of H$_2$.

In Fig.~\ref{fig:h2natorb} we plot the natural orbitals from a Born-Oppenheimer calculation and the MCTDHF calculation with the same number of orbitals for the $\nu = 0$ state, labeled by  their occupations.  Although some differences may be seen, particularly in the sixth $\sigma$ orbital, the overall impression is that the two sets of natural orbitals are remarkably similar.  That similarity suggests that the electron-nuclear correlation is substantially accounted for by the dependence of the orbitals on $R$ via the prolate spheroidal coordinate system, since they have no other $R$ dependence.  A similar conclusion can be drawn from an examination of the natural orbitals for the state averaged calculation (not shown).


\section{Calculation of ionization cross sections}

We calculation ionization probabilities and cross sections using the flux formalism of J\"ackle and Meyer~\cite{jae96:6778}.  Three MCTHDF steps are involved:
(1)  relaxation to the ground initial state,  (2) propagation from $t=-T$ to $t=0$ in which a pulse of duration $T$ is applied, and 
(3) propagation of $\Psi(0)$ forward in time until the ionized portion has been absorbed by complex the ECS grid in $\xi$.   The wave function
propagated during the third step, from $t=0$ onward, is saved and used in the following analysis.

As per Ref.~\cite{jae96:6778}, the total ionized flux at energy $E$ is defined as 
\begin{equation}
\begin{split}
f(E&) = \\
&\int_0^\infty dt \int_0^\infty dt' \ \left\langle \Psi(0) \left\vert e^{i(\hat{H}-E)t'} \hat{F} e^{-i(\hat{H}-E)t} \right\vert \Psi(0) \right\rangle
\end{split}
\end{equation}
The flux operator $\hat{F}$ is defined as the flux through a hypersurface, the region exterior to which corresponds to the breakup
process of interest -- in this case, ionization.
Defining the heaviside operator $\Theta(\vec{r_1}, \vec{r_2}, ...)$ to be  the unit operator in this exterior region and zero within, the
flux operator may be expressed as the commutator of the Hamiltonian with this heaviside function,
\begin{equation}
\hat{F} = i[\hat{H},\Theta] \ .
\end{equation}
Under appropriate assumptions and after some algebra\cite{jae96:6778}, one arrives at
\begin{equation}
f(E) = \int_0^\infty dt \int_0^\infty dt' \ e^{iE(t-t')} \left\langle \Psi(t') \left\vert i(\hat{H}-\hat{H}^\dag) \right\vert \Psi(t) \right\rangle 
\label{eq:flux}
\end{equation}
In this expression, the flux is obtained through matrix elements of the antihermitian part of the Hamiltonian between 
wave functions at different times; in deriving it, we have exploited the fact that the antihermitian part of the Hamiltonian lies only
on the complex part of the ECS contour.  

In the present context, the antihermitian part of the Hamiltonian comes from the exterior complex
scaling in the $\xi$ coordinates, and is nonzero only when at least one electron has reached the FEM-DVR element
that terminates at $\xi_0$, the start of the ECS tail.  As long as $\xi_0$ has been chosen sufficiently large, 
the antihermitian part is nonzero in the region corresponding to single or multiple ionization.  In the MCTDH implementation for heavy particle motion, 
complex absorbing potentials (CAPs) are used instead of ECS in the exterior region to absorb the outgoing part of the
wave function, and the validity of the above equations for the flux depend on the CAP being weak enough not to perturb
the wave function in the inner region.  In contrast, ECS is an analytic continuation of the Hamiltonian to complex coordinates,
and does not perturb the solution in the inner region at all, unless a significant basis set error is present.

In terms of the flux, the integral photoionization cross section (summed over final channels) is 
\begin{equation}
\sigma(E) = \frac{2\pi\alpha \, \omega}{\vert \mathcal{F}(E)\vert^2}f(E) \label{eq:sigma}
\end{equation}
where $\alpha$ is the fine structure constant,  $\omega$ is the photon energy, $\omega = E-E_0$ where $E_0$ is the ground state energy,  and $\mathcal{F}(E)$ is  
the Fourier transform of the pulse from a length gauge calculation, for example,
\begin{equation}
\mathcal{F}(E) = \int_{-T}^0 dt \ \mathcal{E}(t) e^{i(E-E_0)t} \, .
\end{equation}

To evaluate Eq.(\ref{eq:flux}) it is necessary to evaluate the matrix element of $\hat{H}-\hat{H}^\dag$
between wave functions at two different times, comprised of two different sets of orbitals, in an efficient manner.  Although for the present applications to H$_2$ simpler methods would suffice, we use an approach that will be applicable to larger systems.  To that end, we transform to a biorthogonal
set of orbitals, in the spirit of the treatment of Malmqvist~\cite{malmqvist}, after which we may evaluate matrix elements of arbitrary operators, for instance the flux operator,
via Slater's rules for zero, single and double excitations just as in the usual, orthogonal case.
Thus, given orbitals and A-vectors at $t$ and $t'$, we first transform the orbitals $\phi(t')$ into a new set $\varphi(t')$ which obey a biorthonormality
relationship to $\phi(t)$: $\langle \varphi_i(t') \vert \phi_j(t) \rangle  = \delta_{ij}$.  Whereas the MCTDHF orbitals $\phi$ are themselves orthonormal at all times, 
the $\varphi$ functions  alone obey no such relationship,
\begin{equation}
\bm{s}_{ij} = \langle \phi_i(t) \vert \phi_j(t') \rangle \qquad \varphi_i(t') = \sum_j (\bm{s^{-1}})_{ji} \phi_j(t') \, .\\
\end{equation}
The full wave function at time $t'$ has a new A-vector of configuration coefficients -- which we denote as the B-vector, $\vec{B}$ -- corresponding to its expansion in the new biorthogonal orbitals $\varphi(t')$.
In our notation the A-vector corresponds to the configuration basis $\vert \vec{n}(t') \rangle$, and we denote the configurations made from the $\varphi(t')$ orbitals as
$\vert \vec{m} \rangle$, where $\langle \vec{n}(t) \vert \vec{m}(t') \rangle = \delta_{\vec{n}\vec{m}}$.  We solve for $\vec{B}$ via
\begin{equation}
\begin{split}
\Psi(t') &  = \sum_{\vec{m}\vec{n}} B_{\vec{m}}(t') \vert \vec{n} \rangle \langle \vec{n}  \vert \vec{m} \rangle = \sum_{\vec{n}} A_{\vec{n}}(t')  \vert \vec{n} \rangle \\
\vec{A}(t') & = \bm{S}(t') \vec{B}(t') \qquad \bm{S}_{\vec{n}\vec{m}} = \langle \vec{n}(t') \vert \vec{m}(t') \rangle \\
\end{split}
\end{equation}

To solve these equations we must first construct $\bm{S}_{\vec{n}\vec{m}}$, the overlap between configurations defined in terms of nonorthogonal sets of orbitals,  a task which becomes increasingly more demanding as the number of electrons increases.  
We can take advantage of the remarkable fact that for full CI wave functions, although the matrix  $\bm{S}$ is dense,  its logarithm,  $\bm{\ln S}$,  has sparse representations.  In fact, the matrix $\bm{\ln S}$  is not unique, for the same reason that the multibranched complex function $\ln(z)$ is not unique, and it has both sparse and nonsparse  representations.

\begin{figure}
\begin{center}
\begin{tabular}{c}
\resizebox{0.95\columnwidth}{!}{\includegraphics{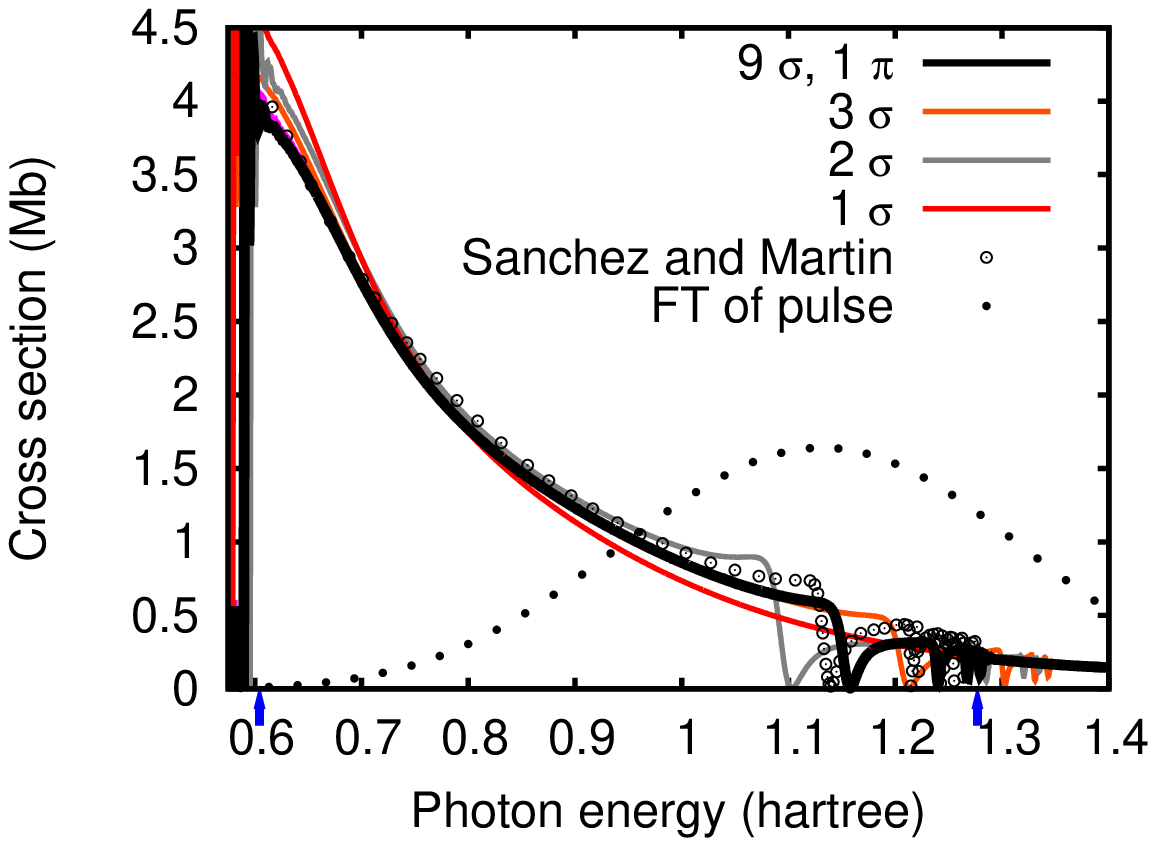}}\\
\resizebox{0.95\columnwidth}{!}{\includegraphics{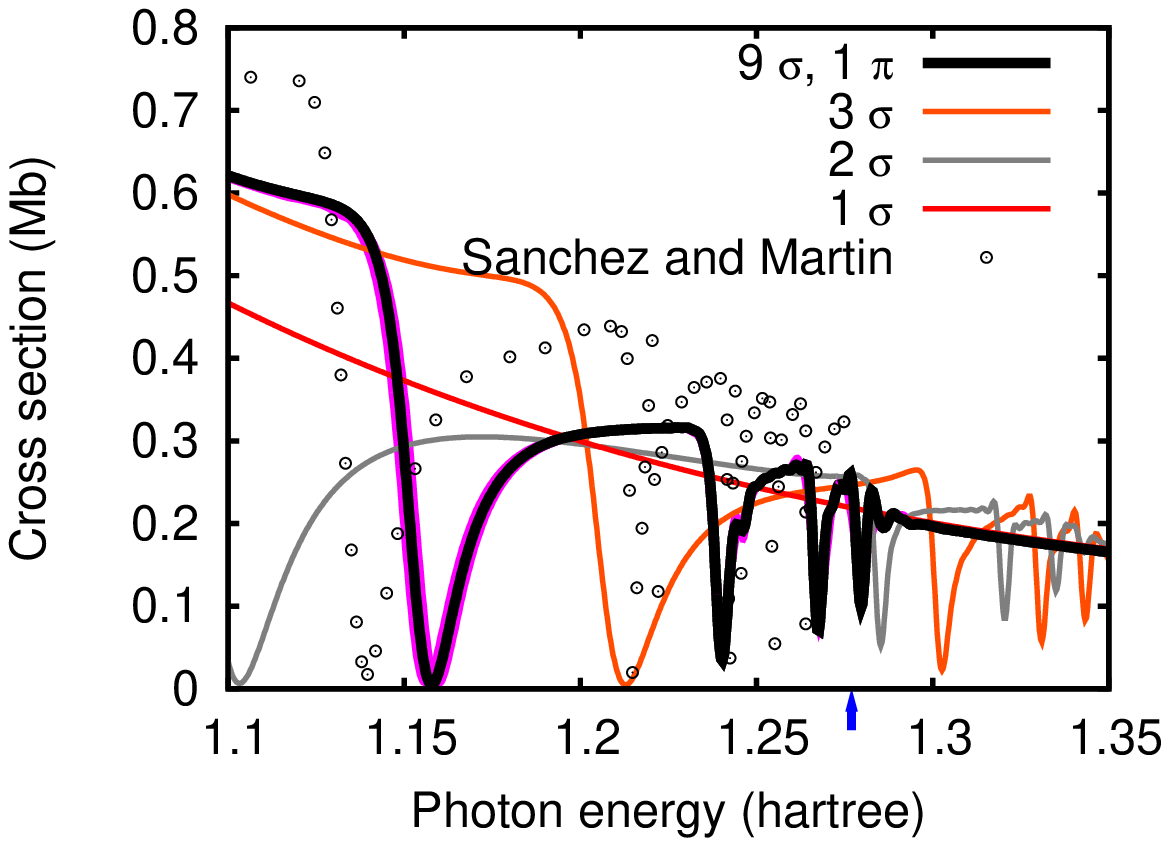}} 
\end{tabular}
\end{center}
\caption{(Color online) Fixed-nuclei ionization cross section of H$_2$, $\Sigma$ symmetry, calculated using the full wave function with one $\pi$ and nine $\sigma$ orbitals, with  
a single pulse of intensity 1$\times 10^{-13}$ W cm$^{-2}$, frequency 1.1 hartree, and  duration 0.5 fs in the length gauge.  Top: cross section from
threshold (0.60449 hartree) to the H$_2^+$ 1 $\Sigma_u$ threshold at 1.28 hartree.  The thresholds are marked with arrows and the Fourier transform of the pulse is plotted with arbitrary units as
dots.  Also shown are the results of Sanchez and Mart\'\i{}n~\cite{sanchezmartin}.  Bottom: magnification of resonance region, including result using only three sigma orbitals.  Several other results are plotted that nearly coincide: lower intensity (1 $\times 10^{-15}$ W cm$^{-2}$); velocity gauge; and more orbitals of sigma, pi, and delta symmetry.   \label{fig:h2ion}}
\end{figure}

The nonzero matrix elements of a sparse branch of $\bm{\ln S}$ may be
calculated by applying Slater's rules for the matrix elements of a
one-electron operator such as the kinetic energy, which usually are
applied to a determinant basis constructed of (bi)orthonormal
orbitals.  The matrix element $\bm{ \ln S }_{\vec{n}\vec{m}}$ with respect to configurations
$\vert \vec{n}_a(t)\rangle$ and $\vert \vec{m}(t') \rangle$ is zero if these configurations differ by more than one
index, and otherwise is given by Slater's other rules for a one
electron operator as applied to any branch of the matrix logarithm
$\bm{\ln s}$.


For full CI wave functions, which are employed in the present work,
the solution of the linear equation $\vec{A} = \bm{S} \vec{B}$ can be done in sparse arithmetic using the Krylov-space \verb#expokit#~\cite{expokit} subroutine
\verb#ZGEXPV#, which performs a matrix exponential onto a vector.  The solution is thereby expressed as
\begin{equation}
\vec{B} = \bm{\exp(-\ln S)} \vec{A} \ .
\end{equation}

The transformation to the biorthogonal basis being done, we proceed to evaluate matrix elements of the antihermitian parts of the Hamiltonian operators appearing due to 
our use of exterior complex scaling, calculating orbital matrix elements and assembling them into the configuration matrix elements in the same
manner as in constructing the A-vector Hamiltonian matrix $\bm{H}$.  

The results for ionization of H$_2$ in $\Sigma$ symmetry (polarization parallel to the bond axis) are shown in Fig.~\ref{fig:h2ion}.  We find that calculation is converged  at a total of one $\pi$ and nine $\sigma$ orbitals.  The other parameters of the calculation 
are given in the caption to Fig.~\ref{fig:h2ion}.  In obtaining Eq.~\ref{eq:flux} we assumed that $\langle \psi(0) \vert \psi(t) \rangle = \langle \psi(0) \vert \psi(-t) \rangle^*$,
and via backwards time propagation, we have verified that this identity is obeyed in general for these MCTDHF wave functions, at least to one part in 10$^{-4}$.
To eliminate the oscillations of the Gibbs phenomenon in the Fourier transform over a finite interval we additionally multiply Eq.(\ref{eq:flux}) by a sinusoids in $t$ and $t'$, as $\cos \frac{t\pi}{2T}$, where T is the time for which
we propagate the wave function after the pulse.  The result is converged to visual accuracy within approximately 1600 atomic time units.

\begin{figure}
\begin{center}
\resizebox{0.9\columnwidth}{!}{\includegraphics{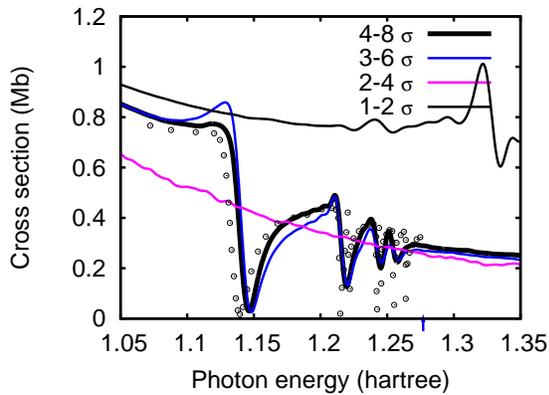}}
\end{center}
\caption{Fixed-nuclei ionization cross section of H$_2$, $\Sigma$ symmetry, using the treatment of 
Eq.(\ref{eq:psiprime}) to solve for only the perturbation to the wave function: results for one through four initial state $\sigma$ orbitals, with twice that number for the pulse and flux
steps, using the same pulse parameters as in Fig.~\ref{fig:h2ion}. Circles: results of Sanchez and Mart\'\i{}n~\cite{sanchezmartin}.
\label{fig:psiprime}}
\end{figure}

In Fig.~\ref{fig:h2ion} we also plot the results for one, two, and three $\sigma$ orbitals only.  All reproduce the overall magnitude and shape of the cross section,
but are incorrect in the energy range where the autoionizing resonances appear.  The one-orbital treatment is featureless there, as the parent H$_2^+$ ($\Sigma_u$) state is not represented in the basis,
but otherwise correct.   The two and three-orbital treatments reproduce the Fano lineshapes of the resonances, but place them at incorrect energies,  and additionally their locations do not appear to converge until the nine $\sigma$, one $\pi$ result shown in black.

However, the result converges to a cross section slightly different than the accurate results of Sanchez and Mart\'\i{}n~\cite{sanchezmartin}.  We may be able to understand this numerical behavior by realizing that at these intensities, only 
about ${1}/{1000}$th of the wave function has been ionized, and that these calculations have not converged that portion.   This slow convergence behavior would seem to be a problem for the utility of the MCTDHF method for describing photoionization or other perturbative processes in general.  One would like to treat perturbative problems just as well as nonperturbative ones, but the variational ansatz of the MCTDHF wave function
will use the variational flexibility in the calculation to optimize the larger, unperturbed (initial state) portion of the wave function at the expense of the smaller components in which we are more interested.  

In the limit of a large number of orbitals, the MCTDHF equation should converge to the exact result.
It is likely that this number is much larger than  we have used in the calculations shown in Fig.~\ref{fig:h2ion},  as additional orbitals are likely to mostly further optimize the correlation within the ground state 
until enough have been added so that the occupation numbers of the natural orbitals describing ground state correlation have fallen at least two orders of magnitude.  There is, however, an alternative approach.

\section{More efficient calculation of ionization}

The problem that additional orbitals mostly serve to improve the description of the initial state  is particular to the present application to calculate a perturbative result, and 
more intense-field applications would not suffer from it.  This state of affairs in unsatisfactory, but fortunately 
there is a straightforward solution to this problem.  We can calculate not $\Psi(t)$, the full wave function, but the quantity we label $\Psi'(t)$, the change in the wave function due to the pulse:
\begin{equation}
\begin{split}
\Psi(t) & = e^{-iE_0t} \Psi(0) + \Psi'(t) \\
i \frac{\partial}{\partial t} \Psi'(t) & = H(t) \Psi'(t) + V(t) e^{-iE_0t} \Psi(0) \label{eq:psiprime}
\end{split}
\end{equation}
where $E_0$ is the initial state eigenvalue and $V(t)$ is the perturbation.  
This formulation modifies the MCTDHF working equations, Eqs.(\ref{eq:orbEOM}) and (\ref{eq:Aequation}), introducing driving terms to each, but we were not immediately 
able to implement the orbital driving term in a numerically stable way.  We thus double the orbital dimension at the start of the pulse, generating additional orbitals by operating
with the dipole operator upon the occupied orbitals of the initial state, and causing the orbital driving term  to equal zero at the start of the pulse.  We then calculate $\Psi^\circ(t) \equiv e^{iE_0t}\Psi'(t)$
by modifying Eq.(\ref{eq:Aequation}) as
\begin{equation}
i \frac{\partial}{\partial t} \vec{A^\circ}(t) =  ({\bf H}(t) -  E_0)  \vec{A^\circ}(t)  + {\bf V}(t) \vec{A}(0) \, ,
\end{equation}
where the matrix ${\bf V}$ is the matrix of the dipole operator in the nonorthogonal basis of orbitals at times $t$ and 0, ${\bf V}_{\vec{n}\vec{n'}}=\langle \vec{n}(t) \vert \hat{\mu} \vert \vec{n}'(0)\rangle$.  
This equation is solved with routine \verb#ZGPHIV# in the \verb#expokit# package\cite{expokit}.

The results are significantly better than in our first treatment, calculating the entire wave function, although we find that we cannot accurately calculate
the entire range between the H$_2$ ionization thresholds with the single pulse we used for the calculation of the full wave function $\Psi(t)$.   We focus on the resonance region, where
our $\hbar \omega$ = 1.1 hartree, 0.5 fs pulse is centered.  The results are again insensitive to intensity across the whole energy range.  

In Fig.~\ref{fig:psiprime} we show that the ionization
cross section converges to essentially its correct value using $\sigma$ orbitals only.  In our treatment we double the number of orbitals going from the ground initial state to the propagation steps of the calculation; thus the minimum is two. 
We can see that
the minimum one orbital (Hartree-Fock) ground state, two propagation orbital treatment yields an ionization cross section without the correct resonance features; the two orbitals of the propagation of $\Psi'(t)$ correspond to the ground H$_2^+$ 1$\sigma_g$ cation state and the wavepacket ionized in its field.  The resonances, which are based on the 1$\sigma_u$ cation state, are thus not represented.
initial, six propagation orbitals, the resonances  appear  in essentially their correct locations.  Two additional propagation (one additional initial) orbitals are all that is needed to give good agreement with the calculations of  Sanchez and Mart\'\i{}n~\cite{sanchezmartin}.

\section{Conclusion and outlook \label{sec9}}

We have explored the formulation of the MCTDHF approach both for fixed nuclei and including nuclear motion for application to any diatomic molecule within the nonrelativistic approximation.  Furthermore methods for overcoming several important technical barriers to such calculations have been demonstrated.  The use of prolate spheroidal coordinates, and an expansion of the wave function including nuclear motion in terms of configurations of orbitals which depend on $R$ only through the dependence of the underlying coordinates on internuclear distance are crucial parts of the strategy described  here.  We have demonstrated that the use of such orbitals gives a rapidly convergent representation of the vibrational states of diatomics in calculations that avoid the Born-Oppenheimer approximation.    Furthermore, we have shown that photoionization cross sections can be extracted from the MCTDHF wave functions in full dimensionality, and demonstrated that accurate photoionization cross sections can be calculated using a method for solving for only the perturbation caused by a time-dependent potential.   The methods we describe here should be immediately applicable to calculation of the single (or total) ionization of larger molecules, as well as studies of Auger spectra with or without nuclear motion.

\begin{acknowledgments}
This work was performed under the auspices of the US Department of Energy
by the University of California Lawrence Berkeley National Laboratory
under Contract DE-AC02-05CH11231 and
was supported by the U.S. DOE Office of Basic Energy
Sciences, Division of Chemical Sciences.
\end{acknowledgments}

\appendix

\section{One electron matrix elements}
\label{sec:1ematrices}

We follow Ref.~\cite{prolate3_2010} in the derivation of the matrix elements of the electronic Hamiltonian in our prolate spheroidal DVR basis, and provide 
additional formulas for the nonadiabatic terms.  For details on the use of FEM-DVR, 
see Ref.~\cite{TopicalReview04}.  We evaluate all the integrals by quadrature; the only non-polynomial terms approximated by quadrature are
the repulsive inverse integer powers that appear, for example, in the centrifugal potentials.

We refer to the matrix elements using the notation of Eq.(\ref{eq:shorthand}) representing the Hamiltonian we use here, which is exact for $J=0$ except for the
omission of the two-electron terms in $\hat{l^2}$.  The Hamiltonian is otherwise accurate except for Coriolis coupling for $J\ne 0$.  In the equations
below, $f_{i\alpha}^\mathcal{M}$ refers to an electronic FEM-DVR product basis function of Eq.(\ref{eq:DVRbasis}), $\chi$ to a 
primitive DVR function of Eq.(\ref{eq:prim_DVR}), and $\chi'$ to its first derivative.  

The matrix elements in $R$ are straightforward,
\begin{equation}
\begin{split}
 \mathbf{T}^R_{ij} & = \left\langle \chi_i(R) \left\vert - \frac{1}{2} \frac{\partial^2}{\partial R^2} \right\vert \chi_j(R) \right\rangle \\
 & \approx \frac{1}{2}\sum_k w_k \chi'(i)(R_k) \chi'_j(R_k) \\
\mathbf{D}^R_{ij}   & = \left\langle \chi_i(R) \left\vert  \frac{2}{R} \left( \frac{\partial}{\partial R} \right)_{\xi\eta\phi} - \frac{1}{R^2} \right\vert  \chi_j(R) \right\rangle \\
& \approx \frac{1}{R_i}\chi'_j(R_i) - \frac{1}{R_j} \chi'_i(R_j)
\end{split}
\end{equation}
The electronic matrix elements
\begin{equation}
\begin{split}
 \mathbf{T}^{el}_{iajb \mathcal{M}}  & =  \left\langle f_{ia}^\mathcal{M} \left\vert  \frac{-R^2}{2\mu_e} \nabla^2 + \frac{1}{2\mu_R}\left[ -\left(\hat{Y} + {\small \frac{3}{2}}\right)^2 + \hat{l}^2\right] \right\vert f_{jb}^\mathcal{M} \right\rangle \\
\mathbf{D}^{el}_{ia jb \mathcal{M}}  & =  \left\langle f_{ia}^\mathcal{M} \left\vert   \hat{Y}+\frac{3}{2}\right\vert f_{jb}^\mathcal{M} \right\rangle
\end{split}
\end{equation}
involving the operators
{\small
\begin{equation}
\begin{split}
R^2 \delsq &= \frac{4}{(\xi^2-\eta^2)} \left[ \frac{\partial}{\partial \xi} (\xi^2-1)  \frac{\partial}{\partial \xi} + \frac{\partial}{\partial \eta} (1-\eta^2) \frac{\partial}{\partial \eta}\right] \nonumber \\
\hat{Y} &= \frac{(\xi+\alpha\eta)(\xi^2-1)}{\xi^2 - \eta^2} \frac{\partial}{\partial \xi} + \frac{(\eta+\alpha\xi)(1-\eta^2)}{\xi^2 - \eta^2} \frac{\partial}{\partial \eta}  \nonumber \\
\hat{l}^2 &= \mathcal{M}^2 + \frac{1}{2}\left(\hat{l}^+ \hat{l}^- + \hat{l}^- \hat{l}^+ \right) = \mathcal{M}^2 + \frac{1}{2}\left(\hat{l^-}^\dag \hat{l}^- + \hat{l^+}^\dag \hat{l}^+ \right)\nonumber \\
l^\pm &= e^{\pm i\phi} \left( \pm \frac{\rho(\eta + \alpha\xi)}{\xi^2-\eta^2} \frac{\partial}{\partial \xi} \mp \frac{\rho(\xi + \alpha\eta)}{\xi^2-\eta^2}\frac{\partial}{\partial \eta} -  \frac{\xi\eta + \alpha}{\rho} \mathcal{M} \right)\nonumber \\
\end{split}
\end{equation}}
where we use the shorthand $ \rho \equiv \sqrt{(\xi^2-1)(1-\eta^2)}$, may be expressed using the identities
\begin{equation}
\footnotesize \hat{Y} + \frac{3}{2} = \frac{1}{2} \left( \hat{Y} - \hat{Y}^\dag \right) \quad \left( \hat{Y} + \frac{3}{2} \right)^2 = -\left( \hat{Y} + \frac{3}{2} \right)^\dag \left( \hat{Y} + \frac{3}{2} \right) 
\end{equation} 
as
\begin{equation}
\small
\begin{split}
\mathbf{T}^{el}_{ia jb \mathcal{M}}  &
\approx  \delta_{ij} \sum_c 
\chi'_a(\xi_c)  \chi'_b(\xi_c) w_c \times  \\
&\qquad \left[ 
\frac{   \xi_c^2-1}{2\mu_e}+ 
\frac{(\xi_c+\alpha\eta_i)^2(\xi^2-1)^2 + \rho^2 (\eta+\alpha\xi)^2 }{2\mu_R}
\right] \\
& + \delta_{ab} \sum_k  \chi'_i(\eta_k) \chi_j(\eta_k) w_k \times \\
& \qquad \left[  \frac{1-\eta_k^2}{2\mu_e} + 
\frac{(\eta_k+\alpha\xi_a)^2(1-\eta_k^2)^2 + \rho^2 (\xi+\alpha\eta)^2}{2\mu_R}
 \right]  \\
%
%
 - \frac{3}{2}  \delta _{ij}  \Big( & (\xi_a+\alpha\eta_i)(\xi_a^2-1) \chi'_b(\xi_a)  + (\xi_b+\alpha\eta_i)(\xi_b^2-1) \chi'_a(\xi_b) \Big) \\ 
- \frac{3}{2} \delta_{ab} \Big( & (\eta_i+\alpha\xi_a)(1-\eta_i^2)  \chi'_j(\eta_i)  + (\eta_j+\alpha\xi_a)(1-\eta_j^2)  \chi'_i(\eta_j)  \Big) \\
& + \delta_{ij}\delta_{ab} \left( \frac{9}{4} + \left(\mathcal{M} \frac{\xi_a \eta_i + \alpha}{\rho}\right)^2 \right) \\
\end{split}
\end{equation}
\begin{equation}
\begin{split}
\mathbf{D}^{el}_{ia jb \mathcal{M}}  & \approx  \delta _{ij}  \Big( (\xi_a+\alpha\eta_i)(\xi_a^2-1) \chi'_b(\xi_a) \\
  & \qquad \qquad - (\xi_b+\alpha\eta_i)(\xi_b^2-1) \chi'_a(\xi_b) \Big) \\ 
& + \delta_{ab} \Big( (\eta_i+\alpha\xi_a)(1-\eta_i^2)  \chi'_j(\eta_i) \\
& \qquad \qquad + (\eta_j+\alpha\xi_a)(1-\eta_j^2)  \chi'_i(\eta_j)  \Big) \\
\end{split}
\end{equation}
In the present applications the pulse was parallel to the bond axis, and the dipole operator is given in length and velocity gauge
as
\begin{equation}
\left\langle  f_{ia}^\mathcal{M} \left\vert \hat{\mu} \right\vert f_{jb}^\mathcal{M} \right\rangle  
= \left\langle  f_{ia}^\mathcal{M} \left\vert \frac{z}{R}\right\vert f_{jb}^\mathcal{M} \right\rangle  
\approx  \delta_{ij}\delta_{ab} \frac{\xi_a \eta_i + \alpha}{2}
\end{equation}
and
\begin{equation}
\begin{split}
& \left\langle  f_{ia}^\mathcal{M} \left\vert \hat{\mu}\right\vert f_{jb}^\mathcal{M} \right\rangle  = \left\langle f_{ia}^\mathcal{M} \left\vert R \frac{\partial}{\partial z}\right\vert f_{jb}^\mathcal{M} \right\rangle \approx \\
& 2 \delta_{ij}[ (\xi_a^2-1) (\eta_i +\alpha \xi_a) \chi_b'(\xi_a) 
- (\xi_b^2-1) (\eta_i +\alpha \xi_b) \chi_a'(\xi_b) ] \\
& + 2 \delta_{ab}[ (1-\eta_i^2) (\xi_a + \alpha \eta_i) \chi'_j(\eta_i) 
- (1-\eta_j^2) (\xi_b + \alpha \eta_j) \chi'_j(\eta_i) ]
\end{split}
\end{equation}
respectively.

\section{Two-electron matrix elements}
\label{sec:two-elec-ints}

We can follow the method of Refs.~\cite{TopicalReview04,prolate3_2010} to construct the matrix elements of ${1}/{r_{12}}$ in the DVR basis.  We evaluate the matrix elements of a multipole
expansion of this operator by solving Poisson's equation in the radial ($\xi$) coordinate while evaluating the $\eta$ integrals by quadrature.  The resulting matrix
elements retain the sparsity of the DVR representation, being diagonal in $\xi$ and $\eta$ and off-diagonal only in the $\mathcal{M}$ quantum numbers of the
electrons.  This point is crucial for the present time-dependent application, as it means that the two-electron transformations do not dominate the computational time, which is instead primarily determined by the action of the Jacobian of Eq.(\ref{eq:orbEOM}) onto the orbitals within the mean field step.

We begin by defining regular and irregular Legendre functions~\cite{NBShandbook} with an additional normalization factor, 
\begin{equation}
\begin{split}
&\widetilde{P}_{lm}(\xi)  \equiv N_{lm} P_{lm}(\xi)  = \sqrt{\frac{(2l+1)(l-m)!}{2(l+m)!}} P_{lm}(\xi) \\
&\widetilde{Q}_{lm}(\xi)  \equiv N_{lm} Q_{lm} (\xi) = \sqrt{\frac{(2l+1)(l-m)!}{2(l+m)!}} Q_{lm}(\xi) 
\end{split}
\end{equation}
so that the Neumann expansion of ${1}/{r_{12}}$ may be written
\begin{equation}
\begin{split}
\frac{1}{r_{12}} & = \frac{8\pi}{R} \sum_{lm} (-1)^m \frac{2}{2l+1}  \frac{e^{im\phi_1}}{\sqrt{2\pi}}\frac{e^{-im\phi_2}}{\sqrt{2\pi}} \times \\
& \widetilde{P}_{lm}(\xi_<) \widetilde{Q}_{lm}(\xi_>)  \widetilde{P}_{lm}(\eta_1) \widetilde{P}_{lm}(\eta_2) \label{r12eq}
\end{split}
\end{equation}

Thus to compute the two-electron integrals we require the matrix elements of $\widetilde{P}_{lm}(\xi_<) \widetilde{Q}_{lm}(\xi_>)$ in the  DVR basis in $\xi$.  This function is the Green's function for the following equation:
\begin{equation}
\begin{split}
& \left[ \frac{\partial}{\partial \xi_1}(\xi^2-1)\frac{\partial}{\partial \xi_1} - l(l+1) - \frac{m^2}{\xi_1^2-1} \right]
\widetilde{P}_{lm}(\xi_<) \widetilde{Q}_{lm}(\xi_>) \\
& \qquad = (\xi_1^2-1) W\left(\widetilde{P}_{lm}(\xi_1),\widetilde{Q}_{lm}(\xi_1)\right) \delta(\xi_1 - \xi_2) \, ,
\end{split}
\label{eq:satisfies}
\end{equation}
where $W$
is the Wronskian of the two Legendre functions, which has the value
\begin{equation}
W_{lm}(\xi_1) = \frac{(-1)^m 2^{2m} }{ (1-\xi_1^2)}
\frac{ \Gamma\left(\frac{l+m+2}{2}\right)\Gamma\left(\frac{l+m+1}{2}\right)}
{  \Gamma\left(\frac{l-m+2}{2}\right) \Gamma\left(\frac{l-m+1}{2}\right)}
\end{equation}
Expressing the operator in square brackets in Eq.(\ref{eq:satisfies}) in the DVR basis and approximating the matrix elements of both sides of that equation using the DVR quadrature, we arrive at an expression for the  matrix elements of  $\widetilde{P}_{lm}(\xi_<) \widetilde{Q}_{lm}(\xi_>)$,
\begin{equation}
\begin{split}
R^{lm}_{ab} &\equiv \Big\langle \chi_a \chi_b \Big\vert \frac{2}{2l+1} \widetilde{P}_{lm}(\xi_<) \widetilde{Q}_{lm}(\xi_>) \Big\vert \chi_c \chi_d \Big\rangle  \\
&   =
 \delta_{ac}\delta_{bd}  \Bigg[
 \frac{-2}{2l+1} \frac{\widetilde{P}_{lm}(\xi_a) \widetilde{P}_{lm}(\xi_b) \widetilde{Q}_{lm}(\xi_N)}{ \widetilde{P}_{lm}(\xi_N)} \\
+ & (-1)^m \frac{(l-m)!}{(l+m)!}
\frac{  \Gamma\left(\frac{l+m+2}{2}\right)\Gamma\left(\frac{l+m+1}{2}\right)}
{ \Gamma\left(\frac{l-m+2}{2}\right) \Gamma\left(\frac{l-m+1}{2}\right)}
\frac{\left(T_{lm}^{-1}\right)_{ab}}{\sqrt{w_a w_b}}  \, .
\Bigg] \label{radeq}
\end{split}
\end{equation}
In Eq.(\ref{radeq}),  $N$ is the last Gauss-Radau gridpoint in $\xi$, corresponding to a DVR function discarded to enforce the correct boundary condition at the end of the $\xi$ grid on the solution of Eq.(\ref{eq:satisfies}), and $T_{lm}$ is the  the matrix of the operator in square brackets in that equation in terms of the DVR functions $\chi_i(\xi)$ (for even m) or $\sqrt{(\xi^2-1)/(\xi_i^2-1)}\chi_i(\xi)$ (for odd m), which are normalized to one with respect to integration over $\xi$.  Those matrix elements are
\begin{equation}
\begin{split}
(T_{lm})_{ij}  = &
-\delta_{ij}  \left( \frac{m^2}{\xi_i^2-1} + l(l+1) \right)  \\
& + \sum_k \chi'_i(\xi_k) \chi'_j(\xi_k) w_k (\xi_k^2-1)  
\end{split}
\end{equation}

Using the expression for $R^{lm}_{ab}$ in  Eq. (\ref{radeq}) we obtain the final result for the two electron matrix elements in our DVR basis,
\begin{equation}
\begin{split}
& \Big\langle f_{ia}^{\mathcal{M}_1} f_{jb}^{\mathcal{M}_2}  \left\vert \frac{1}{r_{12}} \right\vert
f_{kc}^{\mathcal{M}_1-m}f_{ld}^{\mathcal{M}_2+m} \Big\rangle  \\
& = \delta_{ac}\delta_{bd}\delta_{ik}\delta_{jl}
\frac{4}{R} \sum_{l=0}^{l_{max}} R^{lm}_{ab} \widetilde{P}_{lm}(\eta_i)\widetilde{P}_{lm}(\eta_j)
\end{split}
\end{equation}
which has exactly the form in Eq.(\ref{eq:two_e_int}). This expression depends also on having used a fixed DVR quadrature to approximate the $\eta_1$ and $\eta_2$ integrations, and a given quadrature order cannot be used for arbitrarily large $l$ values in the Neumann expansion in Eq.(\ref{r12eq}).   In our numerical calculations we use $l_{max} = N_\eta$ where $N_\eta$ is the number of DVR functions in $\eta$.  We have found this choice to be optimal when using this algorithm implemented for spherical polar coordinates, and at least near optimal for the present case of prolate spheroidal coordinates.
\\
\

\bibliography{ProMC}

\begin{thebibliography}{74}
\expandafter\ifx\csname natexlab\endcsname\relax\def\natexlab#1{#1}\fi
\expandafter\ifx\csname bibnamefont\endcsname\relax
  \def\bibnamefont#1{#1}\fi
\expandafter\ifx\csname bibfnamefont\endcsname\relax
  \def\bibfnamefont#1{#1}\fi
\expandafter\ifx\csname citenamefont\endcsname\relax
  \def\citenamefont#1{#1}\fi
\expandafter\ifx\csname url\endcsname\relax
  \def\url#1{\texttt{#1}}\fi
\expandafter\ifx\csname urlprefix\endcsname\relax\def\urlprefix{URL }\fi
\providecommand{\bibinfo}[2]{#2}
\providecommand{\eprint}[2][]{\url{#2}}

\bibitem[{\citenamefont{Krausz and Ivanov}(2009)}]{krausz2009}
\bibinfo{author}{\bibfnamefont{F.}~\bibnamefont{Krausz}} \bibnamefont{and}
  \bibinfo{author}{\bibfnamefont{M.}~\bibnamefont{Ivanov}},
  \bibinfo{journal}{Rev. Mod. Phys.} \textbf{\bibinfo{volume}{81}},
  \bibinfo{pages}{163} (\bibinfo{year}{2009}).

\bibitem[{\citenamefont{Sekikawa et~al.}(2004)\citenamefont{Sekikawa, Kosuge,
  Kanai, and Watanabe}}]{Sekikawa}
\bibinfo{author}{\bibfnamefont{T.}~\bibnamefont{Sekikawa}},
  \bibinfo{author}{\bibfnamefont{A.}~\bibnamefont{Kosuge}},
  \bibinfo{author}{\bibfnamefont{T.}~\bibnamefont{Kanai}}, \bibnamefont{and}
  \bibinfo{author}{\bibfnamefont{S.}~\bibnamefont{Watanabe}},
  \bibinfo{journal}{Nature} \textbf{\bibinfo{volume}{432}},
  \bibinfo{pages}{605} (\bibinfo{year}{2004}).

\bibitem[{\citenamefont{{V. Ayvazyan, {\em et al.}}}(2006)}]{Ayvazyan}
\bibinfo{author}{\bibnamefont{{V. Ayvazyan, {\em et al.}}}},
  \bibinfo{journal}{Eur. Phys. J. D} \textbf{\bibinfo{volume}{37}},
  \bibinfo{pages}{297} (\bibinfo{year}{2006}).

\bibitem[{\citenamefont{Ullrich et~al.}(2003)\citenamefont{Ullrich, Moshammer,
  Dorn, Doerner, Schmidt, and Schmidt-Bocking}}]{doernercoltrims}
\bibinfo{author}{\bibfnamefont{J.}~\bibnamefont{Ullrich}},
  \bibinfo{author}{\bibfnamefont{R.}~\bibnamefont{Moshammer}},
  \bibinfo{author}{\bibfnamefont{A.}~\bibnamefont{Dorn}},
  \bibinfo{author}{\bibfnamefont{R.}~\bibnamefont{Doerner}},
  \bibinfo{author}{\bibfnamefont{L.}~\bibnamefont{Schmidt}}, \bibnamefont{and}
  \bibinfo{author}{\bibfnamefont{H.}~\bibnamefont{Schmidt-Bocking}},
  \bibinfo{journal}{Rep. Prog. Phys.} \textbf{\bibinfo{volume}{66}},
  \bibinfo{pages}{143} (\bibinfo{year}{2003}).

\bibitem[{\citenamefont{Lesius et~al.}(2002)\citenamefont{Lesius, Blanchet,
  Ivanov, and Stolow}}]{stolow_nmd}
\bibinfo{author}{\bibfnamefont{M.}~\bibnamefont{Lesius}},
  \bibinfo{author}{\bibfnamefont{V.}~\bibnamefont{Blanchet}},
  \bibinfo{author}{\bibfnamefont{M.}~\bibnamefont{Ivanov}}, \bibnamefont{and}
  \bibinfo{author}{\bibfnamefont{A.}~\bibnamefont{Stolow}},
  \bibinfo{journal}{J. Chem. Phys.} \textbf{\bibinfo{volume}{117}},
  \bibinfo{pages}{1575} (\bibinfo{year}{2002}).

\bibitem[{\citenamefont{Tanaka and Mukamel}(2002)}]{mukamelxray}
\bibinfo{author}{\bibfnamefont{S.}~\bibnamefont{Tanaka}} \bibnamefont{and}
  \bibinfo{author}{\bibfnamefont{S.}~\bibnamefont{Mukamel}},
  \bibinfo{journal}{Phys. Rev. Lett.} \textbf{\bibinfo{volume}{89}},
  \bibinfo{pages}{043001} (\bibinfo{year}{2002}).

\bibitem[{\citenamefont{Virshup et~al.}(2009)\citenamefont{Virshup, Punwong,
  Pogorelov, Lindquist, Ko, and MartiÌnez}}]{martinezrev}
\bibinfo{author}{\bibfnamefont{A.~M.} \bibnamefont{Virshup}},
  \bibinfo{author}{\bibfnamefont{C.}~\bibnamefont{Punwong}},
  \bibinfo{author}{\bibfnamefont{T.~V.} \bibnamefont{Pogorelov}},
  \bibinfo{author}{\bibfnamefont{B.~A.} \bibnamefont{Lindquist}},
  \bibinfo{author}{\bibfnamefont{C.}~\bibnamefont{Ko}}, \bibnamefont{and}
  \bibinfo{author}{\bibfnamefont{T.~J.} \bibnamefont{MartiÌnez}},
  \bibinfo{journal}{J. of Phys. Chem. B} \textbf{\bibinfo{volume}{113}},
  \bibinfo{pages}{3280} (\bibinfo{year}{2009}).

\bibitem[{\citenamefont{Arasaki et~al.}(2010)\citenamefont{Arasaki, Takatsuka,
  Wang, and McKoy}}]{mckoyno2}
\bibinfo{author}{\bibfnamefont{Y.}~\bibnamefont{Arasaki}},
  \bibinfo{author}{\bibfnamefont{K.}~\bibnamefont{Takatsuka}},
  \bibinfo{author}{\bibfnamefont{K.}~\bibnamefont{Wang}}, \bibnamefont{and}
  \bibinfo{author}{\bibfnamefont{V.}~\bibnamefont{McKoy}}, \bibinfo{journal}{J.
  Chem. Phys.} \textbf{\bibinfo{volume}{132}}, \bibinfo{pages}{124307}
  (\bibinfo{year}{2010}).

\bibitem[{\citenamefont{Hudock et~al.}(2007)\citenamefont{Hudock, Levine,
  Thompson, Satzger, Townsend, Gador, Ullrich, Stolow, and
  Martinez}}]{martinezUT}
\bibinfo{author}{\bibfnamefont{H.}~\bibnamefont{Hudock}},
  \bibinfo{author}{\bibfnamefont{B.}~\bibnamefont{Levine}},
  \bibinfo{author}{\bibfnamefont{A.}~\bibnamefont{Thompson}},
  \bibinfo{author}{\bibfnamefont{H.}~\bibnamefont{Satzger}},
  \bibinfo{author}{\bibfnamefont{D.}~\bibnamefont{Townsend}},
  \bibinfo{author}{\bibfnamefont{N.}~\bibnamefont{Gador}},
  \bibinfo{author}{\bibfnamefont{S.}~\bibnamefont{Ullrich}},
  \bibinfo{author}{\bibfnamefont{A.}~\bibnamefont{Stolow}}, \bibnamefont{and}
  \bibinfo{author}{\bibfnamefont{T.}~\bibnamefont{Martinez}},
  \bibinfo{journal}{J. Phys. Chem. A} \textbf{\bibinfo{volume}{111}},
  \bibinfo{pages}{8500} (\bibinfo{year}{2007}).

\bibitem[{\citenamefont{Eroms et~al.}(2009)\citenamefont{Eroms, Vendrell,
  Jungen, Meyer, and Cederbaum}}]{cederbaum_h2o_auger}
\bibinfo{author}{\bibfnamefont{M.}~\bibnamefont{Eroms}},
  \bibinfo{author}{\bibfnamefont{O.}~\bibnamefont{Vendrell}},
  \bibinfo{author}{\bibfnamefont{M.}~\bibnamefont{Jungen}},
  \bibinfo{author}{\bibfnamefont{H.-D.} \bibnamefont{Meyer}}, \bibnamefont{and}
  \bibinfo{author}{\bibfnamefont{L.}~\bibnamefont{Cederbaum}},
  \bibinfo{journal}{J. Chem. Phys.} \textbf{\bibinfo{volume}{130}},
  \bibinfo{pages}{154307} (\bibinfo{year}{2009}).

\bibitem[{\citenamefont{Nakai}(2007)}]{nakai_nomo_2007}
\bibinfo{author}{\bibfnamefont{H.}~\bibnamefont{Nakai}}, \bibinfo{journal}{Int.
  J. Quantum Chem.} \textbf{\bibinfo{volume}{107}}, \bibinfo{pages}{2849}
  (\bibinfo{year}{2007}).

\bibitem[{\citenamefont{Ishimoto et~al.}(2009)\citenamefont{Ishimoto,
  Tachikawa, and Nagashima}}]{nagashima_mcmo_2009}
\bibinfo{author}{\bibfnamefont{T.}~\bibnamefont{Ishimoto}},
  \bibinfo{author}{\bibfnamefont{M.}~\bibnamefont{Tachikawa}},
  \bibnamefont{and}
  \bibinfo{author}{\bibfnamefont{U.}~\bibnamefont{Nagashima}},
  \bibinfo{journal}{Int. J. Quantum Chem.} \textbf{\bibinfo{volume}{109}},
  \bibinfo{pages}{2677} (\bibinfo{year}{2009}).

\bibitem[{\citenamefont{Kreibich et~al.}(2004)\citenamefont{Kreibich, van
  Leeuwen, and Gross}}]{kreibich_leeuwen_gross2004}
\bibinfo{author}{\bibfnamefont{T.}~\bibnamefont{Kreibich}},
  \bibinfo{author}{\bibfnamefont{R.}~\bibnamefont{van Leeuwen}},
  \bibnamefont{and} \bibinfo{author}{\bibfnamefont{E.}~\bibnamefont{Gross}},
  \bibinfo{journal}{Chem. Phys.} \textbf{\bibinfo{volume}{304}},
  \bibinfo{pages}{183} (\bibinfo{year}{2004}).

\bibitem[{\citenamefont{Yonehara et~al.}(2009)\citenamefont{Yonehara,
  Takahashi, and Takatsuka}}]{takatsuka}
\bibinfo{author}{\bibfnamefont{H.}~\bibnamefont{Yonehara}},
  \bibinfo{author}{\bibfnamefont{S.}~\bibnamefont{Takahashi}},
  \bibnamefont{and}
  \bibinfo{author}{\bibfnamefont{K.}~\bibnamefont{Takatsuka}},
  \bibinfo{journal}{J. Chem. Phys.} \textbf{\bibinfo{volume}{130}},
  \bibinfo{pages}{214113} (\bibinfo{year}{2009}).

\bibitem[{\citenamefont{Alon et~al.}(2007)\citenamefont{Alon, Streltsov, and
  Cederbaum}}]{Cederbaum2007}
\bibinfo{author}{\bibfnamefont{O.~E.} \bibnamefont{Alon}},
  \bibinfo{author}{\bibfnamefont{A.~I.} \bibnamefont{Streltsov}},
  \bibnamefont{and} \bibinfo{author}{\bibfnamefont{L.~S.}
  \bibnamefont{Cederbaum}}, \bibinfo{journal}{J. Chem. Phys.}
  \textbf{\bibinfo{volume}{127}}, \bibinfo{pages}{154103}
  (\bibinfo{year}{2007}).

\bibitem[{\citenamefont{Alon et~al.}(2009)\citenamefont{Alon, Streltsov, and
  Cederbaum}}]{Cederbaum2009}
\bibinfo{author}{\bibfnamefont{O.~E.} \bibnamefont{Alon}},
  \bibinfo{author}{\bibfnamefont{A.~I.} \bibnamefont{Streltsov}},
  \bibnamefont{and} \bibinfo{author}{\bibfnamefont{L.~S.}
  \bibnamefont{Cederbaum}}, \bibinfo{journal}{Phys. Rev. A}
  \textbf{\bibinfo{volume}{79}}, \bibinfo{pages}{022503}
  (\bibinfo{year}{2009}).

\bibitem[{\citenamefont{Kitzler et~al.}(2004)\citenamefont{Kitzler,
  Zanghellini, Jungreuthmayer, Smits, Scrinzi, and
  Brabec}}]{Scrinzi_MCTDHF_2004}
\bibinfo{author}{\bibfnamefont{M.}~\bibnamefont{Kitzler}},
  \bibinfo{author}{\bibfnamefont{J.}~\bibnamefont{Zanghellini}},
  \bibinfo{author}{\bibfnamefont{C.}~\bibnamefont{Jungreuthmayer}},
  \bibinfo{author}{\bibfnamefont{M.}~\bibnamefont{Smits}},
  \bibinfo{author}{\bibfnamefont{A.}~\bibnamefont{Scrinzi}}, \bibnamefont{and}
  \bibinfo{author}{\bibfnamefont{T.}~\bibnamefont{Brabec}},
  \bibinfo{journal}{Phys. Rev. A} \textbf{\bibinfo{volume}{70}},
  \bibinfo{pages}{041401} (\bibinfo{year}{2004}).

\bibitem[{\citenamefont{Caillat et~al.}(2005)\citenamefont{Caillat,
  Zanghellini, Kitzler, Koch, Kreuzer, and Scrinzi}}]{Scrinzi_MCTDHF_2005}
\bibinfo{author}{\bibfnamefont{J.}~\bibnamefont{Caillat}},
  \bibinfo{author}{\bibfnamefont{J.}~\bibnamefont{Zanghellini}},
  \bibinfo{author}{\bibfnamefont{M.}~\bibnamefont{Kitzler}},
  \bibinfo{author}{\bibfnamefont{O.}~\bibnamefont{Koch}},
  \bibinfo{author}{\bibfnamefont{W.}~\bibnamefont{Kreuzer}}, \bibnamefont{and}
  \bibinfo{author}{\bibfnamefont{A.}~\bibnamefont{Scrinzi}},
  \bibinfo{journal}{Phys. Rev. A} \textbf{\bibinfo{volume}{71}},
  \bibinfo{pages}{012712} (\bibinfo{year}{2005}).

\bibitem[{\citenamefont{Kato and Kono}(2008)}]{Kato_Kono2008}
\bibinfo{author}{\bibfnamefont{T.}~\bibnamefont{Kato}} \bibnamefont{and}
  \bibinfo{author}{\bibfnamefont{H.}~\bibnamefont{Kono}}, \bibinfo{journal}{J.
  Chem. Phys.} \textbf{\bibinfo{volume}{128}}, \bibinfo{pages}{184102}
  (\bibinfo{year}{2008}).

\bibitem[{\citenamefont{Kato and Yamanouchi}(2009)}]{KatoYamanouchi2009}
\bibinfo{author}{\bibfnamefont{T.}~\bibnamefont{Kato}} \bibnamefont{and}
  \bibinfo{author}{\bibfnamefont{K.}~\bibnamefont{Yamanouchi}},
  \bibinfo{journal}{J. Chem. Phys.} \textbf{\bibinfo{volume}{131}},
  \bibinfo{pages}{164118} (\bibinfo{year}{2009}).

\bibitem[{\citenamefont{Nest}(2006)}]{Nest2006}
\bibinfo{author}{\bibfnamefont{M.}~\bibnamefont{Nest}}, \bibinfo{journal}{Phys.
  Rev. A} \textbf{\bibinfo{volume}{73}}, \bibinfo{pages}{023613}
  (\bibinfo{year}{2006}).

\bibitem[{\citenamefont{Nest et~al.}(2007)\citenamefont{Nest, Padmanaban, and
  Saalfrank}}]{Nest2007}
\bibinfo{author}{\bibfnamefont{M.}~\bibnamefont{Nest}},
  \bibinfo{author}{\bibfnamefont{R.}~\bibnamefont{Padmanaban}},
  \bibnamefont{and}
  \bibinfo{author}{\bibfnamefont{P.}~\bibnamefont{Saalfrank}},
  \bibinfo{journal}{J. Chem. Phys.} \textbf{\bibinfo{volume}{126}},
  \bibinfo{pages}{214106} (\bibinfo{year}{2007}).

\bibitem[{\citenamefont{Nest et~al.}(2008)\citenamefont{Nest, Remacle, and
  Levine}}]{Levine2008}
\bibinfo{author}{\bibfnamefont{M.}~\bibnamefont{Nest}},
  \bibinfo{author}{\bibfnamefont{F.}~\bibnamefont{Remacle}}, \bibnamefont{and}
  \bibinfo{author}{\bibfnamefont{R.~D.} \bibnamefont{Levine}},
  \bibinfo{journal}{New J. Phys.} \textbf{\bibinfo{volume}{10}},
  \bibinfo{pages}{025019} (\bibinfo{year}{2008}).

\bibitem[{\citenamefont{Nest}(2009)}]{Nest2009}
\bibinfo{author}{\bibfnamefont{M.}~\bibnamefont{Nest}}, \bibinfo{journal}{Chem.
  Phys. Letts.} \textbf{\bibinfo{volume}{472}}, \bibinfo{pages}{171}
  (\bibinfo{year}{2009}).

\bibitem[{\citenamefont{Kucar et~al.}(1987)\citenamefont{Kucar, Meyer, and
  Cederbaum}}]{kucar_tdrh}
\bibinfo{author}{\bibfnamefont{J.}~\bibnamefont{Kucar}},
  \bibinfo{author}{\bibfnamefont{H.-D.} \bibnamefont{Meyer}}, \bibnamefont{and}
  \bibinfo{author}{\bibfnamefont{L.~S.} \bibnamefont{Cederbaum}},
  \bibinfo{journal}{Chem.\ Phys.\ Lett.} \textbf{\bibinfo{volume}{140}},
  \bibinfo{pages}{525} (\bibinfo{year}{1987}).

\bibitem[{\citenamefont{Meyer et~al.}(1990)\citenamefont{Meyer, Manthe, and
  Cederbaum}}]{mey90:73}
\bibinfo{author}{\bibfnamefont{H.-D.} \bibnamefont{Meyer}},
  \bibinfo{author}{\bibfnamefont{U.}~\bibnamefont{Manthe}}, \bibnamefont{and}
  \bibinfo{author}{\bibfnamefont{L.~S.} \bibnamefont{Cederbaum}},
  \bibinfo{journal}{Chem.\ Phys.\ Lett.} \textbf{\bibinfo{volume}{165}},
  \bibinfo{pages}{73} (\bibinfo{year}{1990}).

\bibitem[{\citenamefont{Beck and Meyer}(1997)}]{bec97:113}
\bibinfo{author}{\bibfnamefont{M.~H.} \bibnamefont{Beck}} \bibnamefont{and}
  \bibinfo{author}{\bibfnamefont{H.-D.} \bibnamefont{Meyer}},
  \bibinfo{journal}{Z.~Phys.~D} \textbf{\bibinfo{volume}{42}},
  \bibinfo{pages}{113} (\bibinfo{year}{1997}).

\bibitem[{\citenamefont{Beck et~al.}(2000)\citenamefont{Beck, J{\"a}ckle,
  Worth, and Meyer}}]{Meyer_review}
\bibinfo{author}{\bibfnamefont{M.}~\bibnamefont{Beck}},
  \bibinfo{author}{\bibfnamefont{A.}~\bibnamefont{J{\"a}ckle}},
  \bibinfo{author}{\bibfnamefont{G.}~\bibnamefont{Worth}}, \bibnamefont{and}
  \bibinfo{author}{\bibfnamefont{H.-D.} \bibnamefont{Meyer}},
  \bibinfo{journal}{Physics Reports} \textbf{\bibinfo{volume}{324}},
  \bibinfo{pages}{1} (\bibinfo{year}{2000}).

\bibitem[{\citenamefont{Meyer and Worth}(2003)}]{Meyer_feature}
\bibinfo{author}{\bibfnamefont{H.-D.} \bibnamefont{Meyer}} \bibnamefont{and}
  \bibinfo{author}{\bibfnamefont{G.~A.} \bibnamefont{Worth}},
  \bibinfo{journal}{Theor. Chem. Acc.} \textbf{\bibinfo{volume}{109}},
  \bibinfo{pages}{251} (\bibinfo{year}{2003}).

\bibitem[{\citenamefont{Worth et~al.}()\citenamefont{Worth, Beck, J{\"a}ckle,
  and Meyer}}]{MCTDH_package}
\bibinfo{author}{\bibfnamefont{G.~A.} \bibnamefont{Worth}},
  \bibinfo{author}{\bibfnamefont{M.~H.} \bibnamefont{Beck}},
  \bibinfo{author}{\bibfnamefont{A.}~\bibnamefont{J{\"a}ckle}},
  \bibnamefont{and} \bibinfo{author}{\bibfnamefont{H.-D.} \bibnamefont{Meyer}},
  \bibinfo{howpublished}{The {MCTDH} {P}ackage, {V}ersion 8.3, (2002),
  http://www.pci.uni-heidelberg.de/tc/usr/mctdh/}.

\bibitem[{\citenamefont{Tao et~al.}(2009{\natexlab{a}})\citenamefont{Tao,
  McCurdy, and Rescigno}}]{prolate1_2009}
\bibinfo{author}{\bibfnamefont{L.}~\bibnamefont{Tao}},
  \bibinfo{author}{\bibfnamefont{C.~W.} \bibnamefont{McCurdy}},
  \bibnamefont{and} \bibinfo{author}{\bibfnamefont{T.~N.}
  \bibnamefont{Rescigno}}, \bibinfo{journal}{Phys. Rev. A}
  \textbf{\bibinfo{volume}{79}}, \bibinfo{pages}{012719}
  (\bibinfo{year}{2009}{\natexlab{a}}).

\bibitem[{\citenamefont{Tao et~al.}(2009{\natexlab{b}})\citenamefont{Tao,
  McCurdy, and Rescigno}}]{prolate2_2009}
\bibinfo{author}{\bibfnamefont{L.}~\bibnamefont{Tao}},
  \bibinfo{author}{\bibfnamefont{C.~W.} \bibnamefont{McCurdy}},
  \bibnamefont{and} \bibinfo{author}{\bibfnamefont{T.~N.}
  \bibnamefont{Rescigno}}, \bibinfo{journal}{Phys. Rev. A}
  \textbf{\bibinfo{volume}{80}}, \bibinfo{pages}{013402}
  (\bibinfo{year}{2009}{\natexlab{b}}).

\bibitem[{\citenamefont{Tao et~al.}(2010)\citenamefont{Tao, McCurdy, and
  Rescigno}}]{prolate3_2010}
\bibinfo{author}{\bibfnamefont{L.}~\bibnamefont{Tao}},
  \bibinfo{author}{\bibfnamefont{C.~W.} \bibnamefont{McCurdy}},
  \bibnamefont{and} \bibinfo{author}{\bibfnamefont{T.~N.}
  \bibnamefont{Rescigno}}, \bibinfo{journal}{Phys. Rev. A}
  \textbf{\bibinfo{volume}{82}}, \bibinfo{pages}{023423}
  (\bibinfo{year}{2010}).

\bibitem[{\citenamefont{Moiseyev}(1998)}]{moirev}
\bibinfo{author}{\bibfnamefont{N.}~\bibnamefont{Moiseyev}},
  \bibinfo{journal}{Physics Reports} \textbf{\bibinfo{volume}{302}},
  \bibinfo{pages}{211} (\bibinfo{year}{1998}).

\bibitem[{\citenamefont{McCurdy et~al.}(2004)\citenamefont{McCurdy, Baertschy,
  and Rescigno}}]{TopicalReview04}
\bibinfo{author}{\bibfnamefont{C.~W.} \bibnamefont{McCurdy}},
  \bibinfo{author}{\bibfnamefont{M.}~\bibnamefont{Baertschy}},
  \bibnamefont{and} \bibinfo{author}{\bibfnamefont{T.~N.}
  \bibnamefont{Rescigno}}, \bibinfo{journal}{J. Phys. B}
  \textbf{\bibinfo{volume}{37}}, \bibinfo{pages}{R137} (\bibinfo{year}{2004}).

\bibitem[{\citenamefont{Horner et~al.}(2008)\citenamefont{Horner, McCurdy, and
  Rescigno}}]{Horner2008a}
\bibinfo{author}{\bibfnamefont{D.~A.} \bibnamefont{Horner}},
  \bibinfo{author}{\bibfnamefont{C.~W.} \bibnamefont{McCurdy}},
  \bibnamefont{and} \bibinfo{author}{\bibfnamefont{T.~N.}
  \bibnamefont{Rescigno}}, \bibinfo{journal}{Phys. Rev. A}
  \textbf{\bibinfo{volume}{78}}, \bibinfo{eid}{043416} (\bibinfo{year}{2008}).

\bibitem[{\citenamefont{Palacios et~al.}(2008)\citenamefont{Palacios, Rescigno,
  and McCurdy}}]{palacios08}
\bibinfo{author}{\bibfnamefont{A.}~\bibnamefont{Palacios}},
  \bibinfo{author}{\bibfnamefont{T.~N.} \bibnamefont{Rescigno}},
  \bibnamefont{and} \bibinfo{author}{\bibfnamefont{C.~W.}
  \bibnamefont{McCurdy}}, \bibinfo{journal}{Phys. Rev. A}
  \textbf{\bibinfo{volume}{77}}, \bibinfo{eid}{032716} (\bibinfo{year}{2008}).

\bibitem[{\citenamefont{Palacios et~al.}(2009)\citenamefont{Palacios, Rescigno,
  and McCurdy}}]{palacios09}
\bibinfo{author}{\bibfnamefont{A.}~\bibnamefont{Palacios}},
  \bibinfo{author}{\bibfnamefont{T.~N.} \bibnamefont{Rescigno}},
  \bibnamefont{and} \bibinfo{author}{\bibfnamefont{C.~W.}
  \bibnamefont{McCurdy}}, \bibinfo{journal}{Phys. Rev. A}
  \textbf{\bibinfo{volume}{79}}, \bibinfo{eid}{033402} (\bibinfo{year}{2009}).

\bibitem[{\citenamefont{Vanroose et~al.}(2006)\citenamefont{Vanroose, Horner,
  Mart\'\i{}n, Rescigno, and McCurdy}}]{H2_DPI_vanroose}
\bibinfo{author}{\bibfnamefont{W.}~\bibnamefont{Vanroose}},
  \bibinfo{author}{\bibfnamefont{D.~A.} \bibnamefont{Horner}},
  \bibinfo{author}{\bibfnamefont{F.}~\bibnamefont{Mart\'\i{}n}},
  \bibinfo{author}{\bibfnamefont{T.~N.} \bibnamefont{Rescigno}},
  \bibnamefont{and} \bibinfo{author}{\bibfnamefont{C.~W.}
  \bibnamefont{McCurdy}}, \bibinfo{journal}{Phys. Rev. A}
  \textbf{\bibinfo{volume}{74}}, \bibinfo{pages}{052702}
  (\bibinfo{year}{2006}).

\bibitem[{\citenamefont{Yip et~al.}(2010)\citenamefont{Yip, McCurdy, and
  Rescigno}}]{Yip_2010}
\bibinfo{author}{\bibfnamefont{F.~L.} \bibnamefont{Yip}},
  \bibinfo{author}{\bibfnamefont{C.~W.} \bibnamefont{McCurdy}},
  \bibnamefont{and} \bibinfo{author}{\bibfnamefont{T.~N.}
  \bibnamefont{Rescigno}}, \bibinfo{journal}{Phys. Rev. A}
  \textbf{\bibinfo{volume}{81}}, \bibinfo{pages}{063419}
  (\bibinfo{year}{2010}).

\bibitem[{\citenamefont{Wahl et~al.}(1964)\citenamefont{Wahl, Cade, and
  Roothaan}}]{Roothaan1964}
\bibinfo{author}{\bibfnamefont{A.~C.} \bibnamefont{Wahl}},
  \bibinfo{author}{\bibfnamefont{P.~E.} \bibnamefont{Cade}}, \bibnamefont{and}
  \bibinfo{author}{\bibfnamefont{C.~C.~J.} \bibnamefont{Roothaan}},
  \bibinfo{journal}{J. Chem. Phys.} \textbf{\bibinfo{volume}{41}},
  \bibinfo{pages}{2578} (\bibinfo{year}{1964}).

\bibitem[{\citenamefont{E.~A.~McCullough}(1975)}]{McCullough1975}
\bibinfo{author}{\bibfnamefont{J.}~\bibnamefont{E.~A.~McCullough}},
  \bibinfo{journal}{J. Chem. Phys.} \textbf{\bibinfo{volume}{62}},
  \bibinfo{pages}{3991} (\bibinfo{year}{1975}).

\bibitem[{\citenamefont{Esry and Sadeghpour}(1999)}]{Esry1999}
\bibinfo{author}{\bibfnamefont{B.~D.} \bibnamefont{Esry}} \bibnamefont{and}
  \bibinfo{author}{\bibfnamefont{H.~R.} \bibnamefont{Sadeghpour}},
  \bibinfo{journal}{Phys. Rev. A} \textbf{\bibinfo{volume}{60}},
  \bibinfo{pages}{3604} (\bibinfo{year}{1999}).

\bibitem[{pol()}]{polynote}
\bibinfo{note}{The term polyspherical coordinates denotes a set of (Jacobi or
  Radau) spherical polar vectors defining an orthogonal internal coordinate
  system for the nuclei of a molecule in the center of mass frame.}

\bibitem[{\citenamefont{Gatti et~al.}(2001)\citenamefont{Gatti, Munoz, and
  Iung}}]{gatti}
\bibinfo{author}{\bibfnamefont{F.}~\bibnamefont{Gatti}},
  \bibinfo{author}{\bibfnamefont{C.}~\bibnamefont{Munoz}}, \bibnamefont{and}
  \bibinfo{author}{\bibfnamefont{C.}~\bibnamefont{Iung}}, \bibinfo{journal}{J.
  Chem. Phys.} \textbf{\bibinfo{volume}{114}}, \bibinfo{pages}{8275}
  (\bibinfo{year}{2001}).

\bibitem[{\citenamefont{Smith}(1980)}]{smithradau}
\bibinfo{author}{\bibfnamefont{F.~T.} \bibnamefont{Smith}},
  \bibinfo{journal}{Phys. Rev. Lett.} \textbf{\bibinfo{volume}{45}},
  \bibinfo{pages}{1157} (\bibinfo{year}{1980}).

\bibitem[{\citenamefont{McCurdy et~al.}(1991)\citenamefont{McCurdy, Stroud, and
  Wisinski}}]{Stroud_McCurdy1991}
\bibinfo{author}{\bibfnamefont{C.~W.} \bibnamefont{McCurdy}},
  \bibinfo{author}{\bibfnamefont{C.~K.} \bibnamefont{Stroud}},
  \bibnamefont{and} \bibinfo{author}{\bibfnamefont{M.~K.}
  \bibnamefont{Wisinski}}, \bibinfo{journal}{Phys. Rev. A}
  \textbf{\bibinfo{volume}{43}}, \bibinfo{pages}{5980} (\bibinfo{year}{1991}).

\bibitem[{\citenamefont{Tao et~al.}(2009{\natexlab{c}})\citenamefont{Tao,
  Vanroose, Reps, Rescigno, and McCurdy}}]{Tao_ECS}
\bibinfo{author}{\bibfnamefont{L.}~\bibnamefont{Tao}},
  \bibinfo{author}{\bibfnamefont{W.}~\bibnamefont{Vanroose}},
  \bibinfo{author}{\bibfnamefont{B.}~\bibnamefont{Reps}},
  \bibinfo{author}{\bibfnamefont{T.~N.} \bibnamefont{Rescigno}},
  \bibnamefont{and} \bibinfo{author}{\bibfnamefont{C.~W.}
  \bibnamefont{McCurdy}}, \bibinfo{journal}{Phys. Rev. A}
  \textbf{\bibinfo{volume}{80}}, \bibinfo{pages}{063419}
  (\bibinfo{year}{2009}{\natexlab{c}}).

\bibitem[{\citenamefont{Scrinzi}(2010)}]{Scrinzi2010}
\bibinfo{author}{\bibfnamefont{A.}~\bibnamefont{Scrinzi}},
  \bibinfo{journal}{Phys. Rev. A} \textbf{\bibinfo{volume}{81}},
  \bibinfo{pages}{053845} (\bibinfo{year}{2010}).

\bibitem[{\citenamefont{Rescigno and McCurdy}(2000)}]{dvr00}
\bibinfo{author}{\bibfnamefont{T.~N.} \bibnamefont{Rescigno}} \bibnamefont{and}
  \bibinfo{author}{\bibfnamefont{C.~W.} \bibnamefont{McCurdy}},
  \bibinfo{journal}{Phys. Rev. A} \textbf{\bibinfo{volume}{62}},
  \bibinfo{pages}{032706} (\bibinfo{year}{2000}).

\bibitem[{\citenamefont{Koch and Lubich}(2010)}]{kochlubich}
\bibinfo{author}{\bibfnamefont{O.}~\bibnamefont{Koch}} \bibnamefont{and}
  \bibinfo{author}{\bibfnamefont{C.}~\bibnamefont{Lubich}},
  \bibinfo{journal}{IMA Journal of Numerical Analysis}  (\bibinfo{year}{2010}),
  \urlprefix\url{http://imajna.oxfordjournals.org/content/early/2010/01/31/ima%
num.drp040.abstract}.

\bibitem[{\citenamefont{Koch et~al.}(2009)\citenamefont{Koch, Kreuzer, and
  Scrinzi}}]{kochapprox}
\bibinfo{author}{\bibfnamefont{O.}~\bibnamefont{Koch}},
  \bibinfo{author}{\bibfnamefont{W.}~\bibnamefont{Kreuzer}}, \bibnamefont{and}
  \bibinfo{author}{\bibfnamefont{A.}~\bibnamefont{Scrinzi}},
  \bibinfo{journal}{Appl. Math. Comp.} \textbf{\bibinfo{volume}{173}},
  \bibinfo{pages}{960} (\bibinfo{year}{2009}).

\bibitem[{\citenamefont{Lubich}(2004)}]{lubichvar}
\bibinfo{author}{\bibfnamefont{C.}~\bibnamefont{Lubich}},
  \bibinfo{journal}{Appl. Numerical Mathematics} \textbf{\bibinfo{volume}{48}},
  \bibinfo{pages}{355} (\bibinfo{year}{2004}).

\bibitem[{\citenamefont{Sidje}(1998)}]{expokit}
\bibinfo{author}{\bibfnamefont{R.~B.} \bibnamefont{Sidje}},
  \bibinfo{journal}{ACM Trans. Math. Softw.} \textbf{\bibinfo{volume}{24}},
  \bibinfo{pages}{130} (\bibinfo{year}{1998}).

\bibitem[{\citenamefont{Jensen}(1999)}]{jensen1999}
\bibinfo{author}{\bibfnamefont{F.}~\bibnamefont{Jensen}}, \bibinfo{journal}{J.
  Chem. Phys} \textbf{\bibinfo{volume}{110}}, \bibinfo{pages}{6601}
  (\bibinfo{year}{1999}).

\bibitem[{\citenamefont{Thompson and Williams}(1990)}]{ThompsonWilliams}
\bibinfo{author}{\bibfnamefont{J.}~\bibnamefont{Thompson}} \bibnamefont{and}
  \bibinfo{author}{\bibfnamefont{S.}~\bibnamefont{Williams}},
  \bibinfo{journal}{J. Phys. B: At. Mol. Opt.} \textbf{\bibinfo{volume}{23}},
  \bibinfo{pages}{2205} (\bibinfo{year}{1990}).

\bibitem[{\citenamefont{Kobus}(1993)}]{Kobus}
\bibinfo{author}{\bibfnamefont{J.}~\bibnamefont{Kobus}},
  \bibinfo{journal}{Chem. Phys. Lett.} \textbf{\bibinfo{volume}{202}},
  \bibinfo{pages}{7} (\bibinfo{year}{1993}).

\bibitem[{\citenamefont{Meyer et~al.}(2006)\citenamefont{Meyer, {Le Qu\'er\'e},
  L\'eonard, and Gatti}}]{mey06:179}
\bibinfo{author}{\bibfnamefont{H.-D.} \bibnamefont{Meyer}},
  \bibinfo{author}{\bibfnamefont{F.}~\bibnamefont{{Le Qu\'er\'e}}},
  \bibinfo{author}{\bibfnamefont{C.}~\bibnamefont{L\'eonard}},
  \bibnamefont{and} \bibinfo{author}{\bibfnamefont{F.}~\bibnamefont{Gatti}},
  \bibinfo{journal}{Chem. Phys.} \textbf{\bibinfo{volume}{329}},
  \bibinfo{pages}{179} (\bibinfo{year}{2006}).

\bibitem[{\citenamefont{Doriol et~al.}(2008)\citenamefont{Doriol, Gatti, Iung,
  and Meyer}}]{dor08:224109}
\bibinfo{author}{\bibfnamefont{L.~J.} \bibnamefont{Doriol}},
  \bibinfo{author}{\bibfnamefont{F.}~\bibnamefont{Gatti}},
  \bibinfo{author}{\bibfnamefont{C.}~\bibnamefont{Iung}}, \bibnamefont{and}
  \bibinfo{author}{\bibfnamefont{H.-D.} \bibnamefont{Meyer}},
  \bibinfo{journal}{J. Chem. Phys.} \textbf{\bibinfo{volume}{129}},
  \bibinfo{pages}{224109} (\bibinfo{year}{2008}).

\bibitem[{\citenamefont{Lishka et~al.}(2006)\citenamefont{Lishka, Shepard,
  Shavitt, Pitzer, Dallos, Muller, Szalay, Brown, Ahlrichs, Bohm
  et~al.}}]{col3}
\bibinfo{author}{\bibfnamefont{H.}~\bibnamefont{Lishka}},
  \bibinfo{author}{\bibfnamefont{R.}~\bibnamefont{Shepard}},
  \bibinfo{author}{\bibfnamefont{I.}~\bibnamefont{Shavitt}},
  \bibinfo{author}{\bibfnamefont{R.~M.} \bibnamefont{Pitzer}},
  \bibinfo{author}{\bibfnamefont{M.}~\bibnamefont{Dallos}},
  \bibinfo{author}{\bibfnamefont{T.}~\bibnamefont{Muller}},
  \bibinfo{author}{\bibfnamefont{P.~G.} \bibnamefont{Szalay}},
  \bibinfo{author}{\bibfnamefont{F.~B.} \bibnamefont{Brown}},
  \bibinfo{author}{\bibfnamefont{R.}~\bibnamefont{Ahlrichs}},
  \bibinfo{author}{\bibfnamefont{H.~J.} \bibnamefont{Bohm}},
  \bibnamefont{et~al.}, \emph{\bibinfo{title}{Columbus, an ab initio electronic
  structure program}} (\bibinfo{year}{2006}), \bibinfo{note}{release 5.9.1}.

\bibitem[{\citenamefont{T.H.~Dunning}(1989)}]{dunning}
\bibinfo{author}{\bibfnamefont{J.}~\bibnamefont{T.H.~Dunning}},
  \bibinfo{journal}{J. Chem. Phys.} \textbf{\bibinfo{volume}{90}},
  \bibinfo{pages}{1007} (\bibinfo{year}{1989}).

\bibitem[{\citenamefont{Balint-Kurti et~al.}(1990)\citenamefont{Balint-Kurti,
  Moss, Sadler, and Shapiro}}]{Balint-Kurti1990}
\bibinfo{author}{\bibfnamefont{G.~G.} \bibnamefont{Balint-Kurti}},
  \bibinfo{author}{\bibfnamefont{R.~E.} \bibnamefont{Moss}},
  \bibinfo{author}{\bibfnamefont{I.~A.} \bibnamefont{Sadler}},
  \bibnamefont{and} \bibinfo{author}{\bibfnamefont{M.}~\bibnamefont{Shapiro}},
  \bibinfo{journal}{Phys. Rev. A} \textbf{\bibinfo{volume}{41}},
  \bibinfo{pages}{4913} (\bibinfo{year}{1990}).

\bibitem[{\citenamefont{Bubin and Adamowicz}(2003)}]{bubin2003}
\bibinfo{author}{\bibfnamefont{S.}~\bibnamefont{Bubin}} \bibnamefont{and}
  \bibinfo{author}{\bibfnamefont{L.}~\bibnamefont{Adamowicz}},
  \bibinfo{journal}{J. Chem. Phys.} \textbf{\bibinfo{volume}{118}},
  \bibinfo{pages}{3079} (\bibinfo{year}{2003}).

\bibitem[{\citenamefont{Kinghorn and Adamowicz}(2000)}]{kinghorn2000}
\bibinfo{author}{\bibfnamefont{D.}~\bibnamefont{Kinghorn}} \bibnamefont{and}
  \bibinfo{author}{\bibfnamefont{L.}~\bibnamefont{Adamowicz}},
  \bibinfo{journal}{J. Chem. Phys.} \textbf{\bibinfo{volume}{113}},
  \bibinfo{pages}{4203} (\bibinfo{year}{2000}).

\bibitem[{\citenamefont{Stanke et~al.}(2009)\citenamefont{Stanke, Bubin,
  Molski, and Adamowicz}}]{stankeHD2009}
\bibinfo{author}{\bibfnamefont{M.}~\bibnamefont{Stanke}},
  \bibinfo{author}{\bibfnamefont{S.}~\bibnamefont{Bubin}},
  \bibinfo{author}{\bibfnamefont{M.}~\bibnamefont{Molski}}, \bibnamefont{and}
  \bibinfo{author}{\bibfnamefont{L.}~\bibnamefont{Adamowicz}},
  \bibinfo{journal}{Phys. Rev. A} \textbf{\bibinfo{volume}{79}},
  \bibinfo{pages}{032507} (\bibinfo{year}{2009}).

\bibitem[{\citenamefont{Bubin et~al.}(2005)\citenamefont{Bubin, Adamowicz, and
  Molski}}]{bubin2005}
\bibinfo{author}{\bibfnamefont{S.}~\bibnamefont{Bubin}},
  \bibinfo{author}{\bibfnamefont{L.}~\bibnamefont{Adamowicz}},
  \bibnamefont{and} \bibinfo{author}{\bibfnamefont{M.}~\bibnamefont{Molski}},
  \bibinfo{journal}{J. Chem. Phys.} \textbf{\bibinfo{volume}{123}},
  \bibinfo{pages}{134310} (\bibinfo{year}{2005}).

\bibitem[{\citenamefont{Cooper et~al.}(1985)\citenamefont{Cooper, Gerratt, and
  Raimondi}}]{cooperlih}
\bibinfo{author}{\bibfnamefont{D.}~\bibnamefont{Cooper}},
  \bibinfo{author}{\bibfnamefont{J.}~\bibnamefont{Gerratt}}, \bibnamefont{and}
  \bibinfo{author}{\bibfnamefont{M.}~\bibnamefont{Raimondi}},
  \bibinfo{journal}{Chem. Phys. Lett.} \textbf{\bibinfo{volume}{118}},
  \bibinfo{pages}{580} (\bibinfo{year}{1985}).

\bibitem[{\citenamefont{Ekert and Knight}(1995)}]{Ekert_Knight_1995}
\bibinfo{author}{\bibfnamefont{A.}~\bibnamefont{Ekert}} \bibnamefont{and}
  \bibinfo{author}{\bibfnamefont{P.~L.} \bibnamefont{Knight}},
  \bibinfo{journal}{Am. J. Physics} \textbf{\bibinfo{volume}{63}},
  \bibinfo{pages}{415} (\bibinfo{year}{1995}).

\bibitem[{\citenamefont{Peres}(1993)}]{peres_1993}
\bibinfo{author}{\bibfnamefont{A.}~\bibnamefont{Peres}},
  \emph{\bibinfo{title}{Quantum Theory: Concepts and Methods}}
  (\bibinfo{publisher}{Kluwer Academic}, \bibinfo{address}{Boston},
  \bibinfo{year}{1993}).

\bibitem[{\citenamefont{Nielsen and Chuang}(2000)}]{nielsen_chuang}
\bibinfo{author}{\bibfnamefont{M.~A.} \bibnamefont{Nielsen}} \bibnamefont{and}
  \bibinfo{author}{\bibfnamefont{I.~L.} \bibnamefont{Chuang}},
  \emph{\bibinfo{title}{Quantum Computation and Quantum Information}}
  (\bibinfo{publisher}{Cambridge University Press},
  \bibinfo{address}{Cambridge, England}, \bibinfo{year}{2000}).

\bibitem[{\citenamefont{J{\"a}ckle and Meyer}(1996)}]{jae96:6778}
\bibinfo{author}{\bibfnamefont{A.}~\bibnamefont{J{\"a}ckle}} \bibnamefont{and}
  \bibinfo{author}{\bibfnamefont{H.-D.} \bibnamefont{Meyer}},
  \bibinfo{journal}{J.~Chem.\ Phys.} \textbf{\bibinfo{volume}{105}},
  \bibinfo{pages}{6778} (\bibinfo{year}{1996}).

\bibitem[{\citenamefont{Malmqvist}(2004)}]{malmqvist}
\bibinfo{author}{\bibfnamefont{P.}~\bibnamefont{Malmqvist}},
  \bibinfo{journal}{Int. J. Quant. Chem.} \textbf{\bibinfo{volume}{30}},
  \bibinfo{pages}{479} (\bibinfo{year}{2004}).

\bibitem[{\citenamefont{Sanchez and Mart\'\i{}n}(1997)}]{sanchezmartin}
\bibinfo{author}{\bibfnamefont{I.}~\bibnamefont{Sanchez}} \bibnamefont{and}
  \bibinfo{author}{\bibfnamefont{F.}~\bibnamefont{Mart\'\i{}n}},
  \bibinfo{journal}{J. Phys. B} \textbf{\bibinfo{volume}{30}},
  \bibinfo{pages}{679} (\bibinfo{year}{1997}).

\bibitem[{\citenamefont{Abramowitz and Stegun}(1964)}]{NBShandbook}
\bibinfo{author}{\bibfnamefont{M.}~\bibnamefont{Abramowitz}} \bibnamefont{and}
  \bibinfo{author}{\bibfnamefont{I.~A.} \bibnamefont{Stegun}},
  \emph{\bibinfo{title}{Handbook of Mathematical Functions}}
  (\bibinfo{publisher}{National Bureau of Standards, U. S. Government Printing
  Office}, \bibinfo{address}{Washington, D. C.}, \bibinfo{year}{1964}).

\end{thebibliography}

\end{document}